\documentclass[%
reprint,
 amsmath,
 amssymb,
 aps,
 prx,
]{revtex4-2}

\usepackage{xcolor}
\usepackage[colorlinks=true,citecolor=blue]{hyperref}
\usepackage{graphicx}
\usepackage{subfigure}
\usepackage{bookmark}
\usepackage[normalem]{ulem}
\usepackage{dsfont} 

\newcommand{\thetitle}{Dissipative superradiant spin amplifier for enhanced quantum sensing}
\newcommand{\theauthors}{Martin Koppenh\"ofer$^{1\ast}$, Peter Groszkowski$^{1}$, Hoi-Kwan Lau$^2$, and A.\ A.\ Clerk$^1$}
\newcommand{\theaffiliations}{$^1$Pritzker School of Molecular Engineering, University of Chicago, Chicago, IL, USA\\$^2$Department of Physics, Simon Fraser University, Burnaby, BC, Canada\\$^\ast$To whom correspondence should be addressed; E-mail: koppenhoefer@uchicago.edu}

\begin{document}

\title{\thetitle}
\author{\theauthors}
\affiliation{\theaffiliations}
\date{\today}

\begin{abstract}
Quantum metrology protocols exploiting ensembles of $N$ two-level systems and Ramsey-style measurements are ubiquitous. 
However, in many cases excess readout noise severely degrades the measurement sensitivity; in particular in sensors based on ensembles of solid-state defect spins.  
We present a dissipative ``spin amplification'' protocol that allows one to dramatically improve the sensitivity of such schemes, even in the presence of realistic intrinsic dissipation and noise.
Our method is based on exploiting collective (i.e., superradiant) spin decay, an effect that is usually seen as a nuisance because it limits spin-squeezing protocols. 
We show that our approach can allow a system with a highly imperfect spin readout to approach SQL-like scaling in $N$ within a factor of two, without needing to change the actual readout mechanism. 
Our ideas are compatible with several state-of-the-art experimental platforms where an ensemble of solid-state spins (NV centers, SiV centers) is coupled to a common microwave or mechanical mode.
\end{abstract}

\maketitle


\section{Introduction}
\label{sec:Introduction}

The field of quantum sensing seeks to use the unique properties of quantum states of light and matter to develop powerful new measurement strategies. 
Within this broad field, perhaps the most ubiquitous class of sensors are ensembles of two-level systems. 
Such sensors have been realized in a variety of platforms, including atomic ensembles in cavity QED systems \cite{SchleierSmith2010b,Cox2016,Hosten2016}, and collections of defect spins in semiconductor materials  \cite{Acosta2009,Steinert2010,Pham2011,Wolf2015}. 
They have also been employed to measure a multitude of diverse sensing targets, ranging from magnetometry \cite{Taylor2008,Rondin2014} to the sensing of electric fields \cite{Dolde2011} and even temperature \cite{Acosta2010}. 
Finding new general strategies for improving such sensors could thus have an extremely wide impact. 
A general and well-explored method here is to use collective spin-spin interactions to generate entanglement, with the prototypical example being the creation of spin-squeezed states.  The intrinsic fluctuations of such states can be parametrically smaller than those of a simple product state \cite{Kitagawa1993,Ma2011,Pezze2018}, allowing in principle dramatic improvements in sensitivity. 

Spin squeezing ultimately uses entanglement to suppress fundamental spin projection noise. 
However, this is only a useful strategy in settings where the extrinsic measurement noise associated with the readout of the spin ensemble is smaller than the intrinsic quantum noise of the ensemble's quantum state \cite{Degen2017,Pezze2018}. 
While this limit of ideal readout can be approached in atomic platforms, typical solid-state spin sensors [such as ensembles of nitrogen vacancy (NV) defect center spins that are read out using spin-dependent fluorescence] have measurement noise that is orders of magnitude higher than the fundamental intrinsic quantum noise \cite{Barry2020}.
Thus, in solid-state spin sensors with fluorescence readout, both reducing the readout noise down to the standard quantum limit (SQL) and (in a subsequent step) surpassing the SQL (e.g., using spin squeezing) are major open milestones.
Many experimental efforts have been made to achieve the first one by changing the readout mechanism of the spins \cite{Neumann2010,Shields2015,Jaskula2019,Irber2021,Eisenach2021}.
This strategy typically works well for single or few spins, but projection-noise limited readout of a large ensemble still remains an open problem \cite{Barry2020}.
Here, we propose a different method to reach the first milestone in spin ensembles: 
Starting from extremely large readout noise several orders of magnitude above the SQL, our method reduces the effective readout noise down to a factor of two above the SQL, notably \emph{without} changing the actual fluorescence readout protocol.
We stress that this paper is \emph{not} considering spin-squeezed initial states and sensitivities beyond the SQL, although our method could potentially be extended in this way to approach the Heisenberg limit.

In situations where measurement noise is the key limitation, a potentially more powerful approach than spin squeezing is the complementary strategy of \emph{amplification}: before performing readout, increase the magnitude of the ``signal" encoded in the spin ensemble. 
The amplification then effectively reduces the imprecision resulting from any following measurement noise. 
This strategy is well known in quantum optics \cite{Caves1982,Yurke1986} and is standard when measuring weak electromagnetic signals. 
Different amplification mechanisms have been proposed \cite{Jiang2022,McDonald2020}, but amplification was only recently studied in the spin context \cite{Davis2016,Froewis2016,Macri2016,Davis2017,Haine2018,Anders2018,Schulte2020}.   
Davis \emph{et al.} \cite{Davis2016,Davis2017} demonstrated that the same collective spin-spin interaction commonly used for spin squeezing (the so-called one-axis twist (OAT) interaction) could be harnessed for amplification.
In the absence of dissipation, they showed that their approach allowed near Heisenberg-limited measurement despite having measurement noise that was on par with the projection noise of an unentangled state.
This scheme (which can be viewed as a special kind of more general ``interaction-based readout" protocols \cite{Leibfried2004,Leibfried2005,Nolan2017,Anders2018}) has been implemented in cavity QED \cite{Hosten2016}, Bose-condensed cold atom systems \cite{Linnemann2016}, and atoms trapped in an optical lattice \cite{Colombo2021}; a similar strategy was also used to amplify the displacement of a trapped ion \cite{Burd2019}.

Unfortunately, despite its success in a variety of atomic platforms, the amplification scheme of Ref.~\cite{Davis2016} is ineffective in setups where the spin ensemble consists of simple two-level systems that experience even small levels of $T_1$ relaxation (either intrinsic, or due to the cavity mode used to generate collective interactions). 
As analyzed in the Discussion, the $T_1$ relaxation both causes a degradation of the signal gain and causes the measurement signal to be overwhelmed by a large background contribution.  
This is true even if the single-spin cooperativity is larger than unity. 
Consequently, this approach to spin amplification cannot be used in many systems of interest, including almost all commonly studied solid-state sensing platforms.

In this work, we introduce a conceptually new spin amplification strategy for an ensemble of $N$ two-level systems that overcomes the limitation posed by dissipation. 
Unlike previous work on interaction-based measurement, it does not use collective unitary dynamics for amplification, but instead directly exploits cavity-induced dissipation as the key resource.
We show that the collective decay of a spin ensemble coupled to a lossy bosonic mode gives rise to a signal gain $G$ that exhibits the maximum possible scaling of $G \propto \sqrt{N}$. 
Crucially, in the presence of local dissipation, the amplification in our scheme depends only on the collective cooperativity (not on more restrictive conditions in terms of single-spin cooperativity), and this maximum gain can be reached even in regimes where the single-spin cooperativity is much smaller than unity. 
Moreover, our amplification mechanism has an ``added noise" that approaches the quantum limit one would expect for a bosonic phase-preserving linear amplifier. 
In addition, the scheme is compatible with standard dynamical decoupling techniques to mitigate inhomogeneous broadening.
Our scheme has yet another surprising feature:  in principle, it allows one to achieve an estimation error scaling like $1/\sqrt{N}$ even if one only performs a final readout on a small number of spins $N_A \ll N$. 
Finally, unlike existing unitary amplification protocols, which require the signal to be in a certain spin component \cite{Davis2016,Davis2017}, our scheme amplifies any signal encoded in the transverse polarization of a spin ensemble (similar to phase-preserving amplification in bosonic systems \cite{Clerk2010}).

We stress that in contrast to the majority of interaction-based readout protocols, we are \emph{not} aiming to use entangled states to reach the Heisenberg limit (HL). 
Instead, our goal is to approach the standard quantum limit (SQL) using conventional dissipative spin ensembles, in systems where extrinsic readout noise is extremely large compared to spin projection noise.

It is interesting to also consider our ideas in a broader context.  Our scheme represents a previously unexplored aspect of Dicke superradiance \cite{Dicke1954,Andreev1980,Gross1982,Benedict2018}, a paradigmatic effect in quantum optics.
Superradiance is the collective enhancement of the spontaneous emission of $N$ indistinguishable spins interacting with a common radiation field:
if the spins are initialized in the excited state, quantum interference effects will cause a short superradiant emission burst of amplitude $\propto N^2$ instead of simple exponential decay. 
In contrast to most work on superradiance, our focus is not on properties of the emitted radiation \cite{Dicke1954,Agarwal1970,Rehler1971} or optical amplification \cite{Bohnet2012,Bohnet2013}, but rather on the ``back-action" on the spin system itself.  This back-action directly generates the amplification effect we exploit.  
Somewhat surprisingly, we show that our superradiant amplification mechanism continues to be effective in the limit of dissipation-free unitary dynamics, where the collective physics is described by a standard Tavis-Cummings model.

We stress that our work is also completely distinct from spin-amplification protocols in spintronics and nuclear magnetic resonance (NMR) systems: We are not aiming to measure the state of a single spin by copying it to a large ensemble \cite{Cappellaro2006} or to a distant spin which can be read out more easily \cite{Schaffry2011}.
Instead, our goal is to amplify a signal that is already encoded in the entire spin ensemble. 
On the level of a semiclassical description, superradiance is similar to radiation damping in NMR systems \cite{Bloembergen1954}, which has been proposed as a method to amplify and measure small magnetizations in NMR setups \cite{Augustine2000,Walls2007}. 
However, these protocols cannot be used in quantum metrology (where quantum noise is important) and they use a qualitatively and quantitatively distinct sensing scheme from the ideas we present here (see Supplemental \cite{SM}).


\section{Results}


\subsection{Dissipative gain mechanism and basic sensing protocol}
\label{sec:Res:DissipativeGainMechanism}

\begin{figure*}
	\centering
	\subfigure[]{
		\includegraphics[width=\textwidth]{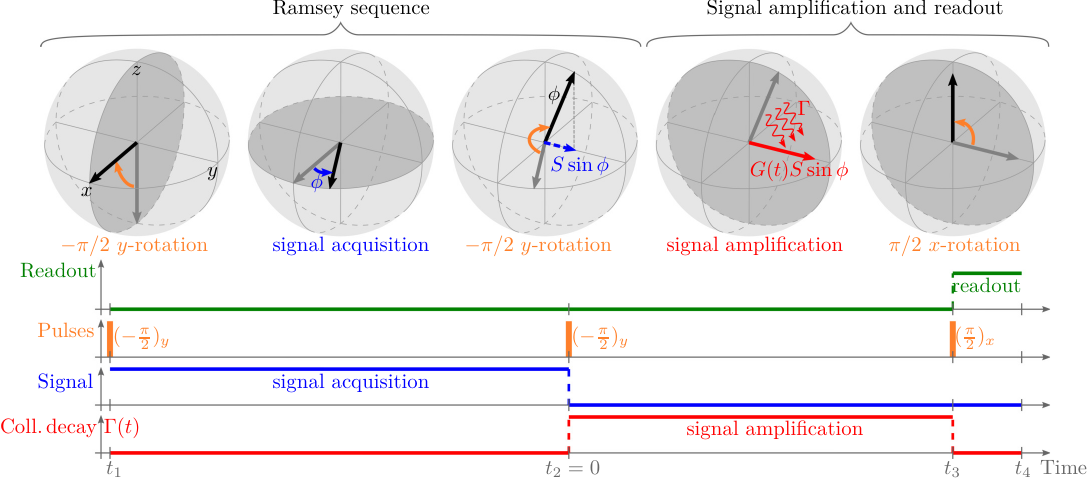}
	}
	\subfigure[]{
        \includegraphics[width=0.48\textwidth]{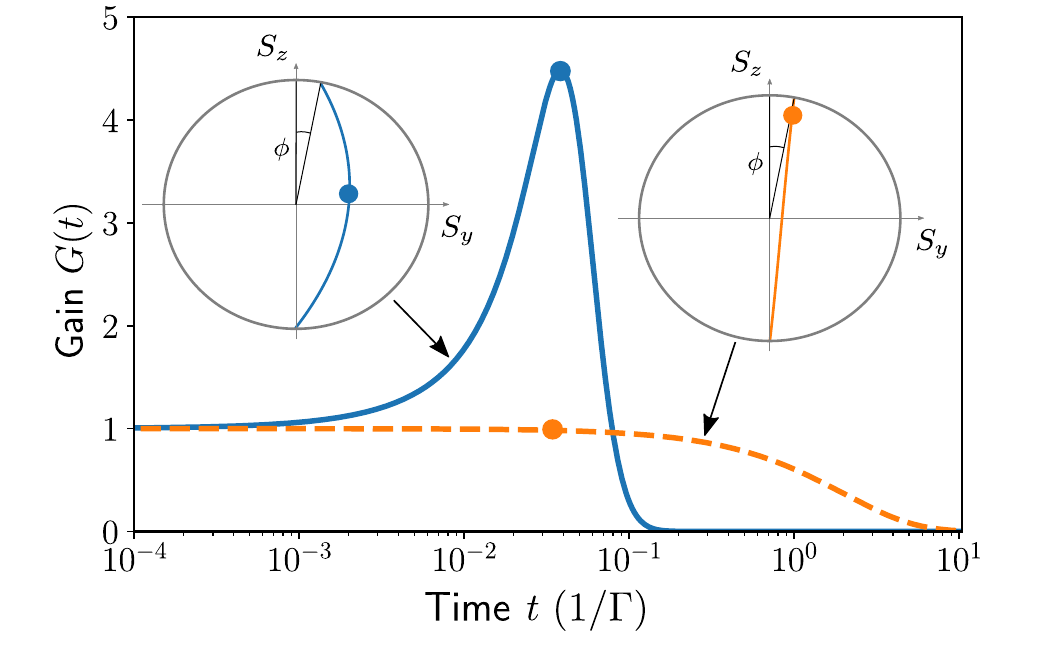}
	}
	\subfigure[]{
        \includegraphics[width=0.48\textwidth]{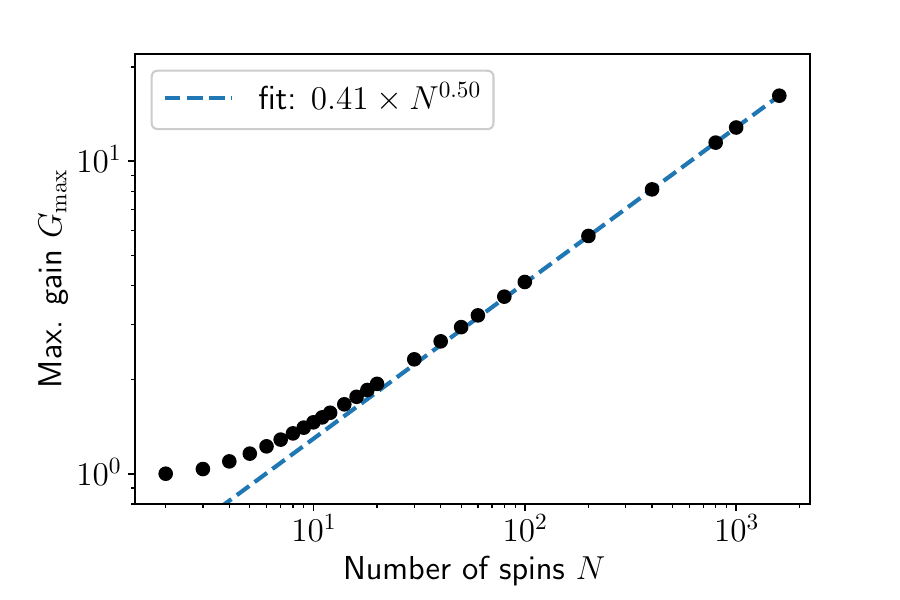}
    }
	\caption{
		\textbf{(a)} 
			Schematic showing the superradiant-amplification-enhanced Ramsey measurement protocol: 
			time evolution of the collective spin vector on the Bloch sphere (top row) and corresponding control operations (bottom row). 
			A standard Ramsey sequence encodes a small phase shift $\phi$ of interest in the $\hat{S}_y$ spin component. 
			The final state of the Ramsey sequence is chosen to be close to the north pole of the collective Bloch sphere to provide energy for the subsequent amplification step. 
			Collective decay [i.e., the second term in Eq.~\eqref{eqn:system:QME_spin_only}] leads to passive transient amplification of the signal by a time-dependent gain factor $G(t)$. 
			The decay is interrupted at the time of maximum gain and a final $\pi/2$ rotation maps the amplified signal onto the $\hat{S}_z$ component for readout.
		\textbf{(b)} 
			Time-dependent gain $G(t)$ (defined in Eq.~\eqref{eqn:amplification:gain}) for collective decay (solid blue curve, numerically exact solution of $\mathrm{d}\hat{\rho}/\mathrm{d}t = \Gamma \mathcal{D}[\hat{S}_-] \hat{\rho}$) vs.~single-spin decay (dashed orange curve, $\mathrm{d} \hat{\rho}/\mathrm{d} t = \Gamma \sum_{j=1}^N \mathcal{D}[\hat{\sigma}_-^{(j)}] \hat{\rho}$).
			Only collective decay leads to transient amplification, i.e., $G(t) > 1$. 
			We use $N=120$ spins and an initial coherent spin state in the $y$-$z$-plane as shown in the third Bloch sphere in (a) with $\phi=10^{-5}$. 
			The insets show sketches of the corresponding trajectories of the spin vector (the initial angle $\phi$ is exaggerated for readability). 
		\textbf{(c)}
			Scaling of the maximum gain $G_\mathrm{max}$ with the number $N$ of spins.
			Black circles are obtained by full numerical solution of the master equation~\eqref{eqn:system:QME_spin_only} for $\gamma_\mathrm{rel}/\Gamma = 0$. 
			The dashed blue line indicates a fit for large $N$.
	}
	\label{fig:protocol}
\end{figure*}

We consider a general sensing setup, where an ensemble of $N$ two-level systems is subject to a global magnetic field whose value we wish to estimate via a standard Ramsey-type measurement protocol (see Fig.~\ref{fig:protocol}(a)). 
This involves starting the ensemble in a fully polarized coherent spin state (CSS) at time $t_1$ and letting it rotate under the signal field by an angle $\phi$, such that the signal $\phi$ is encoded in the value of one component of the collective spin vector (here $\hat{S}_y$) at time $t_2$. 
We assume the standard case of an infinitesimal signal $\phi$, with $\langle \hat{S}_y \rangle$ depending linearly on $\phi$, and we set $t_2 = 0$ for convenience.  
The total error in the estimation of $\phi$ is then given by (see e.g.~\cite{Degen2017,Barry2020}):
\begin{align}
	(\mathbf{\Delta}\phi)^2 
	= (\mathbf{\Delta}\phi)^2_\mathrm{int} + (\mathbf{\Delta}\phi)^2_\mathrm{det} 
	= \frac{(\mathbf{\Delta} S_y)^2 + \Sigma^2_\mathrm{det}}{\vert\partial_\phi \langle \hat{S}_y \rangle\vert^2}~,
	\label{eqn:Protocol:TotalDeltaPhiSquared}
\end{align}
where $(\mathbf{\Delta} S_y)^2 = \langle \hat{S}_y^2 \rangle - \langle \hat{S}_y \rangle^2$. 
The first term $(\mathbf{\Delta}\phi)^2_\mathrm{int}$ is the intrinsic spin-projection noise associated with the quantum state of the ensemble, while the second term $(\mathbf{\Delta}\phi)^2_\mathrm{det}$ describes added noise associated with the imperfect readout of $\hat{S}_y$. 
This additional error can be expressed as an equivalent amount of $\hat{S}_y$ noise, $\Sigma^2_\mathrm{det}$, that is referred back to the signal $\phi$ using the transduction factor $\vert\partial_\phi \langle \hat{S}_y \rangle\vert$.

Consider first the generic situation where the detection noise completely dominates the intrinsic projection noise, $(\mathbf{\Delta}\phi)_\mathrm{det} \gg (\mathbf{\Delta}\phi)_\mathrm{int}$.
This is the typical scenario in many solid-state systems, e.g.~ensembles of NV defects in a diamond crystal whose state is read out using spin-dependent optical fluorescence \cite{Barry2020}. 
The goal is to reduce $(\mathbf{\Delta}\phi)_\mathrm{det}$ without changing the final spin readout mechanism (i.e., $\Sigma_\textrm{det}^2$ remains unchanged).  
The only option available is ``spin amplification", i.e., enhancement of the transduction factor that encodes the sensitivity of the ensemble to $\phi$. 
Specifically, before doing the final readout of $\hat{S}_y$, we want to somehow implement a dynamics that yields
\begin{align}
	\partial_\phi \langle \hat{S}_y(t) \rangle
	= G(t) \partial_\phi \langle \hat{S}_y(0) \rangle ~,
	\label{eqn:Protocol:GainCoefficient}
\end{align}
with a time-dependent gain factor $G(t)$ that is larger than unity at the end of the amplification stage, i.e., $t = t_3$ in Fig.~\ref{fig:protocol}(a).
Achieving large gain will clearly reduce the total estimation error in the regime where measurement noise dominates:  $\mathbf{\Delta} \phi \rightarrow \mathbf{\Delta} \phi / G(t_3)$.
One might worry that in a more general situation, where the intrinsic projection noise is also important, this strategy is not useful, as one might end up amplifying the projection noise far more than the signal. 
We show in Sec.~\ref{sec:Res:ImprovingSensitivity} that this is not the case for our scheme:  
even if we use  the optimal $t_3$ which maximizes the gain $G(t)$, the amplified spin-projection noise referred back to $\phi$ [i.e., $(\mathbf{\Delta} \phi)_{\rm int}$ in Eq.~\eqref{eqn:Protocol:TotalDeltaPhiSquared}] is only approximately twice the value of this quantity in the initial state.  
This is reminiscent of the well-known quantum limit for phase-preserving amplification for bosonic systems \cite{Caves1982,Clerk2010} (see Supplemental 
for a detailed discussion \cite{SM}).

We next focus on what is perhaps the most crucial issue:  how can we  implement amplification dynamics in as simple a way as possible? 
Any kind of amplifier inevitably requires an energy source.  
Here, this will be achieved by preparing the spin ensemble in an excited state.  
For concreteness, we assume that the ensemble has a free Hamiltonian $\hat{H}_0 = \omega \hat{S}_z$, where $\omega > 0$ and $\hbar=1$. 
Hence, at the end of the signal acquisition step at $t=t_2=0$ (see Fig.~\ref{fig:protocol}(a)), we rotate the state such that its polarization is almost entirely in the $+z$ direction (apart from the small rotation caused by the sensing parameter $\phi$), i.e., the ensemble is close to being in its maximally excited state. 
For the following dynamics, we consider simple relaxation of the ensemble towards the ground state of $\hat{H}_0$ (where the net polarization is in the $-z$ direction). 
Consider now a situation where each spin is subject to independent, single-spin $T_1$ relaxation (at rate $\gamma_\mathrm{rel}$) as well as a collective relaxation process (at rate $\Gamma$).
In the rotating frame set by $\hat{H}_0$, the Lindblad master equation governing this dynamics is:
\begin{align}
	\frac{\mathrm{d} \hat{\rho}}{\mathrm{d} t} = 
	\gamma_\mathrm{rel} \sum_{j=1}^N \mathcal{D}[\hat{\sigma}^{(j)}_-] \hat{\rho}
	+ \Gamma \mathcal{D}[ \hat{S}_- ] \hat{\rho}~.
	\label{eqn:system:QME_spin_only}
\end{align}
Here $\hat{S}_- = \sum_{j=1}^N \hat{\sigma}_-^{(j)}$ is the collective spin-lowering operator, 
 $\sigma_-^{(j)} = (\hat{\sigma}_x^{(j)} - i \hat{\sigma}_y^{(j)})/2$ is the lowering operator of spin $j$, 
 $\hat{\sigma}_{x,y,z}^{(j)}$ are the Pauli operators acting on spin $j$, and $\mathcal{D}[\hat{O}]\hat{\rho} = \hat{O}\hat{\rho}\hat{O}^\dagger - \lbrace \hat{O}^\dagger\hat{O},\hat{\rho}\rbrace/2$ is the standard Lindblad dissipation superoperator.

At first glance, it is hard to imagine that such a simple relaxational dynamics will result in anything interesting.  Surprisingly, this is not the case. 
It is straightforward to derive equations of motion that govern the expectation values of $\hat{S}_x$ and $\hat{S}_y$:
\begin{align}
	\frac{\mathrm{d} \langle \hat{S}_{x,y} \rangle}{\mathrm{d} t} 
	= \frac{\Gamma}{2} \left\langle \left\lbrace \hat{S}_z, \hat{S}_{x,y} \right\rbrace - \hat{S}_{x,y} \right\rangle - \frac{\gamma_\mathrm{rel}}{2} \langle \hat{S}_{x,y} \rangle ~.
	\label{eqn:FullEOMSxSy}
\end{align}
Not surprisingly, we see that single-spin relaxation is indeed boring:  it simply causes any initial transverse polarization to decay with time. 
However, the same is not true for the collective dissipation. 
Within a standard mean-field approximation, the first term on the right-hand side of Eq.~\eqref{eqn:FullEOMSxSy} suggests that there will be \emph{exponential growth} of both $\langle \hat{S}_x \rangle$ and $\langle \hat{S}_y\rangle$ at short times if the condition $\langle \hat{S}_z \rangle > 1/2$ holds, i.e., if the spins have a net excitation. 
This is the amplification mechanism that we will exploit, and that we maximize with our chosen initial condition.

The resulting picture is that with collective decay, the relaxation of the ensemble polarization towards the south pole is accompanied (for intermediate times at least) by a \emph{growth} of the initial values of $\langle \hat{S}_{x,y} \rangle$. 
This ``phase-preserving'' (i.e., isotropic in the $S_x$-$S_y$-plane) amplification mechanism will generate a gain $G(t) \geq 1$ that will enhance the subsequent measurement. 
Numerically-exact simulations show that this general picture is correct, see Fig.~\ref{fig:protocol}(b).
One finds that the maximum amplification gain $G(t)$ occurs at a time $t = t_\mathrm{max}$ that approximately coincides with the average polarization vector crossing the equator. 
We stress that the collective nature of the relaxation is crucial: independent $T_1$ decay yields no amplification.  
At a heuristic level, the collective dissipator in Eq.~(\ref{eqn:FullEOMSxSy}) mediates dissipative interactions between different spins, and these interactions are crucial to have gain.

We thus have outlined our basic amplification procedure:  prepare a CSS close to the north pole of the generalized Bloch sphere (with $\phi$ encoded in the small $\hat{S}_x$ and $\hat{S}_y$ components of the polarization), then turn on collective relaxation. 
Stopping the relaxation at time $t = t_\mathrm{max}$ results in the desired amplification of information on $\phi$ in the average spin polarization; this can be then read out as is standard by converting transverse polarization into population differences via a $\pi/2$ rotation, as shown in Fig.~\ref{fig:protocol}(a). 
We stress that the generic ingredients needed here are the same as those needed to realize  OAT spin squeezing and amplification protocols:  a Tavis-Cummings model where the spin ensemble couples to a single, common bosonic mode (a photonic cavity mode \cite{RiedrichMoeller2012,Andrei2012,Lee2012}, or even a mechanical mode \cite{Meesala2016,MacQuarrie2015,Kohler2018,Cady2019}) and time-dependent control over the strength of the collective interaction \cite{Leroux2010,Colombo2021}.
In previously-proposed OAT protocols, cavity loss limits the effectiveness of the scheme, and one thus works with a large cavity-ensemble detuning to minimize its impact.  
In contrast, our scheme utilizes the cavity decay as a key resource, allowing one to operate with a resonant cavity-ensemble coupling.  
In such an implementation, the ability to control the detuning between the cavity and the spin ensemble provides a means to turn on and off the collective decay $\Gamma$.  
This general setup will be analyzed in more in Sec.~\ref{sec:Results:CavityImplementation} and an analysis of timing errors is given in the Supplemental \cite{SM}. 
Alternatively, one can achieve time-dependent control over the collective decay rate by driving Raman transitions in a $\Lambda$-type three-level system \cite{Bowden1978}.

Before proceeding to a more quantitative analysis, we pause to note that, for short times and $\gamma_{\rm rel}=0$, one can directly connect the superradiant spin-amplification physics here to simple phase-preserving bosonic linear amplification. 
Given our initial state, it is convenient to represent the ensemble using a Holstein-Primakoff bosonic mode $\hat{a}$ via $\hat{S}_z \equiv N/2 - \hat{a}^\dagger \hat{a}$.  
For short times, one can linearize the transformation for $\hat{S}_x$ and $\hat{S}_y$, with the result that these are just proportional to the quadratures of $\hat{a}$.  
The same linearization turns the collective decay in Eq.~\eqref{eqn:system:QME_spin_only} into simple bosonic anti-damping:  
$\mathrm{d}\hat{\rho}/\mathrm{d}t \sim \Gamma N \mathcal{D}[\hat{a}^\dagger] \hat{\rho}$. 
This dynamics causes exponential growth of $\langle \hat{a} \rangle$, and describes phase-preserving amplification of a  non-degenerate parametric amplifier in the limit where the idler mode can be adiabatically eliminated \cite{Clerk2010}.
While this linearized picture provides intuition into the origin of gain, it is not sufficient to fully understand our system:  the nonlinearity of the spin system is crucial in determining the non-monotonic behaviour of $G(t)$ shown in Fig.~\ref{fig:protocol}(b), and in determining the maximum gain. 
We explore this more in what follows.

Finally, we note that Eq.~\eqref{eqn:system:QME_spin_only} (with $\gamma_{\rm rel} = 0$) has previously been studied as a spin-only, Markovian description of superradiance, i.e., the collective decay of a collection of two-level atoms coupled to a common radiation field \cite{Agarwal1970,Agarwal1971}. 
The vast majority of studies of superradiance focus on the properties of the radiation emitted by an initially excited collection of atoms. 
We stress that our focus here is very different.
We have no interest in this emission (and will not assume any access to the reservoir responsible for the collective spin dissipation). 
Instead, we use the effective superradiant decay generated by Eq.~\eqref{eqn:system:QME_spin_only} only as a tool to induce nonlinear collective spin dynamics, which can then be used for amplification and quantum metrology.


\subsection{Mean-field theory description of superradiant amplification}
\label{sec:amplification:intuition}

To gain a more quantitative understanding of our nonlinear amplification process, we analyze the dynamics of Eq.~\eqref{eqn:system:QME_spin_only} with $\gamma_{\rm rel} = 0$ using a standard mean-field theory (MFT) decoupling, as detailed in App.~\ref{sec:app:MFT}. 
This analysis goes beyond a linearized bosonic theory obtained from a Holstein-Primakoff transformation and is able to capture aspects of the intrinsic nonlinearity of the spin dynamics. 
We start by using MFT to understand the gain dynamics, which can be determined by considering the evolution of the mean values of the collective spin operator; fluctuations and added-noise physics will be considered in Sec.~\ref{sec:Res:ImprovingSensitivity}.
Note that a simpler approach based on semiclassical equations of motion fails to capture the amplification dynamics correctly, i.e., superradiant amplification is a genuinely quantum effect and quantum fluctuations need to be taken into account (see Supplemental \cite{SM} and Sec.~\ref{sec:Implementations:Undamped}).

As detailed in App.~\ref{sec:app:MFT}, the MFT equation of motion for $S_z \equiv \langle \hat{S}_z \rangle$ in the large-$N$ limit is
\begin{align}
	\frac{\mathrm{d} S_z}{\mathrm{d} t} &= - \Gamma \frac{N^2}{4} - \Gamma S_z (1 - S_z) ~, 
	\label{eqn:amplification:intuition:EomSz} 
\end{align}
where the constant term is obtained by using the fact that the dynamics conserves $\mathbf{\hat{S}}^2$.
Starting from a highly polarized initial state with $S_z(0) = N \cos (\phi)/2$, this equation describes the well-known nonlinear superradiant decay of the $S_z$ component to the steady state $\vert\!\downarrow\rangle^{\otimes N}$ \cite{Rehler1971}. 
The corresponding equations of motion for average values $S_x$ and $S_y$ correspond to the expected decoupling of Eq.~\eqref{eqn:FullEOMSxSy}: 
\begin{align}
	\frac{\mathrm{d} S_{x,y}}{\mathrm{d} t} 
	&= \Gamma \left(S_z - \frac{1}{2} \right) S_{x,y}
	\equiv \lambda(t) S_{x,y} ~,
	\label{eqn:amplification:intuition:EomSy} 
\end{align}
where we have introduced the instantaneous gain rate $\lambda(t)$. 
For $\lambda(t) > 0$ ($\lambda(t) < 0$), any initial polarization component of the collective average Bloch vector in the $x$-$y$ plane will be amplified (damped).  
Without loss of generality, we chose the initial transverse polarization to be entirely in the $y$ direction. 
Thus, the $S_x$ component will always remain zero since the initial state has $S_x(0) = 0$. 
In contrast, the highly polarized initial state $S_z(0) \approx N/2 \gg 1/2$ leads to amplification of the nonzero initial value $S_y(0) = N \sin(\phi)/2$ at short times. 
In the long run, the superradiant decay evolves $S_z(t)$ to its steady-state value $S_z(t \to \infty) = -N/2$.
As a consequence, for sufficiently long times, the time-dependent gain rate $\lambda(t)$ will be reduced and amplification ultimately turns into damping if $S_z(t) < 1/2$. 
The MFT equation of motion~\eqref{eqn:amplification:intuition:EomSy} predicts that maximum amplification of $S_y$ is achieved at the time $t_\mathrm{max}$ where $S_z(t_\mathrm{max}) = 1/2$, which is clearly beyond the regime of applicability of a linearized theory based on the Holstein-Primakoff transformation.
In the large-$N$ limit, the MFT result for $t_\mathrm{max}$ takes the form
\begin{align}
	t_\mathrm{max} = \frac{\ln N}{\Gamma N}~,
	\label{eqn:amplification:intuition:tmax}
\end{align}
which is the well-known delay time of the superradiant emission peak \cite{Rehler1971}.
The short transient period where $\lambda(t)>0$ is enough to yield significant amplification:
\begin{align}
	S_{x,y}(t) 
	&= S_{x,y}(0) e^{\int_0^t \mathrm{d} t' \lambda(t')} \nonumber \\
	&= S_{x,y}(0) \frac{e^{\Gamma t/2} \cosh \left( \frac{1}{2} \ln N \right)}{\cosh \left( \frac{N}{2} \Gamma t - \frac{1}{2} \ln N \right)}~.
	\label{eqn:amplification:Sxydynamics}
\end{align}
Evaluating this at $t=t_\mathrm{max}$ given by Eq.~\eqref{eqn:amplification:intuition:tmax} yields the following MFT result for the maximum value of $S_{x,y}$: 
\begin{align}
	S_{x,y}(t_\mathrm{max}) = \frac{\sqrt{N}}{2} S_{x,y}(0) ~.
	\label{eqn:amplficiation:Sytmax}
\end{align}
Note that the signal gain \emph{increases} with increasing $N$ while the waiting time $t_\mathrm{max}$ required to reach the maximum gain \emph{decreases}, giving rise to very fast amplification. 
Importantly, the optimal amplification time $t_\mathrm{max}$ given in Eq.~\eqref{eqn:amplification:intuition:tmax} is \emph{independent} of the tilt angle $\phi$ in the metrologically relevant limit of $\phi \ll 1$. 
Therefore, the gain $G(t)$ is independent of the signal $\phi$.
The breakdown of this relation defines the dynamic range of the spin amplifier and is analyzed in the Supplemental \cite{SM}. 

\begin{figure*}
	\centering
	\subfigure[]{
		\includegraphics[height=5cm]{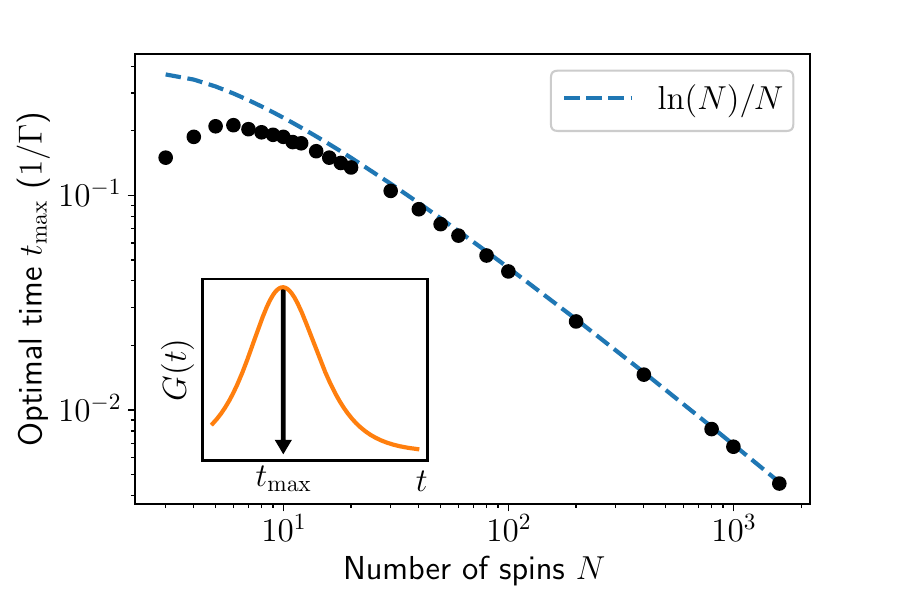}
	}
	\subfigure[]{
		\includegraphics[height=5cm]{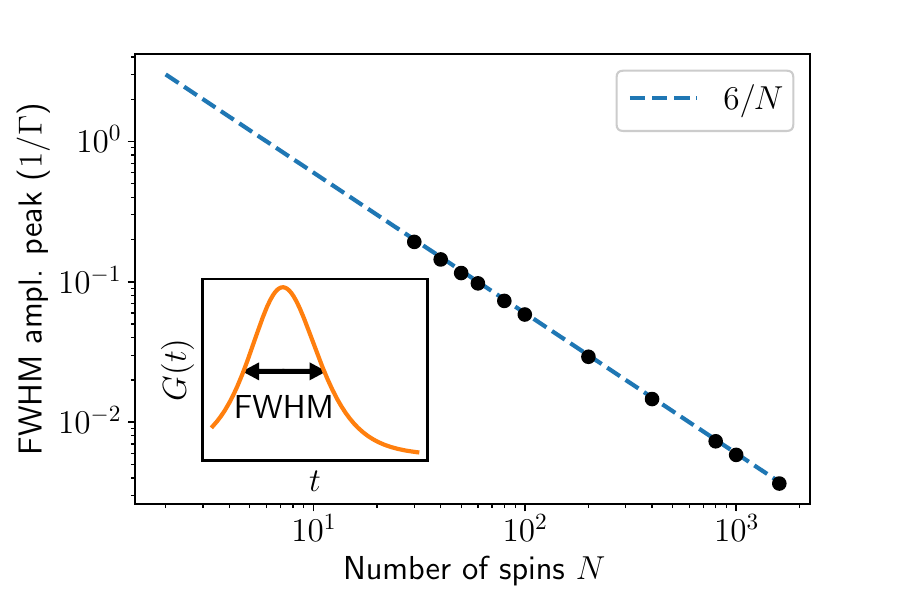}
	}
	\caption{ 
		\textbf{(a)}
			Optimal time $t_\mathrm{max}$ which maximizes the time-dependent gain $G(t)$ defined in Eq.~\eqref{eqn:amplification:gain}. 
		\textbf{(b)}
			Full-width-at-half-maximum of the time-dependent gain $G(t)$. 
		In both panels, the black circles have been obtained by a numerically exact solution of the collective relaxation dynamics generated by the master equation~\eqref{eqn:system:QME_spin_only} for $\gamma_\mathrm{rel}/\Gamma = 0$.
		The blue dashed lines indicate fits of the large-$N$ results based on the scaling laws derived using mean-field theory.
	}
	\label{fig:principle}
\end{figure*}

We now verify this intuitive picture derived from MFT using numerically-exact solutions of Eq.~\eqref{eqn:system:QME_spin_only}.
To analyze the solutions, we define the time-dependent signal gain $G(t)$ as follows:
\begin{align}
	G(t) 
	&= \lim_{\phi \to 0} \frac{\langle \hat{S}_y(t) \rangle}{\langle \hat{S}_y(0) \rangle} \nonumber \\
	G_\mathrm{max} &= \max_{t \geq 0} G(t) = G(t_\mathrm{max}) ~,
	\label{eqn:amplification:gain}
\end{align}
where $t_\mathrm{max}$ is determined numerically. 
Note that this is identical to the definition given in Eq.~\eqref{eqn:Protocol:GainCoefficient}, as  $G(t)$ is independent of $\phi$ for $\phi \ll 1$. 
Combining Eqs.~\eqref{eqn:amplficiation:Sytmax} and~\eqref{eqn:amplification:gain}, we thus expect a scaling $G_\mathrm{max} \propto \sqrt{N}$ based on MFT. 
Numerically-exact master equation simulations shown in Figs.~\ref{fig:protocol}(c) and~\ref{fig:principle} confirm that (up to numerical prefactors) the scaling of $G_\mathrm{max}$ and $t_\mathrm{max}$ predicted by MFT are correct in the large-$N$ limit.

It is also interesting to note that on general grounds, $G_{\rm max} \propto \sqrt{N}$ is the maximal gain scaling that we expect to be possible.  This follows from the fact that we would expect initial fluctuations of $\hat{S}_x$ and $\hat{S}_y$ to be amplified (at least) the same way as the average values of these quantities, and hence expect $(\mathbf{\Delta} S_x)^2 \geq G_\mathrm{max}^2 N/4$, where $N/4$ represents the initial fluctuations of $\hat{S}_x$ in the initial CSS. 
Next, note that because of the finite dimensional Hilbert space, $(\mathbf{\Delta} S_x)^2$ cannot be arbitrarily large and is bounded by $N^2/4$.
This immediately tells us that $G_{\rm max}$ cannot grow with $N$ faster than $\sqrt{N}$.  The gain scaling can also be understood heuristically by using the fact that there is only instantaneous gain for a time $t < t_\mathrm{max} = \ln(N ) / N \Gamma$, and that, during this time period, the instantaneous gain rate is $\lambda(t) \approx  N \Gamma/2$. 
Exponentiating the product of this rate and $t_\mathrm{max}$ again yields a $\sqrt{N}$ scaling.

We stress that the spin-only quantum master equation~\eqref{eqn:system:QME_spin_only} as well as the mean-field results for the behaviour of $S_z$ are well known in the superradiance literature (see e.g.~\cite{Agarwal1970,Agarwal1971,Rehler1971}). 
The new aspect of our work here is to identify the amplification physics associated with superradiant decay, and use MFT to provide a quantiative description of it.  

\subsection{Improving sensitivity and approaching the SQL with extremely bad measurements}
\label{sec:Res:ImprovingSensitivity}

We now discuss how the amplification dynamics can improve the total estimation error $(\mathbf{\Delta}\phi)$ introduced in Eq.~\eqref{eqn:Protocol:TotalDeltaPhiSquared}.
For concreteness, we focus on the general situation where the readout mechanism involves adding independent contributions from each spin in the ensemble, and hence the noise associated with the readout itself scales as $N$: \begin{align}
    \Sigma_\mathrm{det}^2 \equiv \Xi_\mathrm{det}^2 \frac{N}{4}~,
    \label{eqn:Sensitivy:MeasurementNoise}
\end{align}
with $\Xi_\mathrm{det}$ an $N$-independent constant. 
Note that the factor of $1/4$ in the definition is convenient, as $\Xi_\mathrm{det}^2$ directly describes the ratio of readout noise to the intrinsic projection noise.     
Equation~\eqref{eqn:Sensitivy:MeasurementNoise} describes the scaling of readout noise in many practically relevant situations, including standard spin-dependent-fluorescence readout of solid-state spin ensembles \cite{Barry2020} and of trapped ions \cite{Haeffner2008}.  
In this case and for $\phi \ll 1$, one has
\begin{equation}
    \Xi_\mathrm{det}^2 = \frac{1}{\tilde{C}^2 n_\mathrm{avg}}~,
\end{equation}
where $\tilde{C}$ is the fluorescence contrast of the two spin states and $n_\mathrm{avg}$ is the average number of detected photons per spin in a single run of the protocol (see App.~\ref{sec:App:Sensitivity} for details).

\begin{figure}
	\centering
	\subfigure[]{
		\includegraphics[width=0.48\textwidth]{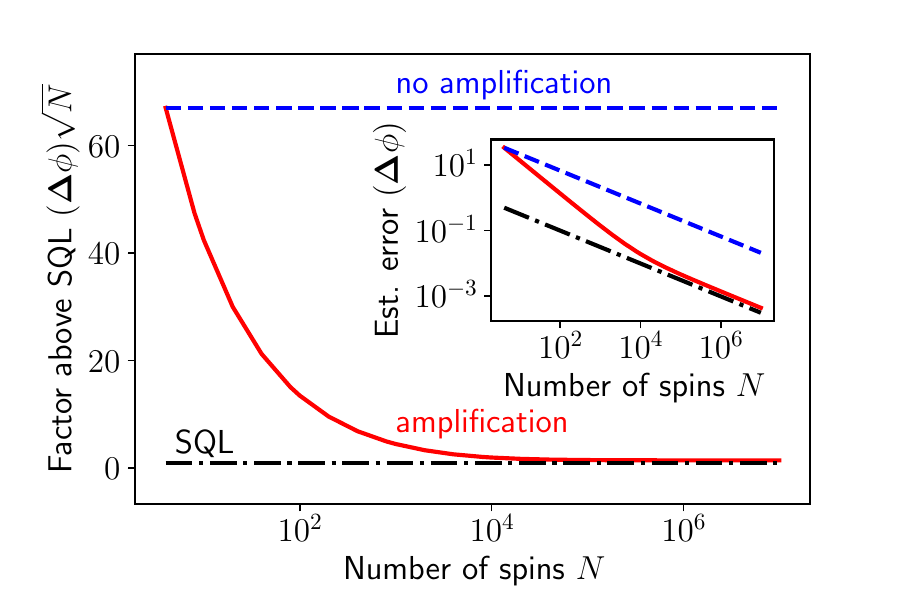}
	}
	\subfigure[]{
		\includegraphics[width=0.48\textwidth]{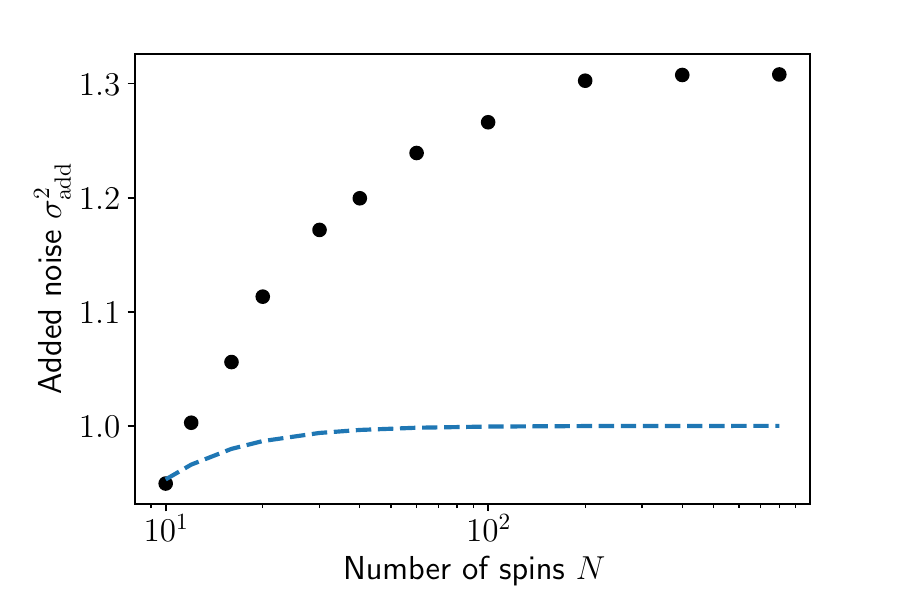}
	}
	\caption{
	\textbf{(a)}
	    Suppression of detection noise by amplification.
	    The best reported fluorescence readout of an NV ensemble is a factor of $\Xi_\mathrm{det} = 1/\sqrt{\tilde{C}^2 n_\mathrm{avg}} = 67$ above the SQL \cite{Barry2020}. 
        We assume the ideal case where this factor is independent of the ensemble size (dashed blue line).
        Amplification suppresses the readout noise (solid red line) and allows one to approach the SQL (dash-dotted black line). 
        The inset shows the scaling of the total estimation error $(\mathbf{\Delta} \phi)^2$ with and without amplification (solid red and dashed blue lines, respectively), and the SQL (dash-dotted black line).
        The curves have been obtained using a MFT analysis of Eq.~\eqref{eqn:system:QME_spin_only} for $\gamma_\mathrm{rel}/\Gamma=0$ and agree qualitatively with numerically exact solutions of the master equation~\eqref{eqn:system:QME_spin_only}, which have been used to calculate the
    \textbf{(b)}
		added noise $\sigma_\mathrm{add}^2$ (defined in Eq.~\eqref{eqn:ReducingEstimationError:GenericModelWithAmplification}) for $\gamma_\mathrm{rel}/\Gamma=0$.
		The dash-dotted blue line illustrates the expected minimum amount of added noise, $1 - 1/G_\mathrm{max}^2 N$, based on a heuristic argument detailed in the Supplemental \cite{SM}. 
		Note that this is not a strict lower bound.
	}
	\label{fig:sensitivity}
\end{figure}

In considering the estimation error, we will also now account for the fact that our amplification mechanism will not only cause $\langle \hat{S}_y \rangle$ to grow, but also cause the variance $(\mathbf{\Delta} S_y)^2$ to grow over its initial CSS value of $N/4$. 
The very best case is that the variance is amplified exactly the same way as the signal, but in general there will be excess fluctuations beyond this. 
This motivates the definition of the \emph{added noise} of our amplification scheme (similar to the definition of the added noise $\sigma_\mathrm{add}$ of a linear amplifier).  
Letting $(\mathbf{\Delta} \hat{S}_y)^2\vert_\mathrm{amp}$ denote the variance of $\hat{S}_y$ in the final, post-amplification state of the spin ensemble after an optimal amplification time, we write: 
\begin{align}
    \left. (\mathbf{\Delta} S_y)^2 \right\vert_\mathrm{amp} \equiv G_\mathrm{max}^2 \frac{N}{4} \left(1 + \sigma^2_\mathrm{add} \right)~.
    \label{eqn:Sensitivity:AddedNoise}
\end{align}
We have normalized $\sigma_\mathrm{add}$ to the value of the CSS variance; hence, $\sigma_\mathrm{add}^2 = 1$ corresponds to effectively doubling the initial fluctuations (once the gain has been included).

For linear bosonic phase-preserving amplifiers, it is well known that the added noise of a phase-preserving amplifier is at best the size of the vacuum noise \cite{Haus1962,Caves1982,Clerk2010}. 
At a fundamental level, this can be attributed to the dynamics amplifying both quadratures of the input signal, quantities that are described by non-commuting operators.  
One might expect a similar constraint here, as our spin amplifier also amplifies two non-commuting quantities (namely $\hat{S}_x$ and $\hat{S}_y$). 
Hence, one might expect that the best we can achieve in our spin amplifier is to have the added noise satisfy $\sigma^2_\mathrm{add} = 1$. 
A heuristic argument that parallels Caves' classic calculation \cite{Caves1982} suggests one indeed has the constraint $\sigma_\mathrm{add}^2 \geq 1 - 1/G^2(T) N$ (see Supplemental \cite{SM}). 
For our system, full master equation simulations let us investigate how the added noise behaves for large $N$ and maximum amplification. 
Remarkably, we find $\sigma_\mathrm{add}^2 \approx 1.3$ in the large-$N$ limit, which is just slightly above the expected level based on the heuristic argument (see Fig.~\ref{fig:sensitivity}(b)). 
This leads to a crucial conclusion:  our amplification scheme is useful even if one cares about approaching the SQL.

Note that the amplified fluctuations in Eq.~\eqref{eqn:Sensitivity:AddedNoise} can at most be $N^2/4$ due to the finite dimensionality of the Hilbert space.
Using the numerical result $G_\mathrm{max} = c_0 \sqrt{N}$ where $c_0 \approx 0.42$ (see Fig.~\ref{fig:protocol}(c)), one can derive an \emph{upper} bound on the added noise:
\begin{align}
    \sigma_\mathrm{add}^2 \leq \frac{1}{c_0^2} - 1 \approx 4.7~.
\end{align}

With the above definitions in hand, we can finally quantify the estimation error in Eq.~\eqref{eqn:Protocol:TotalDeltaPhiSquared} of our amplification-assisted measurement protocol. 
Combining Eqs.~\eqref{eqn:Protocol:GainCoefficient}, \eqref{eqn:Sensitivy:MeasurementNoise}, and~\eqref{eqn:Sensitivity:AddedNoise}, one finds that the general expression applied to our scheme reduces to:
\begin{align}
	(\mathbf{\Delta}\phi)^2 
	    &= \frac{1}{N} \left[ 1 + \sigma_\mathrm{add}^2 +  \frac{\Xi^2_\mathrm{det}}{G_{\rm max}^2} \right] \nonumber \\
	    &= \frac{1}{N} \left[ 1 + \sigma_\mathrm{add}^2 + \frac{ \Xi^2_\mathrm{det}/c_0^2}{N} \right] ~,
	\label{eqn:ReducingEstimationError:GenericModelWithAmplification}
\end{align}
where we have used the large-$N$ scaling of the maximum gain in the last equation:  $G_\mathrm{max} = c_0 \sqrt{N}$ with $c_0 \approx 0.42$.

There are two crucial things to note here:
First, if readout noise completely dominates (despite the amplification), our amplification approach changes the scaling of the estimation error $(\mathbf{\Delta} \phi)$ with the number of spins from $1/\sqrt{N}$ to $1/N$. 
While this scaling is reminiscent of Heisenberg-limited scaling, there is no connection:  in our case, this rapid scaling with $N$ only holds if one is far from the SQL. 
Nonetheless, this shows the potential of amplification to dramatically increase sensitivity in this readout-limited regime.

Second, for very large $N \gg \Xi^2_\mathrm{det}$, the amplification protocol will make the added measurement noise negligible compared to the fundamental noise of the quantum state. 
In this limit, the total estimation error almost reaches the SQL:  it scales as $(\mathbf{\Delta} \phi) \propto \sqrt{(1 + \sigma^2_\mathrm{add})/N} \approx \sqrt{2.3/N}$. 
This is only off by a numerical prefactor $\sqrt{2.3}$ from the exact SQL. 
We thus have established another key feature of our scheme:  using amplification and a large enough ensemble, one can in principle approach the SQL within a factor of two regardless of how bad the spin readout is.  
For a fixed detector noise $\Xi_\mathrm{det}$ , the crossover in the estimation error $(\mathbf{\Delta}\phi)$ from a $1/N$ scaling to a $1/\sqrt{N}$ scaling is illustrated in Fig.~\ref{fig:sensitivity}(a).

\subsection{Enhanced sensitivity despite reading out a small number of spins}

There are many practical situations where, even though the signal of interest $\phi$ influences all $N$ spins in the ensemble, one can only read out the state of a small subensemble $A$ with $N_A \ll N$ spins.
For example, for fluorescence readout of an NV spin ensemble, the spot size of the laser could be much smaller than the spatial extent of the entire ensemble.  
For a standard Ramsey scheme (i.e., no superradiant amplification), there are no correlations between spins, and the  unmeasured $N-N_A$ spins do not help in improving the measurement.
In the best case, the estimation error then scales as $\mathbf{\Delta} \phi \propto 1 / \sqrt{N_A}$.
Surprisingly, the situation is radically different if we first implement superradiant amplification on the full ensemble before reading out the state of the small subensemble.  
In this case, we are able to achieve an SQL-like scaling $\mathbf{\Delta}\phi \propto 1/\sqrt{N}$ even though one measures only $N_A \ll N$ spins.  
This dramatically improved scaling reflects the fact that the superradiant amplification involves a dissipative interaction between all the spins, hence the final state of the small subensemble is sensitive to the total number of spins $N$.

To analyze this few-spin readout scenario, we partition the $N$ spins into two subensembles $A$ and $B$ of size $N_A$ and $N_B \equiv N - N_A$, respectively. 
Without loss of generality, we enumerate the spins starting with subensemble $A$, which allows us to define the subensemble operators $\hat{S}_k^A = \sum_{j=1}^{N_A} \hat{\sigma}_k^{(j)}/2$ and $\hat{S}_k^B = \sum_{j=N_A+1}^N \hat{\sigma}_k^{(j)}/2$, where $k \in \{x,y,z\}$.
Their sum is the spin operator of the full ensemble, $\hat{S}_k = \hat{S}_k^A + \hat{S}_k^B$.  
We now consider a scheme where only the spin state of the $A$ ensemble is measured at the very end of the amplification protocol shown in Fig.~\ref{fig:protocol}.  
The statistics of this measurement are controlled by the operator $\hat{S}^A_y$, with the signal encoded in its average value. 
Note that our ideal amplification dynamics always results in a spin state that is fully permutation-symmetric (i.e., at any instant in time, the average value of single spin operators are identical for all spins).  
It thus follows immediately that the subensemble gain is identical to the gain associated with the full ensemble:
\begin{align}
    \langle \hat{S}^A_y(t_\mathrm{max}) \rangle = G_\mathrm{max} \langle \hat{S}_y^A(0) \rangle = G_\mathrm{max} \frac{N_A}{2} \phi ~,
\end{align}
with $G_\mathrm{max} = c_0 \sqrt{N}$ in the large-$N$ limit (and $c_0 \approx 0.42$).  
We stress that the gain is determined by the size of the full ensemble even though we are only measuring $N_A \ll N$ spins, which can also be seen by inspecting the equations of motion for the transverse components $\hat{\sigma}_{x,y}^{(k)}$ of an arbitrary spin $k$:  
\begin{align}
    \frac{\mathrm{d} \langle \hat{\sigma}_{x,y}^{(k)} \rangle}{\mathrm{d} t} = \frac{\Gamma}{2} \langle \{ \hat{S}_{x,y},  \hat{\sigma}_z^{(k)} \}\rangle - \frac{\Gamma}{2} \langle \hat{\sigma}_{x,y}^{(k)} \rangle~.
\end{align}
The $y$ component of each individual spin is driven by a \emph{collective} spin operator $\hat{S}_{y}$ whose expectation value is proportional to the ensemble size, $\langle \hat{S}_y \rangle = N \phi/2$.

Next, consider the fluctuations in $\hat{S}^A_y$. 
The variance of this operator must be less than $N_A^2/4$ in any state; we thus parameterize these fluctuations by  $(\mathbf{\Delta}S_y^A)^2(t_\mathrm{max}) = q N_A^2/4$ where $q \leq 1$. 
If we now only consider the fundamental spin projection noise (i.e., ignore any additional readout noise), we can combine these results to write the estimation error in $\phi$ as:
\begin{align}
    (\mathbf{\Delta} \phi)^2_\mathrm{int} 
    = \frac{(\mathbf{\Delta} S_y^A)^2(t_\mathrm{max})}{\vert \partial_\phi \langle \hat{S}_y^A(t_\mathrm{max}) \rangle \vert^2} 
    = \frac{q}{G_\mathrm{max}^2}
    = \frac{q/c_0^2}{N}
    \leq \frac{1/c_0^2}{N}~.
\end{align}
We thus have a crucial result:  even in the worst-case scenario $q=1$, for large $N$, our estimation error scales as $1/N$ despite measuring $N_A \ll N$ spins.

We can use a similar analysis to consider the contribution of detection noise to the estimation error in our subensemble readout scheme.  
We again assume (as is appropriate for fluorescence readout) that the detector noise scales with the number spins that are read out, i.e., $\Sigma_\mathrm{det,A}^2 = \Xi_\mathrm{det}^2 N_A/4$. 
We thus obtain the detection-noise contribution to the estimation error:
\begin{align}
    (\mathbf{\Delta}\phi)^2_\mathrm{det} 
    = \frac{\Sigma_\mathrm{det,A}^2}{\vert \partial_\phi \langle \hat{S}_y^A(t_\mathrm{max}) \rangle \vert^2}
    = \frac{\Xi_\mathrm{det}^2}{G^2_\mathrm{max} N_A}
    = \frac{\Xi_\mathrm{det}^2/c_0^2 }{N_A \, N}~,
\end{align}
i.e., the detection noise is again suppressed by a factor of $N$, the size of the full ensemble.

\begin{figure*}
    \centering
    \subfigure[]{
        \includegraphics[width=0.48\textwidth]{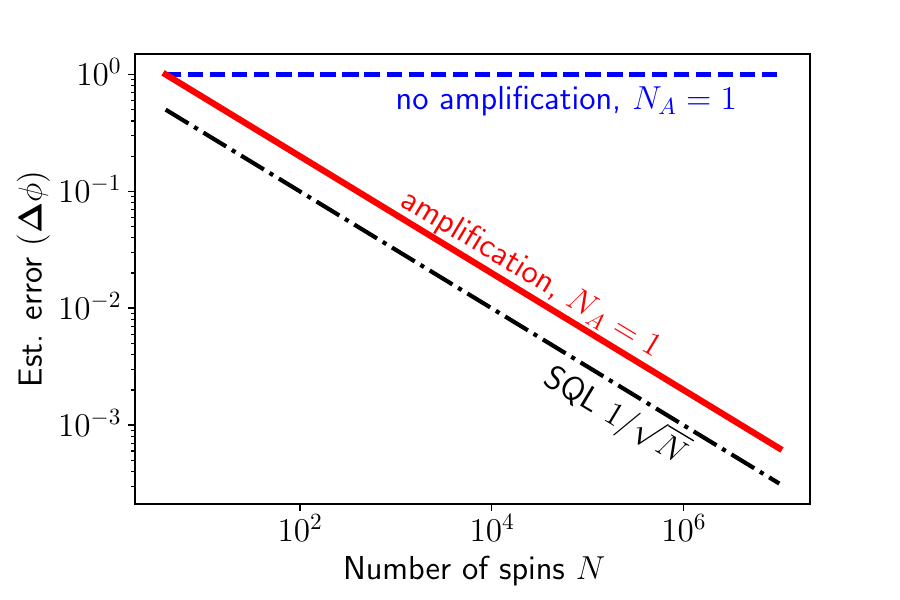}
    }
    \subfigure[]{
        \includegraphics[width=0.48\textwidth]{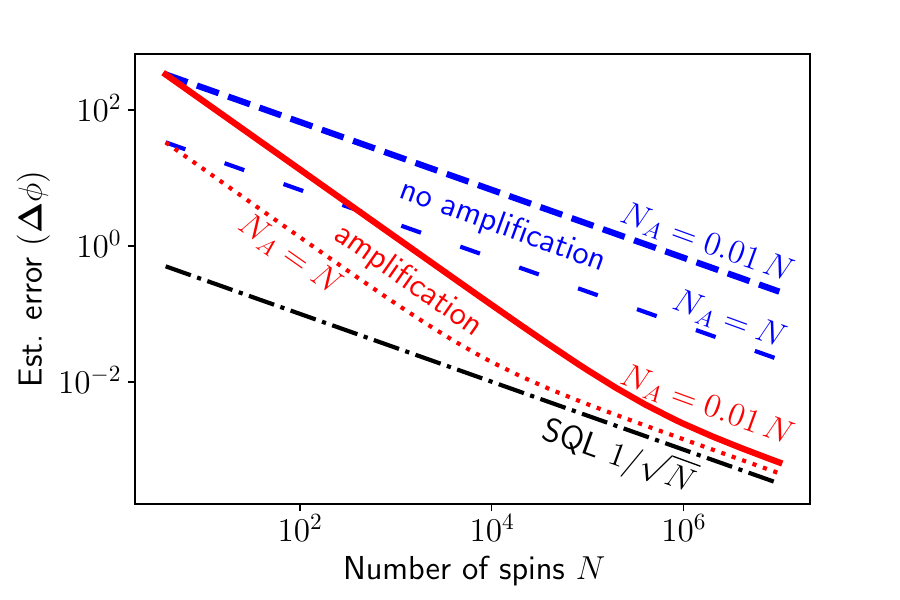}
    }
    \caption{
        Estimation error $(\mathbf{\Delta} \phi)$ if only a subensemble of size $N_A < N$ is read out after the amplification step, see Eq.~\eqref{eqn:subensembles:DeltaPhi}.
        The superradiant decay dynamics involves all $N$ spins, therefore, the gain factor still scales $\propto \sqrt{N}$.
        \textbf{(a)} 
            As a consequence, even if only a \emph{single} spin is measured, $N_A=1$, amplification allows one to reduce the estimation error with a SQL-like scaling $(\mathbf{\Delta}\phi) \propto 1/\sqrt{N}$. 
            Here, we consider the ideal case $\Xi_\mathrm{det}^2 = 0$; adding detection noise will only change a constant prefactor which shifts the dashed blue and solid red curves vertically relative to the dash-dotted SQL curve.
        \textbf{(b)}
            Comparison between a measurement of the full ensemble, $N_A = N$, and a measurement of only a small subensemble, $N_A = 0.01 N$, in the presence of detection noise, $\Xi_\mathrm{det} = 67$. 
            For the subensemble, the initial estimation error is higher due to the smaller number of measured spins but the gain still allows one to reduce the readout noise with a $1/N$-like scaling until intrinsic and added noise become appreciable. 
        The plots are based on Eq.~\eqref{eqn:subensembles:DeltaPhi}, the MFT scaling relations (i.e., $c_0 = 1/2$, $\sigma_\mathrm{add}^2=1$), and a worst-case estimate $q=1$ (equivalent to maximum $(\mathbf{\Delta}S_y^A)^2$ fluctuations of the measured subensemble). 
    }
    \label{fig:subensembles}
\end{figure*}

Combining these results, we find
\begin{align}
    (\mathbf{\Delta}\phi) = \frac{1}{c_0 \sqrt{N}} \sqrt{q + \frac{\Xi_\mathrm{det}^2}{ N_A}}
    \leq
    \frac{1}{c_0 \sqrt{N}} \sqrt{1 + \frac{\Xi_\mathrm{det}^2}{ N_A}}
    ~,
    \label{eqn:subensembles:DeltaPhi}
\end{align}
where $c_0 \approx 0.42$ in the large-$N$ limit.  
We thus find that, in the case where $N_A$ is held fixed while $N$ is increased, our superradiant amplification scheme yields a measurement sensitivity that scales as $(\mathbf{\Delta}\phi) \propto 1 / \sqrt{N}$.  
Surprisingly, it is controlled by the \emph{full} size of the ensemble, and not controlled by the much smaller number of spins that are actually measured, $N_A$. 
We illustrate this in Fig.~\ref{fig:subensembles} for the extreme case of readout of a single spin, $N_A=1$, and for the case of readout of a small fraction of the spin ensemble, $N_A = 0.01 N$.
We stress that the analysis above (like the analysis throughout this paper) is done in the limit of an infinitesimally small signal $\phi$.

\subsection{Impact of single-spin dissipation and finite-temperature in the generic model}

While our superradiant dissipative spin amplifier exhibits remarkable performance in the ideal case where the only dissipation is the desirable collective loss in Eq.~\eqref{eqn:system:QME_spin_only}, it is also crucial to understand what happens when additional unwanted forms of common dissipation are added.

\subsubsection{Local dissipation}

\begin{figure*}
	\centering
	\subfigure[]{
		\includegraphics[width=0.48\textwidth]{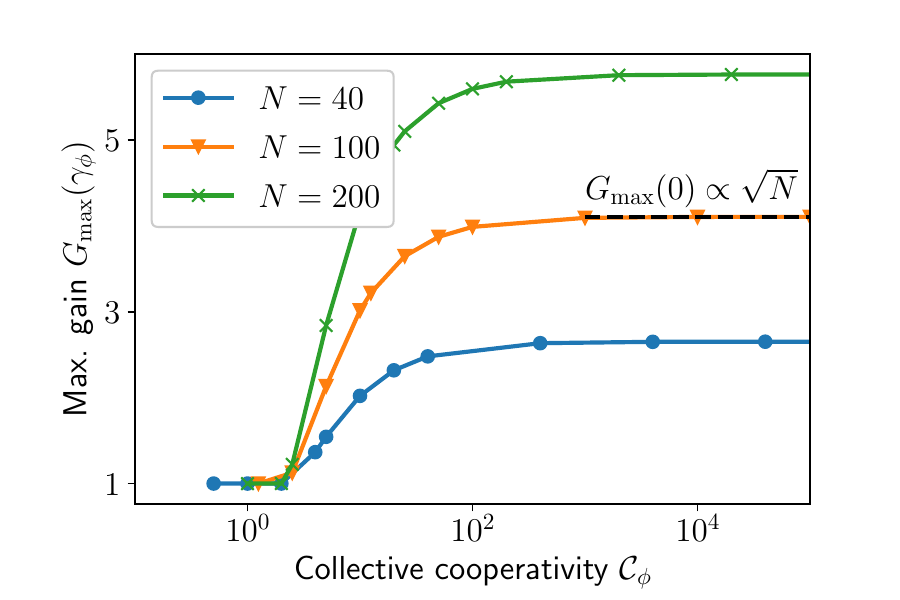}
	}
	\subfigure[]{
		\includegraphics[width=0.48\textwidth]{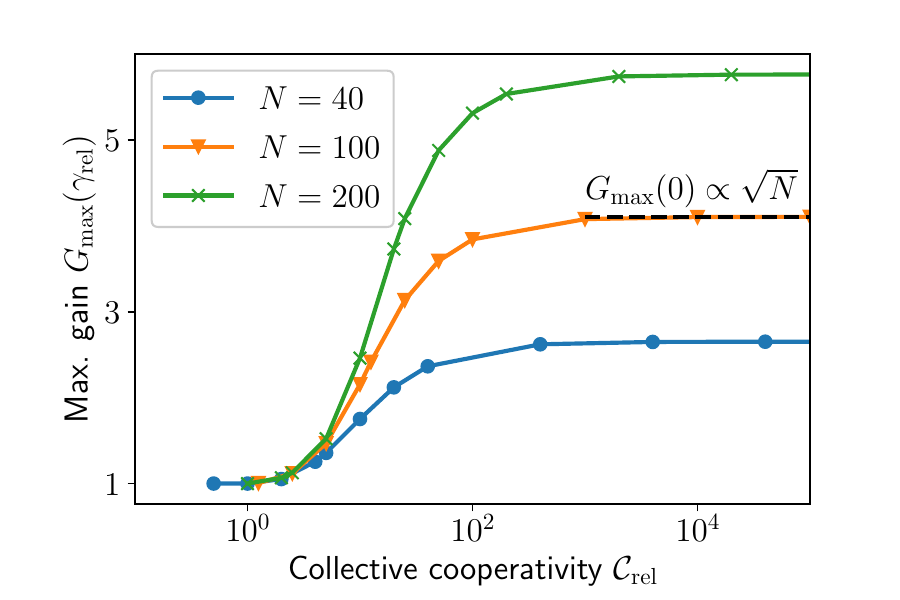}
	}
	\caption{
		\textbf{(a)}
			Maximum gain in the presence of local dephasing, $G_\mathrm{max}(\gamma_\phi)$, as a function of the collective cooperativity $\mathcal{C}_\phi = N \Gamma/\gamma_\phi$ [calculated by numerically exact integration of the master equation~\eqref{eqn:system:QmeCollectiveAndLocalDissipation} with $\gamma_\mathrm{rel} = 0$].
			Each data point is obtained by maximizing the time-dependent gain $G(t)$ over the evolution time $t$.  
			Collective amplification and local dephasing compete and amplification is observed if $\mathcal{C}_\phi \gtrsim 2$, i.e., if the collective amplification rate dominates over local decay. 
		\textbf{(b)}
			Analogous numerical results for maximum gain in the presence of local relaxation, $G_\mathrm{max}(\gamma_\mathrm{rel})$, as a function of the collective cooperativity $\mathcal{C}_\mathrm{rel} = N \Gamma/\gamma_\mathrm{rel}$ (with $\gamma_\phi = 0$).  
			We again see that the collective cooperativity is the relevant parameter for obtaining maximum gain.   
	}
	\label{fig:localDissipation}
\end{figure*}

We first consider the impact of single-spin dissipation, namely Markovian dephasing and relaxation at rates $\gamma_\phi$ and $\gamma_\mathrm{rel}$, respectively. 
The master equation for our spin ensemble now takes the form  
\begin{align}
	\frac{\mathrm{d} \hat{\rho}}{\mathrm{d} t} = \Gamma \mathcal{D} \left[ \hat{S}_- \right] \hat{\rho} + \gamma_\mathrm{rel} \sum_{j=1}^N \mathcal{D} \left[ \hat{\sigma}_-^{(j)} \right] \hat{\rho} + \frac{\gamma_\phi}{2} \sum_{j=1}^N \mathcal{D} \left[ \hat{\sigma}_z^{(j)} \right] \hat{\rho} ~.
	\label{eqn:system:QmeCollectiveAndLocalDissipation}
\end{align}
Numerically exact solutions of Eq.~\eqref{eqn:system:QmeCollectiveAndLocalDissipation}, shown in Fig.~\ref{fig:localDissipation}, demonstrate that an initial signal is still amplified if the collective cooperativities
\begin{align}
	\mathcal{C}_k = \frac{N \Gamma}{\gamma_k} ~, 
	\label{eqn:amplification:disorder:collectiveCooperativities}
\end{align}
with $k \in \{\phi,\mathrm{rel}\}$, exceed a threshold value of the order of unity. 
This is equivalent to the threshold condition for superradiant lasing \cite{Bohnet2012,Meiser2009}. 
Further, we find that achieving the maximum gain $G \propto \sqrt{N}$ does not require strong coupling at the single-spin level:  it only requires a large collective cooperativity, and \emph{not} a large single-spin cooperativity $\eta_k \equiv \mathcal{C}_k / N$.

Note that the dependence of the gain on cooperativity can be understood at a heuristic level by inspecting the MFT equations of motion~\eqref{eqn:amplification:intuition:EomSy}, which now take the form:
\begin{align}
	\frac{\mathrm{d} S_{x,y}}{\mathrm{d} t} = \Gamma \left( S_z - \frac{1}{2} \right) S_{x,y} - \gamma_\phi S_{x,y} - \frac{\gamma_\mathrm{rel}}{2} S_{x,y}~.
	\label{eqn:amplification:disorder:EomSy}
\end{align}
At short times, the collective decay term tends to increase $S_y$ at a rate $N \Gamma$ whereas local dissipation aims to decrease $S_y$ at rates $\gamma_\phi$ and $\gamma_\mathrm{rel}/2$, respectively.
Amplification is only possible if the slope of $S_y$ at $t=0$ is positive, which is equivalent to the conditions $\mathcal{C}_\phi > 1$ and $\mathcal{C}_\mathrm{rel} > 1/2$, respectively. 
For weak local dissipation, i.e., $\mathcal{C}_k \gg 1$, the numerical results shown in Fig.~\ref{fig:localDissipation} are well described by the mean-field result
\begin{align}
	G_\mathrm{max}(\gamma_k) = G_\mathrm{max}(0) \left( 1 - \frac{a_k}{\mathcal{C}_k} \right) ~,
\end{align}
where $a_\phi \approx 3$ and $a_\mathrm{rel} \approx 6$. 
In the opposite limit $\mathcal{C}_k \ll 1$, there is no amplification, $G_\mathrm{max}(\gamma_k) = 1$.

\subsubsection{Finite temperature}

\begin{figure}
	\centering
	\includegraphics[width=0.48\textwidth]{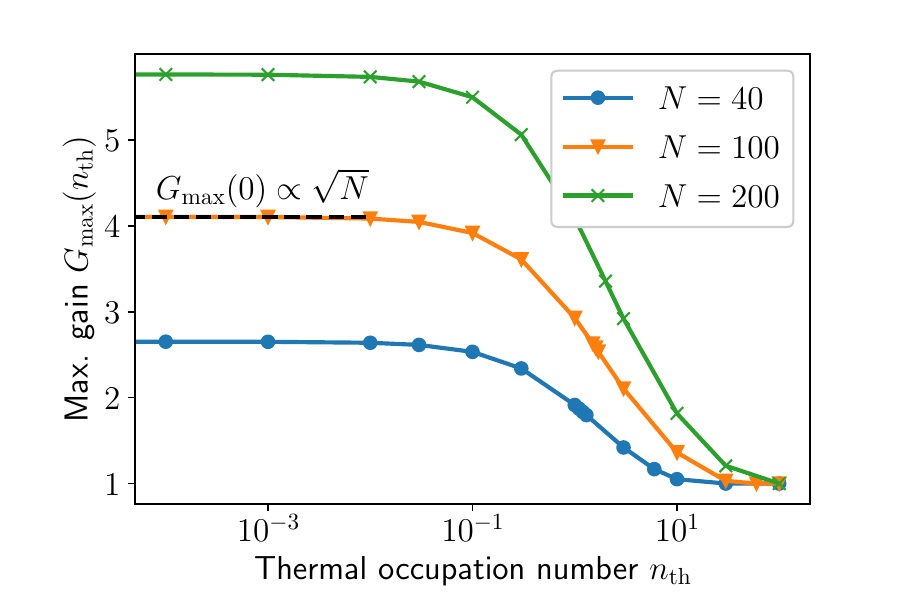}
	\caption{
		Maximum gain in the presence of a non-zero thermal occupancy $n_{\rm th}$ of the environment responsible for collective spin decay, obtained by numerically exact solution of the master equation~\eqref{eqn:QME_finite_temp}. 
		Each data point is obtained by maximizing the gain $G(t)$ over the evolution time $t$. 
		We see that a non-zero bath thermal occupancy rapidly degrades gain.  
		At a heuristic level, the decay of the ensemble's $z$ polarization is seeded by both thermal and quantum bath fluctuations.  
		Non-zero thermal fluctuations hence accelerate the decay, leading to a shorter time interval where the instantaneous gain rate $\lambda(t)$ is positive.  
		This allows one to quantitatively understand the suppression of maximum gain seen here, see Eq.~\eqref{eqn:SpinAmp:FiniteTemp:Gmaxnth}.
	}
	\label{fig:finiteTemperature}
\end{figure}

Another potential imperfection is that the reservoir responsible for collective relaxation may not be at zero temperature, giving rise to an unwanted collective excitation process. 
This could be relevant in setups where collective effects stem from coupling to a mechanical degree of freedom, a promising approach for ensembles of defect spins in solids \cite{Meesala2016,MacQuarrie2015,Cady2019}. 
In this general case, the master equation takes the form
\begin{align}
	\frac{\mathrm{d} \hat{\rho}}{\mathrm{d} t} = \Gamma (n_\mathrm{th} + 1) \mathcal{D}[\hat{S}_-] \hat{\rho} + \Gamma n_\mathrm{th} \mathcal{D}[\hat{S}_+] \hat{\rho}~.
	\label{eqn:QME_finite_temp}
\end{align}
The parameter $n_\mathrm{th}$ determines the relative strength between the collective decay and excitation rates and can be interpreted as an effective thermal occupation of the bath generating the collective decay. 
The gain as a function of the effective thermal occupation number $n_\mathrm{th}$ based on numerically exact solution of the full quantum master equation~\eqref{eqn:QME_finite_temp} is shown in Fig.~\ref{fig:finiteTemperature}(a).
A nonzero $n_\mathrm{th}$ reduces the gain as compared to the ideal gain $G_\mathrm{max}$ obtained for $n_\mathrm{th} = 0$, and ultimately prevents any amplification in the limit $n_\mathrm{th} \gg 1$.

MFT again allows one to develop an intuitive picture of how a bath temperature degrades amplification dynamics.  
In the presence of finite temperature and for large $N$, the mean-field equations of motion~\eqref{eqn:amplification:intuition:EomSz} and~\eqref{eqn:amplification:intuition:EomSy} read
\begin{align}
	\frac{\mathrm{d} S_{x,y}}{\mathrm{d} t} &= \Gamma \left( S_z - \frac{1}{2} - n_\mathrm{th} \right) S_{x,y} ~, 
	\label{eqn:SpinAmp:FiniteTemp:EomSy} \\
	\frac{\mathrm{d} S_z}{\mathrm{d} t} &= - \Gamma \frac{N^2}{4} - \Gamma S_z (1 + 2 n_\mathrm{th} - S_z) ~.
	\label{eqn:SpinAmp:FiniteTemp:EomSz}
\end{align}
The impact of finite temperature $n_\mathrm{th} > 0$ is thus twofold. 
First, the time-dependent gain factor in Eq.~\eqref{eqn:SpinAmp:FiniteTemp:EomSy} is shut off at an earlier time, namely, if the condition $S_z(t) = 1/2 + n_\mathrm{th}$ holds. 
This implies that no amplification will occur if $n_\mathrm{th} > N/2$. 
If this were the only effect, the generation of gain would be largely insensitive to thermal occupancies $n_{\rm th} \ll N$.  Unfortunately, there is a second, more damaging mechanism.  
As the above equations show, the instantaneous gain rate $\lambda(t)$ is controlled by $S_z(t)$.  
The decay of of this polarization is seeded by both quantum and thermal fluctuations in the environment.  
Hence, a non-zero $n_\mathrm{th}$ accelerates this decay, leading to a more rapid decay of polarization, and a shorter time interval where the instantaneous gain rate is positive.  
This ultimately suppresses the maximum gain.

The above argument can be made quantitative if we expand $S_z$ for short times around its initial value, $S_z = N/2 - \delta$, where $\delta \ll 1$. 
To leading order in $N$ and $\delta$, the equation of motion of the deviation $\delta$ is $\mathrm{d} \delta/\mathrm{d} t = N \Gamma  (1 + n_\mathrm{th}) + N \Gamma \delta$, where the first term 
shows explicitly that both bath vacuum fluctuations and thermal fluctuations drive the initial decay of polarization.  
As a consequence, the superradiant emission occurs faster and, in the limit $N \gg 1 + 2 n_\mathrm{th}$, the time to reach maximum amplification is
\begin{align}
	\Gamma t_\mathrm{max} = \frac{1}{N} \ln \frac{N - n_\mathrm{th}}{n_\mathrm{th} + 1}~.
\end{align}
In the same limit, the maximum gain is given by
\begin{align}
	G_\mathrm{max}(n_\mathrm{th}) = \frac{G_\mathrm{max}(0)}{\sqrt{1 + n_\mathrm{th}}} ~,
	\label{eqn:SpinAmp:FiniteTemp:Gmaxnth}
\end{align}
which shows that a thermal occupation of $n_\mathrm{th} = 3$ will decrease the gain by $3\,\mathrm{dB}$. 
Note that $G_\mathrm{max}(n_\mathrm{th})$ still scales $\propto \sqrt{N}$, i.e., for a fixed value of $n_\mathrm{th}$, the reduction can be compensated by increasing the number of spins. 
The experimental demonstration of superradiance in NV-center spins by Angerer \emph{et al.} \cite{Angerer2018} has been performed at $25\,\mathrm{mK}$.
The spins were resonant with a microwave cavity at a frequency of about $3\,\mathrm{GHz}$, which corresponds to a thermal occupation of $n_\mathrm{th} \approx 0.002 \ll 1$.

\subsection{Implementation using cavity-mediated dissipation}
\label{sec:Results:CavityImplementation}

\begin{figure}
	\centering
	\includegraphics[width=0.35\textwidth]{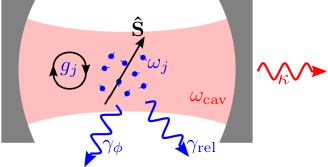}
	\caption{
		The collective decay described by Eqs.~\eqref{eqn:system:QME_spin_only} and~\eqref{eqn:system:QmeCollectiveAndLocalDissipation} can be implemented experimentally by coupling $N$ spin-$1/2$ systems (level splittings $\omega_j$) to a strongly damped bosonic mode (frequency $\omega_\mathrm{cav}$ and single-spin coupling strengths $g_j$). The mode is depicted here as a resonant mode of a photonic cavity, but one could use a wide variety of systems (e.g.~microwave or mechanical modes).
		The energy decay rate of the bosonic mode is $\kappa$ and each spin may undergo local relaxation or dephasing processes at rates $\gamma_\mathrm{rel}$ or $\gamma_\phi$, respectively. 
	}
	\label{fig:sketch}
\end{figure}

While there are many ways to engineer the collective relaxation that powers our superradiant amplifier, we specialize here to a ubiquitous realization that allows the tuneability we require:  couple the spin ensemble to a common lossy bosonic mode.  
To that end, we consider a setup where $N$ spin-$1/2$ systems are coupled to a damped bosonic mode $\hat{a}$ by a standard Tavis-Cummings coupling (see Fig.~\ref{fig:sketch}): 
\begin{align}
	\hat{H} = \omega_\mathrm{cav} \hat{a}^\dagger \hat{a} + \sum_{j=1}^N \omega_j \frac{\hat{\sigma}_z^{(j)}}{2} + \sum_{j=1}^N g_j \left( \hat{\sigma}_-^{(j)} \hat{a}^\dagger + \hat{\sigma}_+^{(j)} \hat{a} \right)~.
	\label{eqn:system:H_spin_and_cavity}
\end{align}
Here, $\omega_\mathrm{cav}$ and $\omega_j$ denote the frequencies of the bosonic mode and the spins, respectively, and $g_j$ denotes the coupling strength of spin $j$ to the bosonic mode.
The bosonic mode is damped at an energy decay rate $\kappa$ and the entire system is thus described by the quantum master equation 
\begin{align}
	\frac{\mathrm{d} \hat{\rho}}{\mathrm{d} t} 
	= - i \left[ \hat{H}, \hat{\rho} \right] 
	+ \kappa \mathcal{D} \left[ \hat{a} \right] \hat{\rho} ~.
	\label{eqn:system:QME_spin_and_cavity}
\end{align} 
For collective phenomena, we ideally want all atoms to have the same frequency $\omega_j = \bar{\omega}$ and be equally coupled to the cavity, $g_j = g$. 
For superradiant decay, we further want the spins to be resonant with the cavity, i.e., have zero detuning $\omega_\mathrm{cav} - \bar{\omega} = 0$.  
If, in addition, the bosonic mode is strongly damped, $\kappa \gg \sqrt{N} g$, the $\hat{a}$ mode can be eliminated adiabatically, which gives rise to the spin-only master equation~\eqref{eqn:system:QME_spin_only} with a collective decay rate 
\begin{align}
	\Gamma = \frac{4 g^2}{\kappa}
	\label{eq:GammaRate}
\end{align}
and $\gamma_\mathrm{rel} = 0$.

Returning to Fig.~\ref{fig:protocol}, note that a crucial part of our protocol is the ability to turn on and off the collective dissipation on demand (i.e., to start the amplification dynamics at the appropriate point in the measurement sequence, and then turn it off once maximum gain is reached). 
This implementation provides a variety of means for doing this. 
Perhaps the simplest is to control the spin-cavity detuning $\Delta$ by, e.g.~changing the applied $z$ magnetic field on the spins.  
In the limit of an extremely large detuning, the superradiant decay rate $\Gamma$ is suppressed compared to Eq.~\eqref{eq:GammaRate} by the small factor $\kappa^2 / (\kappa^2 + 4 \Delta^2) \ll 1$.

In the following, we separately analyze the impact of coupling inhomogeneities, $g_j \neq g$, and of inhomogeneous broadening, $\omega_j \neq \bar{\omega}$.

\subsubsection{Non-uniform single-spin couplings}
\label{sec:Res:NonuniformSinglespinCouplings}

To analyze the impact of inhomogeneous coupling parameters $g_j$, we follow the standard approach outlined in Ref.~\cite{Agarwal1970}.
It uses an expansion of the mean-field equations to leading order in the deviations $\delta_j = g_j - \bar{g}$ of the average coupling $\bar{g} = \sum_{j=1}^N g_j/N$ and retains only leading-order terms in the equations of motion for $S_x$ and $S_y$.
The impact of inhomogeneous couplings is then to reduce the effective length of the collective spin vector associated with the ensemble by the factor
\begin{align}
	\mu = \frac{1}{N} \frac{\sum_{k=1}^N \sum_{l = 1, l \neq k}^N g_k g_l}{\sum_{k=1}^N g_k^2}~,
	\label{eqn:Res:MuNonUniformSingleSpinCouplings}
\end{align}
i.e., the maximum gain and the optimal time are now given by $G_\mathrm{ci} = \sqrt{\mu N}/2$ and $t_\mathrm{max}^\mathrm{ci} = \ln(\mu N)/\gamma_0 \mu N$, respectively, where we defined $\gamma_0 = \sum_{k=1}^N 4 g_k^2/\kappa N$. 
Hence, the maximum gain $G_\mathrm{max}$ is reduced by a disorder-dependent prefactor, but the fundamental scaling is retained.

\subsubsection{Inhomogeneous broadening}
\label{sec:res:InhomogeneousBroadening}

\begin{figure}[t]
	\centering
	\includegraphics[width=0.48\textwidth]{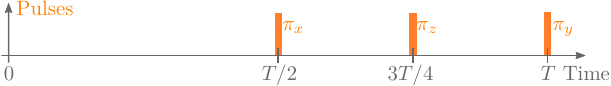}
	\caption{ 
		Dynamical decoupling sequence to cancel inhomogeneous broadening in the Hamiltonian~\eqref{eqn:system:H_spin_and_cavity}.
		The $\pi$ pulse about the $x$ axis cancels disorder in the spin frequencies $\omega_j$.
		The subsequent $\pi$ pulses about the $z$ and $y$ axis compensate unwanted interaction terms generated by the first $\pi$ pulse.
		The overall pulse sequence is applied repeatedly and generates the average Hamiltonian~\eqref{eqn:dyndec:Haverage} if the repetition rate $1/T$ is much larger than the standard deviation of the distribution of the spin transition frequencies $\omega_j$.
	}
	\label{fig:dyndec}
\end{figure}

Inhomogeneous broadening can be canceled by the dynamical decoupling sequence introduced recently in Ref.~\cite{Groszkowski2021}, which is summarized in Fig.~\ref{fig:dyndec}. 
Different spin transition frequencies $\omega_j$ in Eq.~\eqref{eqn:system:H_spin_and_cavity} lead to a dephasing of the individual spins in the ensemble, which can be compensated by a $\pi$ pulse about the $x$ axis halfway through the sequence. 
However, this pulse will modify the interaction term in Eq.~\eqref{eqn:system:H_spin_and_cavity} and will turn collective decay into collective excitation. 
This can be compensated by a $\pi$ pulse about the $z$ axis at time $3 T/4$, which changes the sign of the coupling constants $g_j$.
Note that such a pulse can be generated using a combination of $x$ and $y$ rotations. 
The final $\pi$ pulse about the $y$ axis at time $T$ reverts all previous operations and restores the original Hamiltonian~\eqref{eqn:system:H_spin_and_cavity}. 
The average Hamiltonian of this pulse sequence in a frame rotating at $\omega_0$ is
\begin{align}
    \bar{H} = \sum_{j=1}^N \frac{g_j}{2} \left( \hat{\sigma}_+^{(j)} \hat{a} + \hat{\sigma}_-^{(j)} \hat{a}^\dagger \right)
	\label{eqn:dyndec:Haverage}
\end{align}
if the repetition rate $1/T$ of the decoupling sequence is much larger than the standard deviation of the distribution of the frequencies $\omega_j$. 
More details on the derivation of this decoupling sequence are provided in a recent publication \cite{Groszkowski2021}.
If one chooses not to use dynamical decoupling, the analysis outlined in Sec.~\ref{sec:Res:NonuniformSinglespinCouplings} can be adapted to estimate the effect of inhomogeneous broadening on the superradiant decay dynamics \cite{Agarwal1971}.


\subsubsection{Limit of an undamped cavity}
\label{sec:Implementations:Undamped}

\begin{figure*}
	\centering
	\subfigure[]{
		\includegraphics[width=0.48\textwidth]{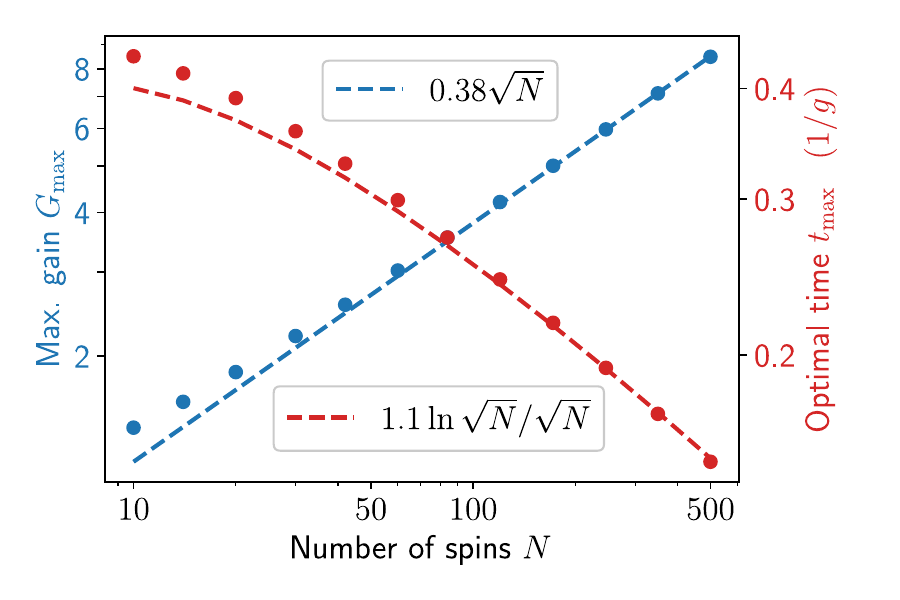}
	}
	\subfigure[]{
		\includegraphics[width=0.48\textwidth]{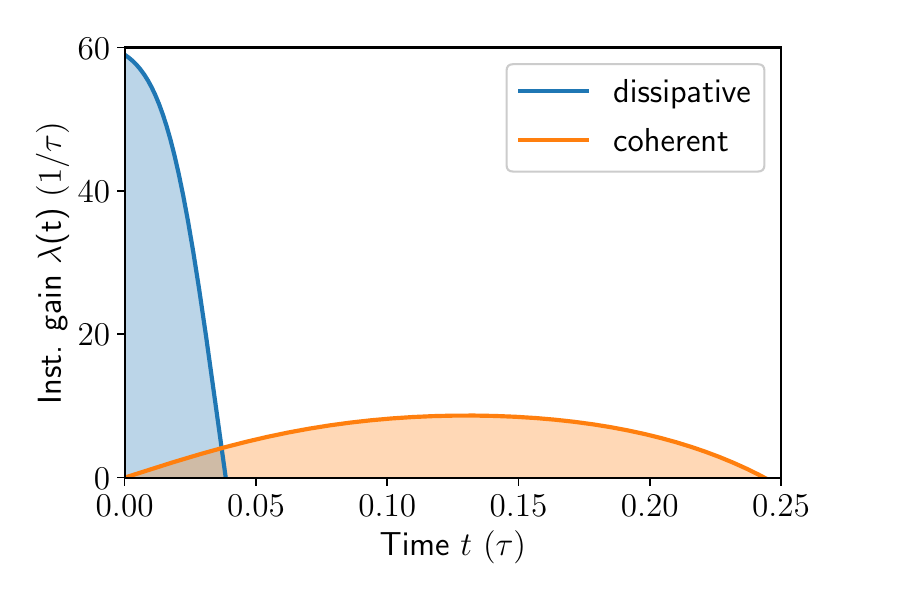}
	}
    \caption{
            Coherent spin amplification in a cavity-spin system described by the resonant Tavis-Cummings Hamiltonian~\eqref{eqn:UnitaryLimit:TCHamiltonian} with $\omega_\mathrm{cav} = \omega$ and  initial state $e^{i \phi \hat{S_x}} \vert\!\!\uparrow \dots \uparrow \rangle \otimes \vert 0 \rangle$.
        \textbf{(a)} 
            The maximum gain $G_\mathrm{max}$ follows the same $\sqrt{N}$ scaling as in the dissipative case, while the optimal evolution time $t_\mathrm{max} \propto \ln {\sqrt{N}}/ \sqrt{N}$, has a different $N$-dependence than the corresponding time $t_\mathrm{max} \propto \ln N/N$ in the dissipative case (cf.~Fig.~\ref{fig:principle}(a)). 
            Colored dots correspond to data obtained from solving the Schr\"odinger equation, while the dashed curves show the corresponding large-$N$ scaling behavior.
        \textbf{(b)} 
            Comparison of the instantaneous gain $\lambda(t) = [\mathrm{d} \langle \hat{S}_{y}(t) \rangle/\mathrm{d} t] / \langle \hat{S}_{y}(t) \rangle$ as a function of time for coherent (orange) and dissipative (blue) amplification protocols with $N=120$ spins. 
            The time scale $\tau$ (which is $N$-independent) is $1/g$ and $1/\Gamma$ for the coherent and dissipative cases, respectively. 
            The maximum gain $G_\mathrm{max}$ shown in (a) corresponds to the integral of $\lambda(t)$ between $t=0$ and $t=t_\mathrm{max}$ (i.e., the shaded regions), which are nearly equal in both cases.  All numerical results presented here for the unitary scheme are obtained by numerical integration of the Schr\"odinger equation with Hamiltonian~\eqref{eqn:UnitaryLimit:TCHamiltonian}.   
}
	\label{fig:nonmarkovian}
\end{figure*}

Returning to our cavity-based implementation of the superradiant spin amplifier in Eqs.~\eqref{eqn:system:H_spin_and_cavity}  and~\eqref{eqn:system:QME_spin_and_cavity}, one might worry about whether this physics also persists in regime where the cavity damping rate $\kappa$ is not large enough to allow for an adiabatic elimination.  
To address this, we briefly consider the extreme limit of this situation, $\kappa \rightarrow 0$, where we simply obtain a completely unitary dynamics generated by the resonant Tavis-Cummings Hamiltonian
\begin{align}
    \hat{H}_\mathrm{TC} = \omega_\mathrm{cav} \hat{a}^\dagger \hat{a} + \omega \hat{S}_z + g \left( \hat{S}_- \hat{a}^\dagger + \hat{S}_+ \hat{a} \right) ~,
    \label{eqn:UnitaryLimit:TCHamiltonian}
\end{align}
where $\omega_\mathrm{cav} = \omega$.
Figure~\ref{fig:nonmarkovian} shows numerical results for the time-maximized gain $G_\mathrm{max}$ starting from an initial state $e^{i \phi \hat{S}_x} \vert\!\uparrow \dots \uparrow \rangle \otimes \vert 0 \rangle$, where $\vert0\rangle$ denotes the vacuum state of the cavity.
A complementing analysis based on MFT is discussed in the Supplemental \cite{SM}. 
We find that spin amplification dynamics still holds in the unitary regime, with an identical $G_\mathrm{max} \propto \sqrt{N}$ scaling of the maximum gain.  
We stress that realizing this limit of fully unitary collective dynamics is challenging in most spin-ensemble sensing platforms.  
Nonetheless, this limit shows that our amplification dynamics will survive even if the adiabatic elimination condition $\sqrt{N} g \ll \kappa$ that leads to Eq.~\eqref{eqn:system:QME_spin_and_cavity} is not perfectly satisfied.  
This further enhances the experimental flexibility of our scheme.   

Although both the dissipative and the unitary case yield $G_\mathrm{max} \propto \sqrt{N}$, the underlying dynamics is quite different. 
The time $t_\mathrm{max}$ to reach maximum amplification in the coherent case, shown in Fig.~\ref{fig:nonmarkovian}(a), is parametrically longer if we consider the limit of a large number of spins $N$: $t_\mathrm{max} \propto \ln \sqrt{N}/\sqrt{N}$ (as opposed to a $t_\mathrm{max} \propto \ln N / N$ scaling in the dissipative case).
Consequently, the instantaneous gain rates $\lambda(t)$ are also quite different in both cases: whereas dissipative superradiant decay has an almost constant instantaneous gain rate over a very short time, the gain in the Tavis-Cummings model is non-monotonic, starts at zero, and grows at short times, as shown in Fig.~\ref{fig:nonmarkovian}(b).

Note that, for the coherent Tavis-Cummings model, the timescale for maximum amplification is analogous to the timescale that governs quasi-periodic oscillations of excitation number in the large-$N$ limit; this latter phenomenon has been derived analytically in previous work \cite{Andreev2004,Barankov2004,Keeling2009}.  
However, the semiclassical approach used in these works fails to accurately describe the gain physics that is of interest here (see Supplemental \cite{SM}). 
Finally, in the Supplemental, we show that the added noise in the unitary case is also close to the expected quantum limit.
Surprisingly, it is approximately equal to what we have found in the dissipative limit, $\sigma_\mathrm{add}^2 \approx 1.3$.

\section{Discussion}
\label{sec:Discussion}


\subsection{Comparison and advantages over unitary OAT amplification schemes}
\label{sec:Discussion:OAT}

The dissipative spin amplification scheme introduced in this work is effective in the presence of collective loss, and in fact harnesses it as a key resource.  
As discussed in the Introduction, this is in sharp contrast to conventional approaches that use unitary dynamics to improve sensing in the presence of measurement noise:  such approaches become infeasible with even small amounts of $T_1$ relaxation (whether collective or single-spin in nature).  
To illustrate this, we focus on the scheme presented in the seminal work by Davis \emph{et al.} \cite{Davis2016}, where OAT dynamics is used to generate effective spin amplification.  
This scheme involves starting a spin ensemble in a CSS $\vert \psi_0 \rangle$ that is fully polarized in the $x$ direction. 
The protocol then corresponds to the composite unitary evolution
\begin{equation}
    \vert \psi_f \rangle = 
    \hat{U}_\mathrm{amp} \, \hat{R}_y(\phi) \, \hat{U}_\mathrm{sqz} \vert \psi_0 \rangle~.
    \label{eqn:OAT:Scheme}
\end{equation}
The first step corresponds to the generation of squeezing using the OAT Hamiltonian  $\hat{H}_\mathrm{OAT} = \chi \hat{S}_z^2$ for a time $t$, i.e., $\hat{U}_\mathrm{sqz} = \exp(-i \hat{H}_\mathrm{OAT} t)$. 
The next step is signal acquisition:  the state is rotated by a small angle $\phi$ about the $\hat{S}_y$ axis, via the unitary $\hat{R}_y(\phi) = e^{-i \phi \hat{S}_y}$.
Finally, the last step is another evolution under the OAT interaction Hamiltonian, for an identical time $t$ as the first step, but with an opposite sign of the interaction $\chi \rightarrow -\chi$, i.e., $\hat{U}_\mathrm{amp} = \hat{U}_\mathrm{sqz}^{-1}$.

In this scheme, the final signal gain is created entirely by the last OAT evolution step $\hat{U}_\mathrm{amp}$; the first ``pre-squeezing'' step only serves to control the fluctuations in the final state. 
The suppression of the readout-noise term $\Xi_\mathrm{det}^2$ depends only on the maximum gain $G_\mathrm{max}$, as shown in Eq.~\eqref{eqn:ReducingEstimationError:GenericModelWithAmplification}. 
Since we consider a regime where readout noise is dominant, we ignore the initial squeezing step in the following discussion and focus only on the gain of the OAT protocol.

We thus consider a CSS that is almost completely polarized in the $x$ direction, with a small $z$ polarization that encodes the signal rotation $\phi$ of interest. 
Without dissipation, the OAT Hamiltonian leads to the Heisenberg equation of motion  
\begin{align}
    \frac{\mathrm{d} \hat{S}_y}{\mathrm{d} t} &= 2 \chi \hat{S}_z \hat{S}_x~. 
    \label{eqn:OAT:EoMSy}
\end{align}
For short times, we have $S_x \approx N/2$, and the OAT interaction causes the expectation value of $S_y$ to grow linearly in time at a rate set by the initial ``signal'' value of $S_z = N \sin(\phi) / 2$.  
The amplified signal is thus contained in $S_y$, and it is this spin component that is ultimately read out.  
We can thus define the signal gain analogously as in Eq.~\eqref{eqn:amplification:gain},
\begin{align}
    G^\mathrm{OAT}(t) \equiv \lim_{\phi \to 0} \frac{\langle \hat{S}_y(t)\rangle}{\langle \hat{S}_z(0)\rangle}~.
\end{align}
Note that the amplification mechanism here is analogous to bosonic amplification using a QND interaction \cite{Szorkovszky2014,Metelmann2015}.  
In the spin system, nonlinearity eventually causes the the growth of $S_y$ to saturate, leading to a maximum gain at a time $t_\mathrm{max}^\mathrm{OAT} \propto 1/\chi \sqrt{N}$ \cite{Davis2016}.

\begin{figure*}
    \centering
	\subfigure[]{
        \includegraphics[width=0.48\textwidth]{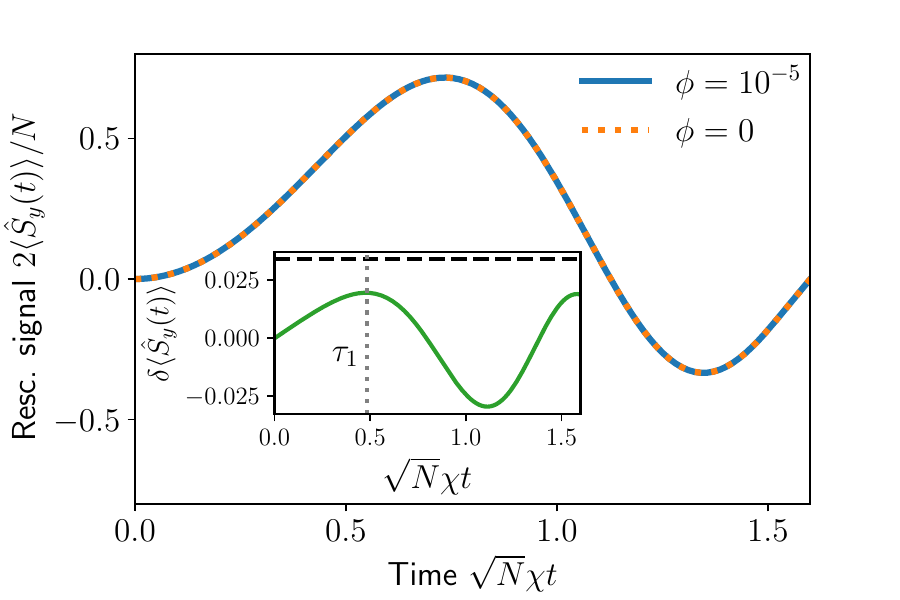}
	}
    \subfigure[]{
        \includegraphics[width=0.48\textwidth]{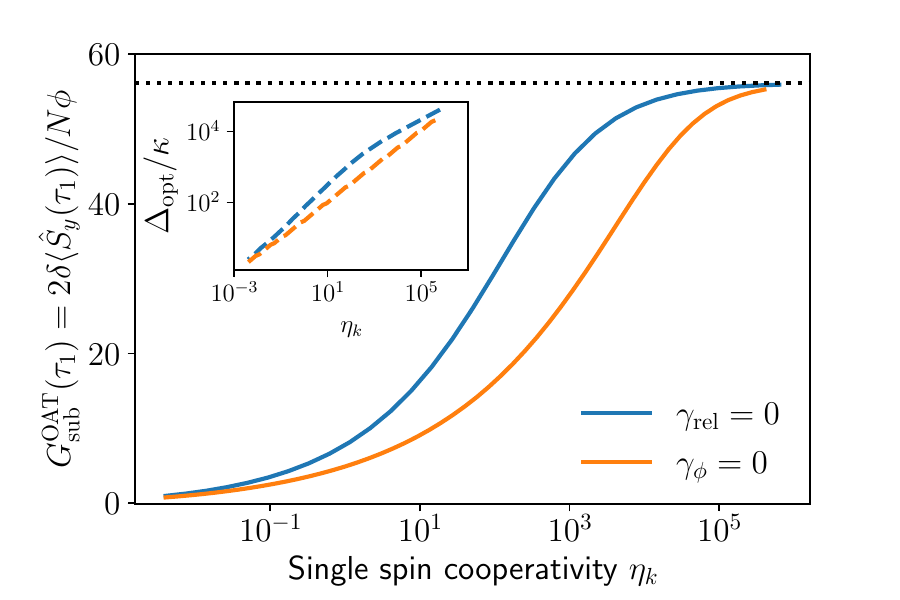}
	}
    \caption{
            Analysis of the OAT spin amplification protocol proposed in Ref.~\cite{Davis2016} for an ensemble of $N$ standard two-level systems  coupled to a detuned bosonic mode via a Tavis-Cummings coupling.
            The initial signal is encoded in $\langle \hat{S}_z \rangle$, whose value is then transduced to $\langle \hat{S}_y \rangle$ with gain, see Eq.~\eqref{eqn:OAT:EoMSy}.
        \textbf{(a)}
            Time evolution of $\langle \hat{S}_y \rangle$, both with (solid blue curve) and without (dashed orange curve) an initial small signal $\phi$ (obtained by numerically exact solution of the master equation~\eqref{eqn:QME_OAT} for
            $N=500$, $\Gamma_\mathrm{coll}/\chi = 0.02$, and $\gamma_\mathrm{rel}/\chi = \gamma_\phi/\chi = 0$). 
            Collective decay leads to a large average value of $\langle \hat{S}_y \rangle$ even without any initial signal $\phi$. 
            The inset shows the tiny signal obtained after subtraction of this background, $\delta \langle \hat{S}_y(t) \rangle = \langle \hat{S}_y(t,\phi)\rangle - \langle \hat{S}_y(t,0) \rangle$.
            The black dashed line is the optimal gain one could reach in the absence of collective dissipation.
        \textbf{(b)}
            Scaling of the gain after background subtraction, $G^\mathrm{OAT}_\mathrm{sub}(\tau_1)$, in the presence of collective decay and either single-spin dephasing (blue curve) or single-spin relaxation (orange curve) as a function of the respective single-spin cooperativity $\eta_k = 4 g^2/\kappa \gamma_k$ with $k \in \{\phi,\mathrm{rel}\}$.
            We evaluate the gain at its first peak at time $t = \tau_1$ (indicated in the inset of (a) by the gray dotted vertical line), which is the time of maximum gain if local dissipation is taken into account. 
            The gain $G^\mathrm{OAT}_\mathrm{sub}(\tau_1)$ is strongly reduced compared to its ideal value in the absence of dissipation (dotted black line; note that MFT predicts this quantity to be slightly smaller than master-equation simulations), unless the \emph{single-spin} cooperativity satisfies $\eta_\phi \gg \sqrt{N}$ or $\eta_\mathrm{rel} \gg N^{0.9}$ (see Supplemental \cite{SM}).
            Simulations were done using MFT for $N=10^4$ spins. 
            For each $\eta_k$, the detuning $\Delta$ was optimized with the optimized values shown in the inset. 
    }
    \label{fig:OAT_plus_decay}
\end{figure*}

A crucial aspect of the OAT gain-generation mechanism is the conservation of $\hat{S}_z$ (analogous to a QND structure in the bosonic system). 
This leaves it vulnerable to any unwanted dissipative dynamics that breaks this conservation law. 
Unfortunately, such symmetry breaking is common in many standard sensing setups. 
Consider perhaps the simplest method for realizing an OAT Hamiltonian, where the spin ensemble is coupled to a bosonic mode (e.g.~a photonic cavity mode or mechanical mode) via the Tavis-Cummings Hamiltonian~\eqref{eqn:UnitaryLimit:TCHamiltonian}.  
Working in the large detuning limit $\vert \Delta \vert = \vert \omega_\mathrm{cav} - \omega \vert \gg g$ allows one to adiabatically eliminate the bosonic mode. 
This results in both the desired OAT Hamiltonian interaction, but also a collective loss dissipator associated with the loss rate $\kappa$ of the cavity mode:
\begin{align}
    \frac{\mathrm{d} \hat{\rho}}{\mathrm{d}t} 
    &= - i \left[ \chi \hat{S}_z^2, \hat{\rho} \right] + \Gamma_\mathrm{coll} \mathcal{D}[\hat{S}_-] \hat{\rho} \nonumber \\
    &\phantom{=}\ + \gamma_\mathrm{rel} \sum_{j=1}^N \mathcal{D}[\hat{\sigma}_-^{(j)}] \hat{\rho} 
    + \frac{\gamma_\phi}{2} \sum_{j=1}^N \mathcal{D}[\hat{\sigma}_z^{(j)}] \hat{\rho}~,
    \label{eqn:QME_OAT}
\end{align}
where the OAT strength is $\chi = g^2/\Delta$ and the collective decay rate is $\Gamma_\mathrm{coll} = \chi \kappa / \Delta$.
We also included the Lindblad terms for single-spin relaxation and dephasing.

We now have an immediate problem:  even with no signal (i.e., $\phi = 0$), the collective loss will cause $S_z$ to grow in magnitude during the amplification part of the protocol. 
This will result in a relatively large contribution to $S_y$ that is indistinguishable from the presence of a signal. 
An approximate mean-field treatment shows that, for $\phi \ll 1$ and short times, the average of $\hat{S}_z$ has the form
\begin{equation}
    S_z = \frac{N}{2} \phi - \frac{N(N+1)}{4} \Gamma_{\rm coll} t~.
\end{equation}
In the limit of interest $\phi \to 0$, the average $z$ polarization induced by relaxation will completely dominate the contribution from the signal $\phi$, which translates into the final measured quantity $\hat{S}_y$ being swamped by a large $\phi$-independent contribution. 
This behaviour is indeed seen in full numerical simulations of the dynamics, as depicted in Fig.~\ref{fig:OAT_plus_decay}(a). 
Note that single-spin relaxation will have an analogous effect here to collective relaxation.

One might think that this problem is merely a technicality, that could be dealt with by simply subtracting off the $\phi$-independent background. 
However, this would require an extremely precise calibration that would be difficult if not impossible to reliably implement in most cases of interest. 
Alternatively, one could try to reduce the deleterious impact of $\Gamma_\mathrm{coll}$ by using a very large detuning $\Delta$ (since $\Gamma_\mathrm{coll} / \chi \propto 1 / \Delta$). 
This strategy is also not effective if there is any appreciable single-spin dissipation. 
Consider for example the case where there is non-zero single-spin dephasing at a rate $\gamma_{\phi}$. 
In this case (and neglecting for a moment collective loss, i.e., $\Gamma=0$), one can show using the exact solution of the master equation~\eqref{eqn:QME_OAT} reported in Refs.~\cite{FossFeig2013,McDonald2021} that the gain $G(t)$ of the OAT protocol is reduced by an exponential factor,       
\begin{align}
    G^\mathrm{OAT}(t) \vert_{\gamma_\phi > 0} 
    = e^{- \gamma_\phi t} G^\mathrm{OAT}(t) \vert_{\gamma_\phi = 0}~.
\end{align}
To obtain a large gain $G^\mathrm{OAT}(t) \propto \sqrt{N}$, it is thus crucial that $t_\mathrm{OAT}$ be at most of the order of $1/\gamma_\phi$, which precludes the use of indefinitely large detuning.

To study the joint impact of both collective and local relaxation in more detail, we use MFT and consider the gain after background subtraction, 
\begin{align}
    G_\mathrm{sub}^\mathrm{OAT}(t) \equiv \lim_{\phi \to 0} \frac{\delta \langle \hat{S}_y(t) \rangle}{\frac{N \phi}{2}} = \frac{2}{N} \partial_\phi \langle \hat{S}_y(t,\phi) \rangle ~,
    \label{eqn:Discussion:GOATsub}
\end{align}
where $\delta \langle \hat{S}_y(t) \rangle = \langle \hat{S}_y(t,\phi) \rangle - \langle \hat{S}_y(t,0) \rangle$ denotes the signal after background subtraction. 
Figure~\ref{fig:OAT_plus_decay}(b) shows $G_\mathrm{sub}^\mathrm{OAT}(t)$ (evaluated at its first peak at time $t = \tau_1$, see inset of Fig.~\ref{fig:OAT_plus_decay}(a)) as a function of the single-spin cooperativity $\eta_k = 4 g^2/\kappa \gamma_k$, where $k \in \{\phi, \mathrm{rel}\}$.
For each data point, we optimize over the detuning $\Delta$ and the optimal values are shown in the inset of Fig.~\ref{fig:OAT_plus_decay}(b). 
While the inset of Fig.~\ref{fig:OAT_plus_decay}(a) suggests that local maxima of $G_\mathrm{sub}^\mathrm{OAT}(t)$ beyond the first peak at $t=\tau_1$ may lead to larger amplification, this is an artifact of having no single-spin dissipation. 
By integrating the quantum master equation~\eqref{eqn:QME_OAT} numerically for $N=20$ spins, we have explicitly verified that the performance of the OAT amplification scheme is not improved by considering time evolution past the first maximum of $G_\mathrm{sub}^\mathrm{OAT}(t)$ if local dissipation is taken into account (i.e., the first gain peak at $t = \tau_1$ is the optimal choice).
In the presence of both collective and local dissipation, we find that amplification in the OAT scheme is strongly reduced unless the \emph{single-spin} cooperativity satisfies $\eta_\phi \gg \sqrt{N}$ or $\eta_\mathrm{rel} \gg N^{0.9}$ (see Supplemental \cite{SM}). 
Note that this condition becomes harder to saturate if the spin number $N$ grows.
This is in sharp contrast to our dissipative amplification scheme, which only requires the \emph{collective} cooperativity to satisfy $\mathcal{C}_k \gg 1$.
We thus find that the OAT amplification scheme is of extremely limited utility in the standard case where dissipative two-level sytems have a Tavis-Cummings coupling to a common bosonic mode: even if one could perform the subtraction of a large $\phi$-independent background, achieving maximum amplification requires an unrealistically large value of the single-spin cooperativity, which is out of reach on solid-state quantum sensing platforms.

We stress that, as already discussed in Ref.~\cite{Davis2016}, one can largely circumvent the above problems by using spin ensembles where each constituent spin has more than two levels. 
For instance, one can then use two extremely long-lived ground-state spin levels for the sensing and generate the OAT interaction using an auxiliary third level of each spin and a driven cavity \cite{Leroux2010}.  
In this case, cavity decay does not lead to a collective relaxation process, only collective dephasing. 
Since there is no net tendency for $S_z$ to relax, one does not need to do a large, calibrated background subtraction. 
Aspects of the effect of the collective dephasing (as well as incoherent spin flips generated by spontaneous emission) were analyzed in Ref.~\cite{Davis2016}.  While this general approach is well suited to several atomic platforms, it is more restrictive than the case we analyze, where we simply require an ensemble of two-level systems.


\subsection{Experimental implementations}

The focus of this paper is not on one specific experimental platform, but is rather to illuminate the general physics of the collective spin amplification process, a mechanism relevant to many different potential systems.  While there are many AMO platforms capable of realizing our resonant, dissipative Tavis-Cummings model, we wish to particularly highlight potential solid-state implementations based on defect spins.  These systems have considerable promise in the context of quantum sensing, but usually suffer from the practical obstacle that the ensemble readout is far above the SQL \cite{Barry2020}.

We start by noting that recent work has experimentally demonstrated superradiance effects in sensing-compatible solid-state spin ensembles \cite{Bradac2017,Angerer2018}. 
Angerer \emph{et al.}\ \cite{Angerer2018} demonstrated superradiant optical emission from $N \approx 10^{16}$ negatively charged NV centers, which were homogeneously coupled to a microwave cavity mode in the fast cavity limit, i.e., with a decay rate $\kappa$ much larger than all other characteristic rates in the system. 
Moreover, improved setups with collective cooperativities larger than unity were reported and ways to increase the collective cooperativities even more have been discussed \cite{Eisenach2021,Ebel2021}. 
The essential ingredients to observe superradiant spin amplification in large ensembles of NV defects coupled to microwave modes have thus been demonstrated experimentally. 
Instead of a microwave cavity mode, the bosonic mode $\hat{a}$ could also be implemented by a mechanical mode that is strain-coupled to defect centers \cite{Lee2017}, e.g.~employing mechanical cantilevers \cite{Meesala2016}, optomechanical crystals \cite{Cady2019}, bulk resonators \cite{MacQuarrie2015}, or surface-acoustic-wave resonators \cite{Golter2016}.
In addition to NV centers, silicon vacancy (SiV) defect centers could be used \cite{Meesala2018,Maity2020}, which offer larger and field-tunable spin-mechanical coupling rates. 
Superradiant amplification could then pave a way to dramatically reduce the detrimental impact of detection noise and to approach SQL scaling.


\section{Conclusion}
\label{sec:Conclusion}

In this work, we have proposed and analyzed a simple yet powerful protocol to reduce the detrimental impact of readout noise in quantum metrology protocols.
Unlike previous ideas for spin amplification, our protocol is effective for dissipative ensembles of standard two-level systems, and does not require a large single-spin cooperativity.  
It allows a system with a highly inefficient spin readout to ultimately reach the SQL within a factor of two.

Our protocol uses the well-known physics of superradiant decay for a new task, namely, amplification of a signal encoded initially in any transverse component of a spin ensemble.
In contrast to usual treatments of superradiance, we are not interested in the emitted radiation. 
Instead, we use superradiance as a tool to induce nonlinear amplification dynamics in the spin system.
The gain factor of our protocol achieves the maximum possible scaling, $G_\mathrm{max} \propto \sqrt{N}$ in the large-$N$ limit.
The added noise associated the amplification is close to the  minimum allowed value one would expect for a quantum-limited bosonic amplifier.
While single-spin dissipation and finite temperature do reduce the gain, they do no change the fundamental scaling $\propto \sqrt{N}$. 
In the case of single-spin dissipation, we stress that maximum gain can be achieved by having a large collective cooperativity, i.e., one does \emph{not} need a large single-spin cooperativity.  
Our protocol is compatible with standard dynamical decoupling techniques to mitigate inhomogeneous broadening effects. 
Note that another unique aspect of our scheme is that it amplifies all spin directions perpendicular to the $z$ axis equally (as opposed to only amplifying a single direction in spin space).  
This could potentially be a useful tool in measurement schemes beyond generalized Ramsey protocols.

Our work also suggests several fruitful directions for future work. 
It would be interesting to combine the dissipative amplification mechanism introduced here with dissipative spin squeezing to achieve near-Heisenberg-limited sensitivity in systems with highly imperfect spin readout. 
On a fundamental level, the intrinsic nonlinearity of spin systems requires generalizations of the existing bounds on added noise of phase-preserving amplifiers.
The fact the amount of added noise is very similar both in the purely dissipative case and in the coherent $\kappa \to 0$ limit may  hint at a more fundamental reason to explain the numerically found level of $\sigma_\mathrm{add}^2 \approx 1.3$.
Regarding experimental platforms for quantum sensing, it would also be interesting to study the dynamics and utility of dissipative spin amplification in ensembles where intrinsic dipolar spin-spin interactions are strong.

\begin{acknowledgments}
We acknowledge discussions with A.\ Bleszynski Jayich, J.\ V.\ Cady, C.\ Padgett, V.\ Dharod, H.\ Oh, and Y.\ Tsaturyan.
This work was supported by the DARPA DRINQS program (Agreement D18AC00014). 
We also acknowledge support from the DOE Q-NEXT Center (Grant No. DOE 1F-60579), and from the Simons Foundation (Grant No. 669487, A. C.)
\end{acknowledgments}

\appendix 

\section{Details on the sensitivity analysis for fluorescence readout}
\label{sec:App:Sensitivity}

In this Appendix, we provide details on the fluorescence readout \cite{Barry2020}, which we model in terms of a positive operator-valued measurement (POVM). 
A more general derivation, which does not use the language of POVMs and keeps the readout method general, is given in the Supplemental \cite{SM}.

Fluorescence readout measures each of the $N$ spins in the $z$ basis, i.e., $\hat{\sigma}_z^{(j)} \vert \sigma_j \rangle = \sigma_j \vert \sigma_j \rangle$ with $\sigma_j = \pm 1$ and $j \in \{1, \dots, N\}$. 
Each spin emits $n_j$ photons independently of the state of other spins in the sample, with a state-dependent Poissonian probability distribution $\mathcal{P}_{\sigma_j}(n_j)$. 
Mean and variance of $\mathcal{P}_{\sigma_j}(n_j)$ are given by $n_\mathrm{b}$ ($n_\mathrm{d}$) if the spin is in the bright $\sigma_j=-1$ (dark $\sigma_j=+1$) state. 
Thus, the readout of each single spin can be modeled by a POVM with measurement operator $\hat{M}_{n_j,\sigma_j} = \sqrt{\mathcal{P}_{\sigma_j}(n_j)} \vert \sigma_j \rangle \langle \sigma_j \vert$ and effect operator $\hat{E}_{n_j} = \sum_{\sigma_j} \mathcal{P}_{\sigma_j}(n_j) \vert \sigma_j \rangle \langle \sigma_j \vert$ defining the probability that spin $j$ emits $n_j$ photons.

The photodetector only measures the total number of photons $n = \sum_{j=1}^N n_j$ emitted by the entire ensemble. 
The corresponding POVM measurement operator describing independent emission by $N$ spins is $\hat{M}_{\{n_j\},\{\sigma_j\}} = \otimes_{j=1}^N \sqrt{\mathcal{P}_{\sigma_j}(n_j)} \vert \sigma_j \rangle \langle \sigma_j \vert$, and the effect operator is 
\begin{align}
	\hat{E}_n 
	&= \sum_{\{n_j\}} \sum_{\{\sigma_j\}} \hat{M}_{\{n_j\},\{\sigma_j\}}^\dagger \hat{M}_{\{n_j\},\{\sigma_j\}} \delta\left(\sum_{j=1}^N n_j - n \right) \nonumber \\
	&= \sum_{\{\sigma_j\}} \mathcal{P}_{\sigma_1,\dots,\sigma_N}(n) \vert \sigma_1, \dots, \sigma_N \rangle \langle \sigma_1, \dots, \sigma_N \vert ~.
	\label{eqn:Methods:POVM:EffectOperator}
\end{align}
Here, $\mathcal{P}_{\sigma_1,\dots,\sigma_N}(n)$ is the convolution of all $N$ single-spin probability distributions $\mathcal{P}_{\sigma_j}(n_j)$, i.e., a Poissonian distribution with mean and variance $N_\mathrm{b} n_\mathrm{b} + N_\mathrm{d} n_\mathrm{d}$, where $N_\mathrm{b}$ ($N_\mathrm{d})$ is the number of spins in the bright (dark) state.

It is now convenient to switch to a basis $\vert j,m \rangle$ of simultaneous eigenstates of the collective operators $\mathbf{\hat{S}}^2$ and $\hat{S}_z$, such that $N_\mathrm{b}$ and $N_\mathrm{d}$ are simply related to the $\hat{S}_z$ quantum number, 
\begin{align}
	N_\mathrm{b} = \frac{N}{2} - m ~, 
	\qquad 
	N_\mathrm{d} = \frac{N}{2} + m ~.
\end{align}
Note that the permutation invariance of the spins allows us to focus on an effective basis which averages over the degeneracy of the total-angular-momentum subspaces with $j < N/2$ \cite{Chase2008}. 
In this basis, we can rewrite the effect operator~\eqref{eqn:Methods:POVM:EffectOperator} as
\begin{align}
	\hat{E}_n = \sum_{j,m} \mathcal{P}_m(n) \vert j,m \rangle \langle j,m \vert ~,
\end{align}
where $P_m(n)$ is a Poissonian distribution with mean and variance $N_\mathrm{b} n_\mathrm{b} + N_\mathrm{d} n_\mathrm{d} = N n_\mathrm{avg} [ 1 - 2 \tilde{C} m/N ]$.
Here, we defined the average number of emitted photons per spin $n_\mathrm{avg} = (n_\mathrm{b} + n_\mathrm{d})/2$ and the contrast between the bright and the dark state $\tilde{C} = (n_\mathrm{b} - n_\mathrm{d})/(n_\mathrm{b} + n_\mathrm{d})$  \cite{Taylor2008,Shields2015,Barry2020}.

The average number of detected photons for the state $\hat{\rho}$ is now given by
\begin{align}
	\bar{n} 
	&= \sum_n n \operatorname{Tr} \left[ \hat{E}_n \hat{\rho} \right]  
	= \sum_{n,j,m} n \mathcal{P}_m(n) \langle j,m \vert \hat{\rho} \vert j,m \rangle \nonumber \\
	&=  N n_\mathrm{avg} \left[ 1 - \frac{2}{N} \tilde{C} \langle \hat{S}_z \rangle \right] ~,
	\label{eqn:Methods:AveragePhotonNumber}
\end{align}
and the fluctuations of the photon number are given by
\begin{align}
	(\mathbf{\Delta}n)^2
	&= \sum_n n^2 \operatorname{Tr}\left[ \hat{E}_n \hat{\rho} \right] - \left( \sum_n n \operatorname{Tr} \left[ \hat{E}_n \hat{\rho} \right] \right)^2 \nonumber \\
	&= \sum_{n,j,m} \left[ n^2 \mathcal{P}_m(n) - \left( n \mathcal{P}_m(n) \right)^2 \right] \langle j,m \vert \hat{\rho} \vert j,m \rangle \nonumber \\
	&\phantom{=}\ + \sum_{n,j,m} \left( n \mathcal{P}_m(n) \right)^2 \langle j,m \vert \hat{\rho} \vert j,m \rangle \nonumber \\
	&\phantom{=}\ - \left( \sum_{n,j,m} n \mathcal{P}_m(n) \langle j,m \vert \hat{\rho} \vert j,m \rangle \right)^2 ~,
	\label{eqn:Methods:DeltanSquared}
\end{align}
where we added a zero in the last step. 
This allows us to separate two contributions to the variance of $(\mathbf{\Delta}n)^2$:
The classical noise which is added by the detector due to the fact that the probability distribution $\mathcal{P}_m(n)$ has a finite variance for each basis state $\vert j,m \rangle$ (the term in the square brackets), and the intrinsic quantum fluctuations of the state $\hat{\rho}$ expressed in terms of the measured photon number $n$ (the last two lines). 
Using the explicit expressions for the mean and variance of $\mathcal{P}_m(n)$, we find
\begin{align}
	(\mathbf{\Delta}n)^2
	= N n_\mathrm{avg} \left[ 1 - \frac{2}{N} \tilde{C} \langle \hat{S}_z \rangle \right] + 4 n_\mathrm{avg}^2 \tilde{C}^2 (\mathbf{\Delta} S_z)^2 ~,
\end{align}
which can be referred back to an uncertainty in $\phi$ using the transduction factor $\partial_\phi \bar{n} = - 2 n_\mathrm{avg} \tilde{C} \partial_\phi \langle \hat{S}_z \rangle$, 
\begin{align}
	(\mathbf{\Delta}\phi)^2 
	= \frac{(\mathbf{\Delta}n)^2}{\vert \partial_\phi \bar{n} \vert^2}
	= \frac{\frac{N}{4} \frac{1 - 2 \tilde{C} \langle \hat{S}_z \rangle/N}{n_\mathrm{avg} \tilde{C}^2}}{\vert \partial_\phi \langle \hat{S}_z \rangle \vert^2} + \frac{(\mathbf{\Delta} S_z)^2}{\vert \partial_\phi \langle \hat{S}_z \rangle \vert^2}~.
	\label{eqn:Methods:Sensitivity:DeltaPhiSquared}
\end{align}
For a small signal $\langle \hat{S}_z \rangle = N \phi/2$ with $\phi \ll 1$, one can ignore the second term in the numerator of the detection noise term. 
Note that these equations are given in terms of the basis of the final measurement at times $t_3 < t \leq t_4$ in Fig.~\ref{fig:protocol}(a), which is the $\hat{S}_z$ spin component. 
In the main text, we discuss everything in terms of the final state of the amplification step at $t = t_3$. 
It differs from the measured state by the $\pi/2$ rotation at $t=t_3$, which maps $\hat{S}_y \to \hat{S}_z$ and $(\mathbf{\Delta} S_y)^2 \to (\mathbf{\Delta} S_z)^2$.

To discuss the scaling of the two terms in Eq.~\eqref{eqn:Methods:Sensitivity:DeltaPhiSquared}, we focus on two typical probe states in a Ramsey experiment: CSS and spin-squeezed states.

For a CSS, the slope of the signal depends on the length of the spin vector, $\vert \partial_\phi \langle \hat{S}_z \rangle \vert = N/2$, i.e., the first (detection noise) term has an SQL-like scaling $1/\tilde{C}^2 n_\mathrm{avg} N \propto 1/N$ with a readout-dependent prefactor $1/\tilde{C}^2 n_\mathrm{avg}$. 
The CSS variance is $(\mathbf{\Delta} S_z)^2 = N/4$, therefore, the second (projection-noise) term reduces to $1/N$. 
In the absence of amplification, the measurement error $(\mathbf{\Delta}\phi)^2$ thus has a SQL-like $1/N$ scaling with a readout-noise-dependent prefactor $1 + 1/\tilde{C}^2 n_\mathrm{avg}$.

For a spin-squeezed state, the slope of the signal depends on the effective length of the spin vector along the mean spin direction $\vert \partial_\phi \langle \hat{S}_z \rangle \vert = \vert \langle \hat{S}_\mathrm{msd} \rangle \vert \leq N/2$. 
Since a spin-squeezed state wraps around the Bloch sphere for sufficiently large squeezing, $\vert \langle \hat{S}_\mathrm{msd} \rangle \vert$ decreases with increasing squeezing strength. 
The first (detection-noise) term thus reduces to 
\begin{align}
    \frac{1}{\tilde{C}^2 n_\mathrm{avg} N} \left( \frac{N/2}{\vert \langle \hat{S}_\mathrm{msd} \rangle \vert} \right)^2~, 
\end{align}
i.e., detection noise is \emph{larger} for a spin-squeezed state than for a simple CSS if squeezing is sufficiently strong to reduce $\vert \langle \hat{S}_\mathrm{msd} \rangle \vert$. 
The second (projection-noise) term can be expressed as $\xi_\mathrm{R}^2/N$, where we introduced the Wineland parameter $\xi_\mathrm{R}^2 = N \min_\perp (\mathbf{\Delta}S_\perp)^2/\vert \langle \hat{S}_\mathrm{msd} \rangle \vert^2$ and $\perp$ denotes the directions perpendicular to the mean-spin direction \cite{Wineland1992,Ma2011,Pezze2018}. 
Using spin squeezing, one can push the Wineland parameter below unity such that the projection noise reaches at best a Heisenberg-like $1/N^2$ scaling. 
Note that, in the presence of a very bad readout $1/\tilde{C}^2 n_\mathrm{avg} \gg 1$, this optimizes an almost irrelevant term of the overall measurement error and thus does not improve $(\mathbf{\Delta} \phi)^2$ significantly. 
As a consequence, the overall measurement error $(\mathbf{\Delta} \phi)^2$ still scales $\propto 1/N$ and the loss of signal slope, $\vert \langle \hat{S}_\mathrm{msd} \rangle \vert \to 0$, increases the detrimental impact of detection noise beyond the level one would have observed for a simple CSS probe state. 
Hence, spin squeezing is not a useful strategy if readout noise dominates.

Finally, we give typical values for the readout-noise prefactor $1/\tilde{C}^2 n_\mathrm{avg}$.
For fluorescence readout in trapped-ion setups \cite{Haeffner2008}, the decay of the dark state into the bright state is slow enough to allow for sufficiently long integration times such that $n_\mathrm{avg} \approx 30$ and $\tilde{C} \approx 98\,\%$ \cite{Myerson2008}, yielding a strong suppression of detection noise by a factor of $1/\tilde{C}^2 n_\mathrm{avg} \approx 0.03 \ll 1$.
However, the situation is dramatically different for solid-state defects, e.g.~negatively charged NV defects in diamond \cite{Barry2020}. 
Here, fluorescence readout leads to a rapid polarization of the NV spin into the bright state, such that the best values even for a \emph{single} NV center are $n_\mathrm{avg} \approx 0.3$ and $\tilde{C} \approx 15\,\%$ \cite{Shields2015,Barry2020}.
Therefore, the detection noise dominates the over projection noise by a factor of $1/\tilde{C}^2 n_\mathrm{avg} \approx 150$. 
For ensembles of many NV centers, the detrimental impact of readout noise is even larger: the best reported value is $1/\tilde{C}^2 n_\mathrm{avg} \approx 67^2 \approx 4500$ \cite{LeSage2012,Barry2020}.


\section{Mean-field theory analysis}
\label{sec:app:MFT}

To gain intuition on the amplification dynamics, we use MFT to derive approximate nonlinear equations of motion for the system. 
The differential equations for the spin components $S_k = \langle \hat{S}_k \rangle$, where $k \in \{x,y,z\}$, generate an (infinite) hierarchy of coupled differential equations for higher-order spin correlation functions, which we truncate and close by performing a second-order cumulant expansion \cite{Kubo1962}. 
This treatment is exact if the state is Gaussian.

We start with the quantum master equation
\begin{align}
	\frac{\mathrm{d} \hat{\rho}}{\mathrm{d} t} 
	&= \Gamma (n_\mathrm{th} + 1) \mathcal{D}[\hat{S}_-] \hat{\rho} + \Gamma n_\mathrm{th} \mathcal{D}[\hat{S}_+] \hat{\rho} \nonumber \\
	&\phantom{=}\ + \gamma_\mathrm{rel} \sum_{j=1}^N \mathcal{D}[\hat{\sigma}_-^{(j)}] \hat{\rho} + \frac{\gamma_\phi}{2} \sum_{j=1}^N \mathcal{D}[\hat{\sigma}_z^{(j)}] \hat{\rho}~.
	\label{eqn:MFT:QME}
\end{align}
Without loss of generality, we take the initial state to be $e^{i \phi \hat{S}_x} \vert \!\! \uparrow \dots \uparrow \rangle$.
This initial state has $S_x(t) = C_{xy}(t) = C_{xz}(t) = 0$, where the covariances are defined by $C_{kl} \equiv \langle (\hat{S}_k \hat{S}_l + \hat{S}_l \hat{S}_k)\rangle/2 - \langle \hat{S}_k \rangle \langle \hat{S}_l \rangle$ for $k,l \in \{x,y,z\}$. 
Since we are interested in the limit of a very small signal, $\phi \ll 1$, we drop all terms of the order $\phi^2$ in the initial conditions and in the equations of motion, which implies $C_{xx}(t) = C_{yy}(t)$. 
The mean-field equations are then given by
\begin{align}
	\dot{S}_y 
		&= - \left[ \frac{\Gamma}{2} + \Gamma n_\mathrm{th} - \Gamma S_z + \gamma_\phi + \frac{\gamma_\mathrm{rel}}{2} \right] S_y + \Gamma C_{yz} ~, 
		\label{eqn:MFT:Sy} \displaybreak[1]\\
	\dot{S}_z 
		&= - \Gamma (1 + 2 n_\mathrm{th}) S_z - 2 \Gamma C_{xx} - \gamma_\mathrm{rel} \left( S_z + \frac{N}{2} \right) ~, 
		\label{eqn:MFT:Sz} \displaybreak[1]\\ 
	\dot{C}_{xx} 
		&= - \left[ \Gamma (1 + 2 n_\mathrm{th}) + 2 \gamma_\phi + \gamma_\mathrm{rel} - 2 \Gamma S_z \right] C_{xx} \nonumber \\
		&\phantom{=}\ + \Gamma (1 + 2 n_\mathrm{th}) S_z^2 - \frac{\Gamma}{2} S_z + \frac{N}{2} \left( \gamma_\phi + \frac{\gamma_\mathrm{rel}}{2} \right) ~, 
		\label{eqn:MFT:Cxx} \displaybreak[1]\\
	\dot{C}_{zz}
		&= 
		+ 2 \Gamma (1 + 2 n_\mathrm{th}) C_{xx} + (\Gamma + \gamma_\mathrm{rel}) S_z + \frac{\gamma_\mathrm{rel}}{2} N ~, \\
	\dot{C}_{yz}
		&= - \left[ \frac{5}{2} \Gamma (1 + 2 n_\mathrm{th}) - \Gamma S_z + \gamma_\phi + \frac{3}{2} \gamma_\mathrm{rel} \right] C_{yz} \nonumber \\
		&\phantom{=}\ - \left[ \Gamma (1 + 2 n_\mathrm{th}) S_z + 2 \Gamma C_{xx} - \frac{\Gamma}{4} - \frac{\gamma_\mathrm{rel}}{2} \right] S_y ~. 
		\label{eqn:MFT:Cyz}
\end{align}

For simplicity, we ignore local dissipation for now, $\gamma_\mathrm{rel} = \gamma_\phi = 0$. 
Then, the equations of motion conserve total angular momentum, 
\begin{align}
	\frac{N}{2} \left(\frac{N}{2} +1 \right) = 2 C_{xx} + S_z^2 + C_{zz}~,
	\label{eqn:MFT:TotalAngularMomentumConservation}
\end{align}
where $C_{zz}$ is found to be suppressed by an order of $N$ compared to $C_{xx}$ and $S_z^2$. 
We thus drop $C_{zz}$ and use Eq.~\eqref{eqn:MFT:TotalAngularMomentumConservation} to eliminate the covariance $C_{xx}$ from the mean-field equations. 
This step decouples the equation of motion for $S_z$ from the rest of the system, but the equation of motion of $S_y$ still depends on the covariance $C_{yz}$. 
At short times, $C_{yz}$ is suppressed compared to the $S_z S_y$ term by a factor of $N$, therefore, we drop it from the equation of motion. 
In this way, we obtain a very simple set of equations of motion for $S_y$ and $S_z$:
\begin{align}
	\dot{S}_y 
		&= - \Gamma \left[ \frac{1}{2} + n_\mathrm{th} - S_z \right] S_y ~, \\
	\dot{S}_z
		&= - \Gamma \left[ \frac{N}{2} \left( \frac{N}{2}+1 \right) + S_z (1 + 2 n_\mathrm{th} - S_z) \right] ~.
		\label{eqn:MFT:EoMSzBasis}
\end{align}
The solutions predict the exact dynamics (determined by numerically exact solution of the quantum master equation~\eqref{eqn:MFT:QME}) qualitatively correct, i.e., they allow us to derive the scaling laws in $N$ up to numerical prefactors.
Note that Eq.~\eqref{eqn:MFT:EoMSzBasis} for $n_\mathrm{th} = 0$ has already been obtained in the literature on superradiance using other derivations \cite{Agarwal1970,Rehler1971}.


\bibliography{scibib}



\newpage 
\clearpage
\thispagestyle{empty}
\onecolumngrid
\begin{center}
\textbf{\large Supplemental Material for\\\thetitle}
\end{center}

\begin{center}
\theauthors\\
\emph{\theaffiliations}\\
(Dated: \today)
\end{center}

\setcounter{equation}{0}
\setcounter{figure}{0}
\setcounter{table}{0}
\setcounter{page}{1}
\makeatletter
\renewcommand{\theequation}{S\arabic{equation}}
\renewcommand{\thefigure}{S\arabic{figure}}
\renewcommand{\bibnumfmt}[1]{[S#1]}
\renewcommand{\citenumfont}[1]{#1}

\twocolumngrid


\subsection{Heuristic argument on added noise}
\label{sec:SM:AddedNoise}

In this section, we adapt Caves' derivation of the quantum limit on the added noise of a bosonic linear amplifier \cite{Caves1982} to the phase-preserving spin amplifier considered in this work.  
We start by writing down a minimal description (in terms of Heisenberg equations of motion) of a spin amplifier that amplifies any polarization transverse to the $z$ direction.
The amplification process translates an input state (as encoded in the initial-time $t=0$ collective spin operators $\hat{S}_\alpha \equiv \hat{S}_\alpha(0)$ for $\alpha \in \{x,y,z\}$), to an output state (similarly encoded in the final time $t=T$ Heisenberg-picture spin operators 
$ \hat{S}_\alpha(T) \equiv \hat{T}_\alpha$).  
Assuming linear amplification dynamics suggests writing the the solution of the Heisenberg equations of motion in the form:
\begin{align}
	\hat{T}_x &= G(T) \hat{S}_x + \hat{F}_x ~, 
	\label{eqn:Methods:Heuristic:Tx}\\
	\hat{T}_y &= G(T) \hat{S}_y + \hat{F}_y ~.
	\label{eqn:Methods:Heuristic:Ty}
\end{align}
The first term in each equation captures the linear amplification dynamics, with $G(T)$ denoting the gain. 
The remaining Hermitian operators $\hat{F}_\alpha$ describe all additional terms arising from solving the Heisenberg equations.  
Note that at this stage, we do not make any assumptions about the dynamics of $\hat{S}_z$ and the value of $\hat{S}_z(T) \equiv \hat{T}_z$.

We next \emph{assume} that the average values of the final-time spin operators are fully described by the linear gain terms (e.g.~as is seen in our system for initial states with small transverse polarizations).  As such, the $\hat{F}_\alpha$ can be viewed as zero-mean operators that describe added noise associated with the amplification dynamics.  
We further assume that these noise operators are uncorrelated with the initial spin state, i.e., $\langle \hat{S}_\alpha \hat{F}_\beta \rangle = 0$ and $[\hat{S}_\alpha,\hat{F}_\beta] = 0$.  
In reality, the $\hat{F}_\alpha$ also describe the nonlinear response of our system; we are implicitly assuming that the initial state of our system lets us safely ignore such terms (i.e., the initial transverse polarization is small).

With these assumptions in hand, we can now construct a bound on the size of the added noise.  
We first use the fact that the final-time spin operators must obey canonical spin commutation relations, and hence we must have $[ \hat{T}_x,\hat{T}_y] = i \hat{T}_z$.  This results in the constraint
\begin{align}
	\hat{T}_z = G^2(T) \hat{S}_z - i [\hat{F}_x,\hat{F}_y]~.
	\label{eqn:Methods:HeuristicArgument:KommutatorFxFy}
\end{align}
The fluctuations in the final-time transverse spin operators are given by: 
\begin{align}
	(\mathbf{\Delta} T_{x,y})^2 
	\equiv \langle \hat{T}_{x,y}^2 \rangle - \langle \hat{T}_{x,y} \rangle^2 
	= G^2(T) (\mathbf{\Delta} S_{x,y})^2 + \langle \hat{F}_{x,y}^2 \rangle ~.
\end{align}
Since the amplifier acts identically on the $x$ and $y$ components, $\langle\hat{F}_x^2\rangle$ and $\langle\hat{F}_y^2\rangle$ are identical and we can write
\begin{align}
	\langle \hat{F}_{x,y}^2 \rangle
	&= \frac{1}{4} \langle \{ \hat{F}_+, \hat{F}_- \} \rangle
	\geq \frac{1}{4} \vert \langle [ \hat{F}_+, \hat{F}_- ] \rangle \vert \nonumber \\
	&= \frac{1}{2} \vert \langle [ \hat{F}_x, \hat{F}_y ] \rangle \vert 
	= \frac{1}{2} \vert [G^2(T) \langle \hat{S}_z \rangle - \langle \hat{T}_z \rangle ] \vert~,
\end{align}
where we defined $\hat{F}_\pm = \hat{F}_x \pm i \hat{F}_y$ and used Eq.~\eqref{eqn:Methods:HeuristicArgument:KommutatorFxFy} in the last step.
Using $\vert \partial_\phi \langle \hat{T}_{x,y} \rangle \vert = G(T) \vert \partial_\phi \langle \hat{S}_{x,y} \rangle \vert$, we thus find
\begin{align}
	\frac{(\mathbf{\Delta} T_{x,y})^2}{\vert \partial_\phi \langle \hat{T}_{x,y} \rangle \vert^2} 
	\geq \frac{(\mathbf{\Delta} S_{x,y})^2}{\vert \partial_\phi \langle \hat{S}_{x,y} \rangle \vert^2} + \frac{1}{2 G^2(T)} \frac{\vert G^2(T) \langle \hat{S}_z \rangle - \langle \hat{T}_z \rangle \vert}{\vert \partial_\phi \langle \hat{S}_{x,y} \rangle \vert^2}~.
\end{align}

In our specific spin amplification protocol, we start with a CSS close to a maximally polarized state, i.e., $\langle \hat{S}_z \rangle = 2 (\mathbf{\Delta} S_{x,y})^2 = N/2$, and we interrupt the amplification dynamics at the time $T = t_\mathrm{max}$ when the condition $\langle \hat{T}_z \rangle = 1/2$ holds. 
This yields
\begin{align}
    \frac{(\mathbf{\Delta} T_{x,y})^2}{\vert \partial_\phi \langle \hat{T}_{x,y} \rangle \vert^2} 
    \geq \frac{1}{N} \left[ 2 - \frac{1}{G^2(T) N} \right]~,
\end{align}
i.e., the spin amplification process doubles the input spin-projection noise in the large-$N$ limit.

Note that this argument is \emph{not} a strict theoretical lower bound on the noise that is added during the amplification step. 
Instead, it is a heuristic argument that helps one to develop a sense how much added noise can be expected due to amplification. 
Even though our amplification scheme is conceptually very simple, we numerically find that it does not saturate the prediction of this heuristic argument (see Figs.~\ref{fig:sensitivity}(b) of the main text and Fig.~\ref{fig:nonmarkovianAddedNoise}).
Hence, an interesting question for future research is to refine this argument to see if the actual lower bound on the added noise in the large-$N$ limit is larger than predicted here.

\begin{figure*}
    \centering
    \subfigure[]{
        \includegraphics[width=0.48\textwidth]{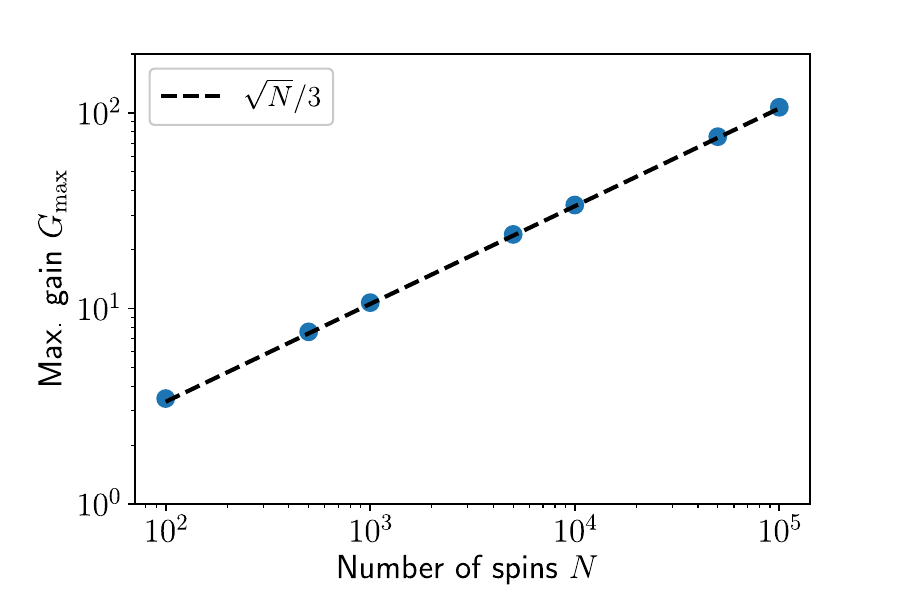}
    }
    \subfigure[]{
        \includegraphics[width=0.48\textwidth]{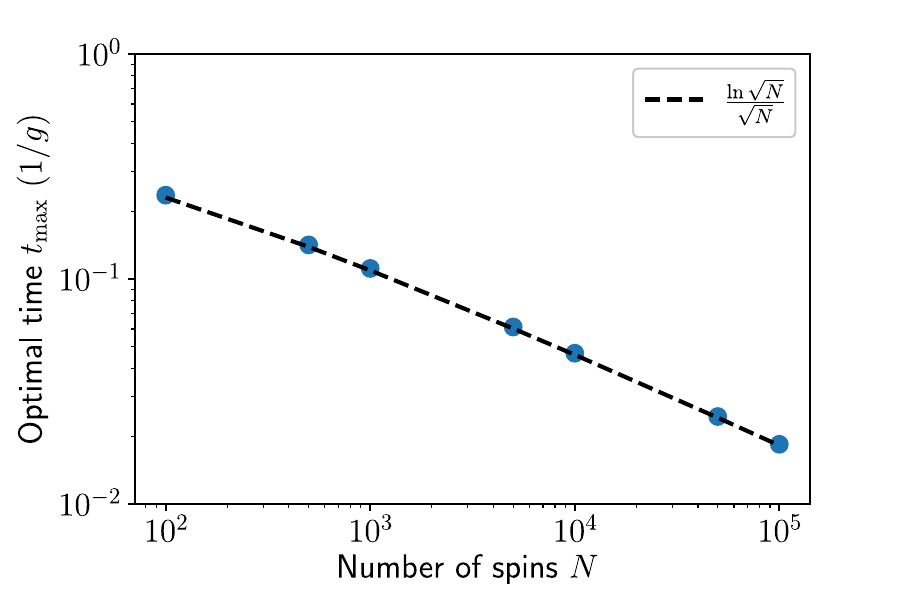}
    }
    \caption{
        Mean-field theory analysis of the coherent amplification protocol (i.e., in the limit of an undamped cavity).
        Scaling of
        \textbf{(a)} 
            the maximum gain $G_{\rm max}$ and 
        \textbf{(b)}
            the corresponding time $t_\mathrm{max}$ as a function of spin number $N$. 
        Results have been obtained by numerical solution of the MFT equations~\eqref{eq:mftCoherentEqFirst} to~\eqref{eq:mftCoherentEqLast}.
        Blue dots correspond to simulation data, while dashed black lines show the respective scaling behavior. 
    }
    \label{fig:mftCoherentScaling}
\end{figure*}

\subsection{Additional details on the dissipative amplification protocol}

\subsubsection{Signal-dependence of the amplification time and dynamic range}
\label{sec:SM:DynamicRange}

In this section, we discuss the dependence of the amplification tiem $t_\mathrm{max}$ and the maximum gain $G_\mathrm{max}$ on the initial tilt angle $\phi$, and we provide additional information on the dynamic range of the spin amplifier.

\begin{figure*}
	\centering
	\subfigure[]{
		\includegraphics[width=0.48\textwidth]{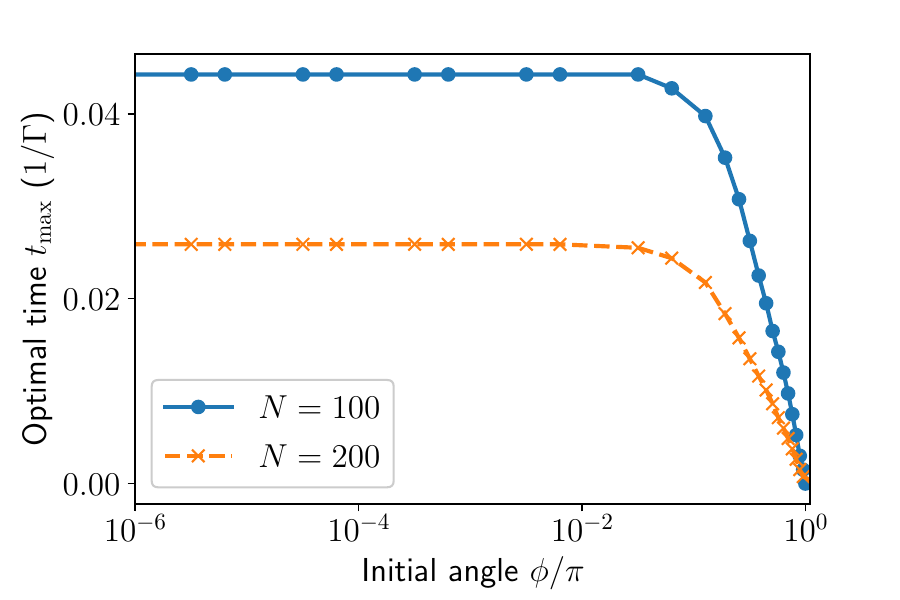}
	}
	\subfigure[]{
		\includegraphics[width=0.48\textwidth]{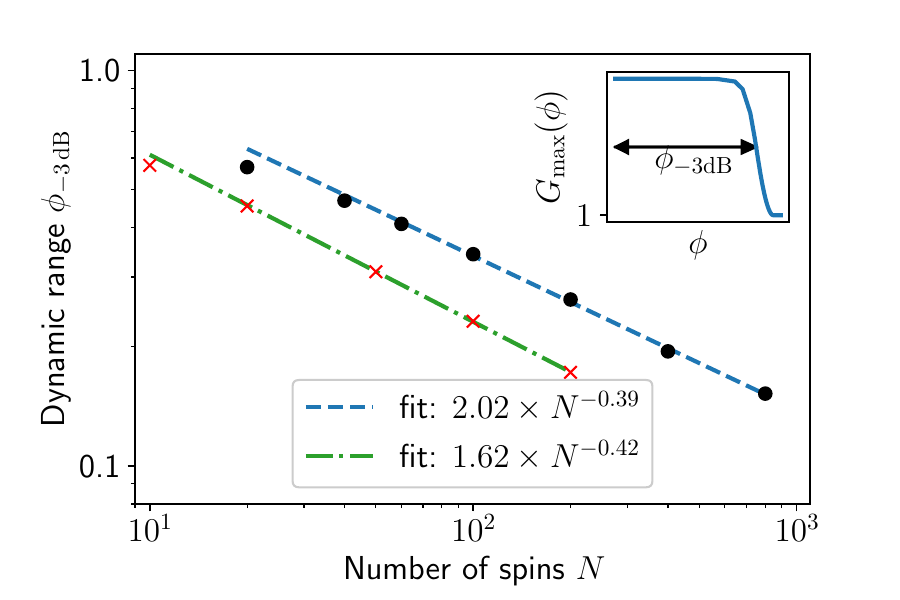}
	}
	\caption{
		(a) Time $t_\mathrm{max}$ to reach maximum gain as a function of the initial tilt angle $\phi$ away from the north pole of the collective Bloch sphere, obtained by a numerically exact integration of Eq.~\eqref{eqn:system:QME_spin_only} of the main text for $\Gamma=1$ and $\gamma_\mathrm{rel} =0$.  
			In the limit $\phi \to 0$, $t_\mathrm{max}$ tends towards a constant value given by Eq.~\eqref{eqn:amplification:intuition:tmax} of the main text. 
		(b) Scaling of the dynamic range $\phi_{-3\mathrm{dB}}$ (see Sec.~\ref{sec:SM:DynamicRange} for a definition) with the number of spins $N$ for the dissipative spin amplification scheme (black circles and dashed blue fit line) and the OAT-twist-untwist amplification scheme (red crosses and dash-dotted green fit line). 
			The two data points with smallest $N$ have been excluded from the fit. 
	}
	\label{fig:SM:DynamicRange}
\end{figure*}

If all $N$ spins are initialized in the excited state, (quantum) fluctuations will seed the superradiant decay towards the collective ground state. 
If the initialization to the excited state is imperfect but the deviation of the collective spin vector from the north pole of the Bloch sphere is smaller than the level of fluctuations, the seeding of the superradiant decay is still dominated by fluctuations.
As a consequence, the time $t_\mathrm{max}$ to reach maximum gain becomes independent of the signal angle $\phi$ for sufficiently small $\phi$ [as shown in Eq.~\eqref{eqn:amplification:intuition:tmax} of the main text] and it depends only on the collective decay rate $\Gamma$ and the number of spins $N$ [as shown in Fig.~\ref{fig:principle}(a) of the main text].
Quantum metrology protocols operate in this regime, $\phi \ll 1$.
Figure~\ref{fig:SM:DynamicRange}(a) shows how $t_\mathrm{max}$ ultimately becomes $\phi$ dependent and decreases to zero as $\phi$ increases.

In the regime where $t_\mathrm{max}$ is independent of $\phi$, the maximum gain $G_\mathrm{max}$ defined in Eq.~\eqref{eqn:amplification:gain} of the main text is also independent of $\phi$.
This is important because it establishes a simple linear relation between the amplified signal and the initial small value of $\phi$ after a constant amplification time $t_\mathrm{max}$. 
For larger values of $\phi$, $G_\mathrm{max}$ decreases as a function of $\phi$, as sketched in the inset of Fig.~\ref{fig:SM:DynamicRange}(b).
The dynamic range of an amplifier quantifies the range of $\phi$ values over which $G_\mathrm{max}$ is constant and a linear amplification relation is obtained. 
We determine the dynamic range of the spin amplifier numerically by rescaling the gain $G_\mathrm{max}(\phi)$ to the range $[0,1]$ by defining
\begin{align*}
	\delta G_\mathrm{max}(\phi) = \frac{G_\mathrm{max}(\phi=0) - G_\mathrm{max}(\phi)}{G_\mathrm{max}(\phi=0) - 1}
\end{align*}
and numerically finding the angle $\phi_{-3\mathrm{dB}}$ where $1 - \delta G_\mathrm{max}(\phi)$ has decreased to $1/2$. 
The dynamic range $\phi_{-3\mathrm{dB}}$ is plotted in the main panel of Fig.~\ref{fig:SM:DynamicRange}(b).  

\subsubsection{Calibration of $t_\mathrm{max}$ and $G_\mathrm{max}$}
\label{sec:SM:Calibration}
Note that precise knowledge of the number $N$ of spins is \emph{not} required to calibrate the maximum amplification time $t_\mathrm{max}$ and the maximum gain $G_\mathrm{max}$. 
Instead, these quantities can be experimentally determined by the following measurement: 
The spin system is repeatedly prepared in a coherent spin state in the $y$-$z$ plane, which is tilted away from the $z$ axis by a small angle $\phi < \phi_{-3\mathrm{dB}}$. 
This can be achieved by standard control pulses. 
One then switches on the superradiant decay for different time delays $t$ and measures the final transverse $S_y$ polarization. 
In this way, one maps out $S_y(t)$ as a function of time, which allows one to determine the time-dependent gain $G(t)$ defined in Eq.~\eqref{eqn:amplification:gain} of the main text [see Fig.~\ref{fig:protocol}(b) of the main text]. 
From this emasurement, one can determine $t_\mathrm{max}$ and $G_\mathrm{max}$.

\subsubsection{Analysis of timing errors in the amplification step}
\label{sec:SM:TimingErrors}

\begin{figure}
	\centering
	\includegraphics[width=0.48\textwidth]{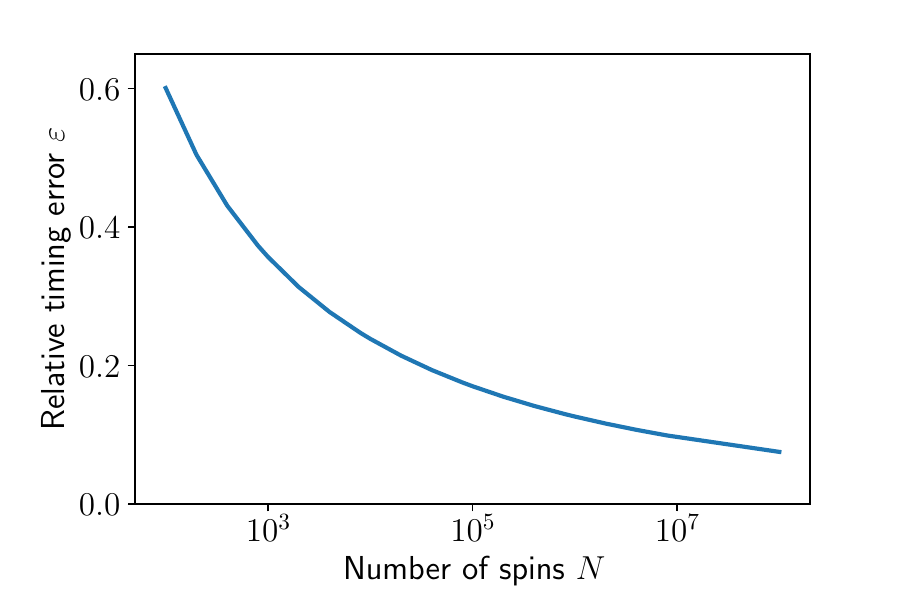}
	\caption{
		Maximum relative timing error $\varepsilon = \delta t/t_\mathrm{max}$ given in Eq.~\eqref{eqn:SM:WorstCaseTimingErrorAnalysis} with the number of spins $N$.
		For a relative timing error below the blue line, the uncertainty $(\mathbf{\Delta} \phi)_{\delta t}^2$ due to imperfect control of the amplification time $t_\mathrm{max}$ is smaller than the uncertainty due to spin-projection noise. 
	}
	\label{fig:SM:TimingErrors}
\end{figure}

The amplification step of the modified Ramsey sequence shown in Fig.~\ref{fig:protocol}(a) of the main text requires time-dependent control of the collective decay rate $\Gamma(t)$. 
Both the optimal amplification time $t_\mathrm{max}$ and the FWHM of the gain peak decrease with increasing $N$, as shown in Figs.~\ref{fig:principle}(a) and (b) of the main text, respectively.
Therefore, one may worry that timing errors in the control of $\Gamma(t)$ could limit the performance of our scheme. 
This is not the case.

Timing errors will contribute an additional term to the uncertainty budget in Eq.~\eqref{eqn:Protocol:TotalDeltaPhiSquared} of the main text, which can be estimated using mean-field theory.
Using Eq.~\eqref{eqn:amplification:Sxydynamics} of the main text, we find in the limit $N \gg 1$
\begin{align*}
	G(t_\mathrm{max} + \delta t) 
	\approx G(t_\mathrm{max}) \left[ 1 - \frac{\Gamma^2}{8} N^2 \delta t^2 \right] ~,
\end{align*}
i.e., we are insensitive to timing fluctuations $\delta t$ to first order in $\delta t$, but they decrease the gain quadratically.
The corresponding uncertainty in the estimation of $\phi$ is
\begin{align*}
	(\mathbf{\Delta}\phi)^2_{\delta t} 
	= \left(- \frac{\phi}{G(t_\mathrm{max})} \frac{\mathrm{d} G}{\mathrm{d} \delta t} \right)^2
	= \phi^2 \frac{N^4 \Gamma^4}{64} \delta t^4~,
\end{align*}
and the overall uncertainty is
\begin{align*}
	(\mathbf{\Delta}\phi)^2 
	= \frac{1 + \sigma_\mathrm{add}^2}{N} + \phi^2 \frac{N^4 \Gamma^4}{64} \delta t^4 + \frac{\Xi_\mathrm{det}^2}{G(t_\mathrm{max})^2 N}~.
\end{align*}
The third, detection-noise term can be ignored if $N$ and, thus, $G(t_\mathrm{max})$ are large enough. 
The second, timing fluctuations term vanishes in the metrologically relevant limit $\phi \to 0$.

As a worst-case estimate, we will now assume that $\phi$ does not vanish but is given by the dynamic range of the amplifier discussed in Sec.~\ref{sec:SM:DynamicRange}, $\phi \approx 1/N^{0.4}$. 
From Fig.~\ref{fig:principle}(a) of the main text, we know that one must be able to switch $\Gamma(t)$ on a timescale $t_\mathrm{max} \propto \ln(N)/\Gamma N$. 
Assuming there is a relative error $\varepsilon$ in the timing, $\delta t = \varepsilon t_\mathrm{max}$, the overall uncertainty can be rewritten as
\begin{align*}
	(\mathbf{\Delta}\phi)^2
	\approx \frac{1 + \sigma_\mathrm{add}^2}{N} + \frac{1}{N^{0.8}} \frac{\ln^4(N)}{64} \varepsilon^4~.
\end{align*}
The first, projection-noise term dominates if the relative timing error satisfies
\begin{align}
	\varepsilon \lesssim \sqrt[4]{\frac{64 (1 + \sigma_\mathrm{add}^2) N^{0.8}}{N \ln^4(N)}}~,
	\label{eqn:SM:WorstCaseTimingErrorAnalysis}
\end{align}
which puts very moderate requirements on the timing error, as shown in Fig.~\ref{fig:SM:TimingErrors}.

\subsubsection{Impact of the signal field during the amplification step}
\label{sec:SM:SignalField}

In the idealized protocol shown in Fig.~\ref{fig:protocol}(a) of the main text, the signal is imprinted to the sensing state only during the signal acquisition interval from $t_1$ to $t_2$, and it is switched off during the subsequent amplification step from $t_2$ to $t_3$. 
In practice, it may not be possible to switch off the signal. 
Therefore, we consider a modified quantum master equation
\begin{align*}
	\frac{\mathrm{d}}{\mathrm{d} t} \hat{\rho} = - i \left[ \omega_\mathrm{sig} \hat{S}_z , \hat{\rho} \right] + \Gamma \mathcal{D}[\hat{S}_-] \hat{\rho} ~,
\end{align*}
where the Hamiltonian $\hat{H} = \omega_\mathrm{sig} \hat{S}_z$ represents the collective rotation about the $z$ axis due to the signal to be sensed. 
The effect of the (unknown) signal can be gauged away by switching to a rotating frame $\hat{\chi}(t) = e^{i \omega_\mathrm{sig} t \hat{S}_z} \hat{\rho}(t) e^{-i \omega_\mathrm{sig} t \hat{S}_z}$, in which we recover the pure superradiant decay dynamics described by Eq.~\eqref{eqn:system:QME_spin_only} of the main text (for $\gamma_\mathrm{rel} = 0$), 
\begin{align*}
	\frac{\mathrm{d}}{\mathrm{d} t} \hat{\chi} = \Gamma \mathcal{D}[\hat{S}_-] \hat{\chi} ~.
\end{align*}
Maximum gain $G_\mathrm{max}$ and the corresponding amplification time $t_\mathrm{max}$ are thus unaffected by the presence of the signal. 
However, the final state in the lab frame after the amplfication step will be rotated by an angle $\phi_\mathrm{sig} = \omega_\mathrm{sig} t_\mathrm{max}$ as compared to the unperturbed dynamics, 
\begin{align*}
	\hat{\rho}(t_\mathrm{max}) = e^{-i \omega_\mathrm{sig} t_\mathrm{max} \hat{S}_z} \hat{\chi}(t_\mathrm{max}) e^{+i \omega_\mathrm{sig} t_\mathrm{max} \hat{S}_z} ~.
\end{align*}
Its collective spin vector will be rotated in the equatorial plane away from the $y$ axis by an angle $\phi_\mathrm{max}$.

In the typical quantum metrology setting, the precession frequency is very small, $\omega_\mathrm{sig} \to 0$, and the additional rotation during the amplification step can be ignored since $\phi_\mathrm{sig} \ll 1$. 
In particular, for our scheme, the amplification time \emph{decreases} if the number of spins is increased, $t_\mathrm{max} \propto \ln(N)/N \approx 1/N$, which suppresses $\phi_\mathrm{sig}$ even more.

Note that, even if $\phi_\mathrm{sig}$ was not negligible, it could be easily determined by repeating the protocol shown in Fig.~\ref{fig:protocol}(a) of the main text twice, measuring $S_x(t_\mathrm{max}) \propto \sin (\phi_\mathrm{sig})$ and $S_y(t_\mathrm{max}) \propto \cos(\phi_\mathrm{sig})$ in the two runs, respectively.

\subsection{Mean-field theory in the limit of an undamped cavity}
\label{sec:SM:TavisCummings}

As discussed in the main text, amplification can also be achieved in a regime where the cavity degree of freedom cannot be eliminated adiabatically, i.e., $\sqrt{N} g \gg \kappa$. 
In this section, we use MFT to analyze the coherent limit of Eq.~\eqref{eqn:system:QME_spin_and_cavity} of the main text (i.e., $\kappa \to 0$).
MFT lets us explore substantially larger system sizes than direct numerical simulation of the Schr\"odinger equation (which were used to generate the data shown in Fig.~\ref{fig:nonmarkovian} of the main text).
We consider the resonant ($\omega_\mathrm{cav} = \omega$) Tavis-Cummings Hamiltonian~\eqref{eqn:UnitaryLimit:TCHamiltonian} of the main text, which reads in a frame rotating at the cavity frequency
\begin{align}
    \hat{H}_{\rm TC} &=  g\left( \hat{a}^{\dagger} \hat{S}_{-} +  \hat{a}\hat{S}_{+} \right)~,
\end{align}
and we consider a separable initial state consisting of the cavity mode in a vacuum, and the spins maximally polarized in a state $e^{i \phi \hat{S}_{x}} \vert \!\!\uparrow \dots \uparrow \rangle$.
Using a second-order cumulant expansion \cite{Kubo1962}, we can derive a closed set of equations of motion (EOMs) for the spin-cavity system. 
Introducing the cavity quadrature operators $\hat{Q} = \left( \hat{a}^{\dagger} + \hat{a} \right)/\sqrt{2}$, and $\hat{P} = i \left( \hat{a}^{\dagger} - \hat{a} \right)/\sqrt{2}$ as well as the notation $Q = \langle \hat{Q} \rangle$, $C_{Px} = \langle (\hat{P} \hat{S}_x + \hat{S}_x \hat{P})\rangle/2 - \langle \hat{P} \rangle \langle \hat{S}_x \rangle$, etc., we can readily write down the set of MFT equations of motion:
\begin{align} 
    \dot Q  &= -\sqrt{2} g S_y~,  \label{eq:mftCoherentEqFirst} \displaybreak[1]\\
    \dot S_y  &= -\sqrt{2} g \left(C_{Qz} +S_z  Q \right)~,\displaybreak[1]\\
    \dot S_z  &= +\sqrt{2} g \left(C_{Px} +C_{Qy} +S_y  Q \right)~,\displaybreak[1]\\
    \dot C_{QQ}  &= -2 \sqrt{2} g C_{Qy} ~,\displaybreak[1]\\
    \dot C_{Qy}  &= -\sqrt{2} g \left(C_{Qz}  Q +C_{QQ}  S_z +C_{yy} \right)~,\displaybreak[1]\\
    \dot C_{Qz}  &= +\sqrt{2} g \left(C_{Qy}  Q +C_{QQ}  S_y -C_{yz} \right)~,\displaybreak[1]\\
    \dot C_{PP}  &= -2 \sqrt{2} g C_{Px} ~,\displaybreak[1]\\
    \dot C_{Px}  &= -\sqrt{2} g \left(C_{PP}  S_z +C_{xx} \right) ~,\displaybreak[1]\\
    \dot C_{xx}  &= -2 \sqrt{2} g C_{Px}  S_z~,\displaybreak[1]\\
    \dot C_{yy}  &= -2 \sqrt{2} g \left(C_{yz}  Q +C_{Qy}  S_z \right) ~,\displaybreak[1]\\
    \dot C_{yz}  &= \sqrt{2} g \left[ C_{Qy}  S_y +\left(C_{yy} -C_{zz} \right) Q -C_{Qz}  S_z \right]~,\displaybreak[1]\\
    \dot C_{zz}  &= 2 \sqrt{2} g \left(C_{yz}  Q +C_{Qz}  S_y \right)~.
    \label{eq:mftCoherentEqLast}
\end{align}
All expectation values (within the second-order cumulant expansion approximation) which are not explicitly shown above have only a trivial evolution, i.e., they remain zero.
As an aside, we note that there are two constraints that are also satisfied by our system. 
The first one is total-angular-momentum conservation, which lets us write 
\begin{align}
    S_x^{2} + C_{xx} +   S_y^{2} + C_{yy} +   S_z^{2} + C_{zz}  &=  \frac{N}{2} \left( \frac{N}{2} + 1 \right)~,
    \label{eq:mftCoherentConst1}
\end{align}
and the second one is conservation of the total excitation number, which yields
\begin{align}
    \frac{1}{2} \left( C_{QQ} +  Q^{2} + C_{PP} + P^{2}  - 1  \right)  + S_{z}  &=  \frac{N}{2}~.
    \label{eq:mftCoherentConst2}
\end{align}
Either (or both) of the above constraints could be used to further reduce the full set of EOMs shown above.

\subsubsection{Scaling of the gain}

Solving the system of equations~\eqref{eq:mftCoherentEqFirst} to~\eqref{eq:mftCoherentEqLast} numerically, we can study the scaling of the maximum gain $G_\mathrm{max}$ as well as the the corresponding time scale $t_\mathrm{max}$ for systems with very large spin number $N$, a regime which is inaccessible by numerical integration of the Schr\"odinger equation. 
Figure~\ref{fig:mftCoherentScaling}(a) shows the scaling of $G_\mathrm{max}$, while panel (b) displays the corresponding time $t_\mathrm{max}$ required to reach the optimal gain value, both as a function of spin number $N$. 
We observe scaling behavior that very closely resembles the one observed for smaller-$N$ Schr\"odinger-equation simulations shown in Fig.~\ref{fig:nonmarkovian} of the main text. 
Like in the dissipative version of our amplification protocol, the gain scales $\propto \sqrt{N}$, while $t_\mathrm{max}$ has a parametrically slower $N$-scaling, $t_\mathrm{max} \propto \ln \sqrt{N} /\sqrt{N}$.

\subsubsection{Semi-classical theory}

\begin{figure}
    \centering
    \includegraphics[width=0.5\textwidth]{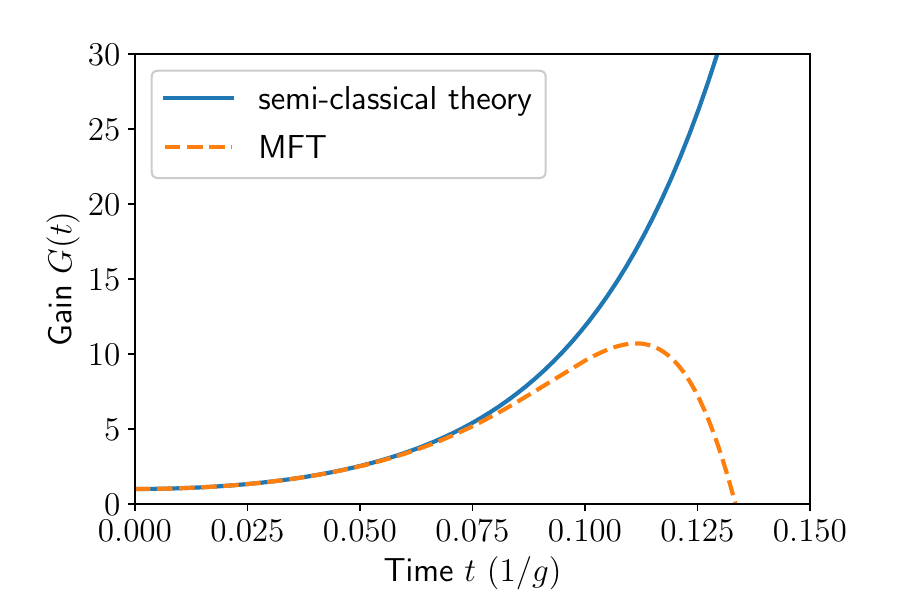}
    \caption{
        Comparison between mean-field theory and a semiclassical approximation of the amplification dynamics.
        Gain as a function of time obtained from MFT (given by Eqs.~\eqref{eq:mftCoherentEqFirst} to~\eqref{eq:mftCoherentEqLast}) as well as a semiclassical approximation (given by Eqs.~\eqref{eq:mftCoherentEqSimple1} to~\eqref{eq:mftCoherentEqSimple2}) for $N=1000$. 
        The semiclassical approximation cannot be used to properly describe the evolution of the cavity-spin system. 
    }
    \label{fig:mftCoherentFullVsClass}
\end{figure}

Following work in \cite{Keeling2009}, one might hope that much of the core physics or the amplification in the coherent limit of our amplification protocol could be captured by a semiclassical approximation, where all fluctuations (i.e., the covariances) are neglected. 
Such a case would let us simplify the set of Eqs.~\eqref{eq:mftCoherentEqFirst} to~\eqref{eq:mftCoherentEqLast} to
\begin{align} 
    \dot Q  &= -\sqrt{2} g S_y ~,\\
    \dot S_y  &= -\sqrt{2} g S_z  Q~, \\
    \dot S_z  &= +\sqrt{2} g S_y  Q ~,
\end{align}
which, using Eq.~\eqref{eq:mftCoherentConst2} could be further reduced to
\begin{align} 
    \dot Q  &= -\sqrt{2} g S_y~,  \label{eq:mftCoherentEqSimple1} \\
    \dot S_y  &= -\sqrt{2} g \left(\frac{N}{2} - \frac{1}{2} \left( Q^{2} - 1 \right)\right) Q~.
\label{eq:mftCoherentEqSimple2}
\end{align}
These semiclassical equations indeed predict that $S_y$ will grow at short time.
However, solving Eqs.~\eqref{eq:mftCoherentEqSimple1} and~\eqref{eq:mftCoherentEqSimple2} numerically, one finds that $S_y$ increases monotonically over a time scale much longer than $t_\mathrm{max}$ obtained from MFT, see Fig.~\ref{fig:mftCoherentFullVsClass}. 
Therefore, it is clear that one must include the effects of fluctuations to properly describe the amplification physics in the coherent $\kappa \rightarrow 0$ limit.

\subsection{Added noise $\sigma_\mathrm{add}^2$ in the limit of an undamped cavity}

\begin{figure}
	\centering
        \includegraphics[width=0.47\textwidth]{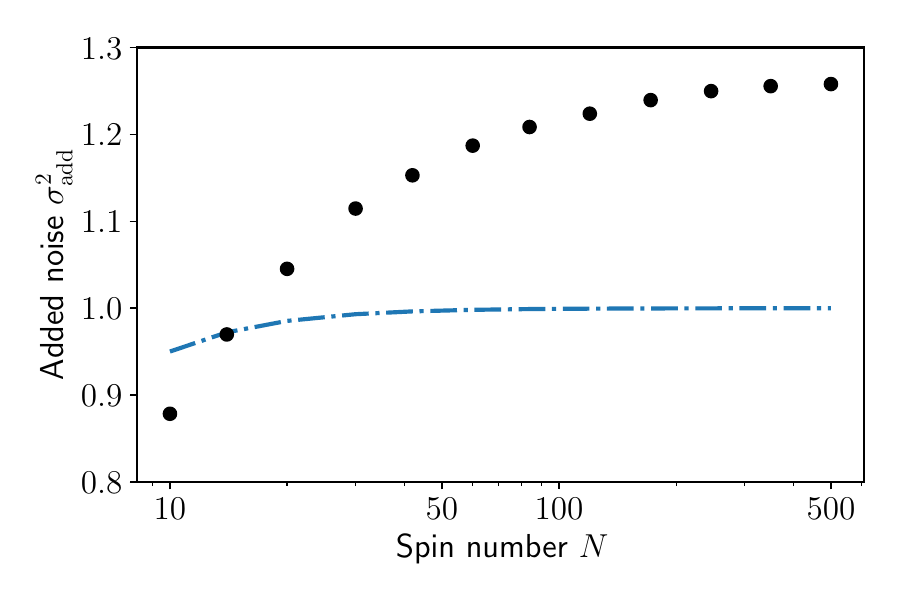}        
    \caption{
        Added noise $\sigma_\mathrm{add}^2$ in the coherent spin-amplification protocol, calculated by numerical integration of the Schr\"odinger equation using the resonant Tavis-Cummings Hamiltonian given in Eq.~\eqref{eqn:UnitaryLimit:TCHamiltonian} of the main text with $\omega_\mathrm{cav} = \omega$.
        The quantity $\sigma_\mathrm{add}^2$ at the optimal evolution time $t_\mathrm{max}$ (data points) is close to the amount of noise $1 - 1/G_\mathrm{max}^2 N$ that is expected based on a heuristic argument valid in the limit $N \gg 1$ (dashed-dotted line). 
}
	\label{fig:nonmarkovianAddedNoise}
\end{figure}

Above, we gave a heuristic argument (assuming linear amplification dynamics) which showed that the added noise $\sigma_\mathrm{add}^2$ (defined in Eq.~\eqref{eqn:Sensitivity:AddedNoise} of the main text), should approximately follow the relation 
\begin{align}
    \sigma_\mathrm{add}^2 \ge 1 - \frac{1}{G_\mathrm{max}^2 N}
\end{align}
in the limit of large spin number $N$. 
For dissipative amplification, we found numerically that the added noise stays close to this heuristic expectation and tends to $\sigma_\mathrm{add}^2 \approx 1.3$ in the large-$N$ limit (see Fig.~\ref{fig:sensitivity}(b) of the main text). 
Here, we show that similar behavior is also present in the case of purely coherent evolution, where $\kappa \to 0$.
Specifically, Fig.~\ref{fig:nonmarkovianAddedNoise} shows $\sigma_\mathrm{add}^2$ as a function of spin number $N$. 
Black dots correspond to data obtained from solving the Schr\"odinger equation numerically, while the dashed blue line indicates $1 - 1/G_\mathrm{max}^2 N$. 
Curiously, we see similar behavior to the dissipative case and observe $\sigma_\mathrm{add}^2 \lesssim 1.3$ in the large-$N$ limit. 
Consequently, our amplification protocol could also be useful in the coherent limit if the readout noise is not extremely large and one cares about approaching the SQL. 

\subsection{Comparison between semiclassical and mean-field-theory dynamics of the transverse magnetization}
\label{sec:SM:SemiclassicsVsMFTDissipative}

In this section, we briefly discuss the effect of radiation damping in NMR systems \cite{Bloembergen1954} (which is somewhat related to superradiance) and we show why our dissipative amplification scheme is very different from sensing protocols based on radiation damping \cite{Augustine2000,Walls2007}.

In NMR setups, the time-dependent magnetic field $\vec{M} = (M_x, M_y, M_z)$ of a spin vector precessing about the quantization axis (typically the $z$ axis) induces a current in the measurement coil. 
This current generates a magnetic field which aims to rotate the spin vector back to its stable equilibrium position. 
This process is called radiation damping \cite{Bloembergen1954} and can be described by classical Bloch equations 
\begin{align}
	\frac{\mathrm{d}}{\mathrm{d} t} M_{x,y} &= - \gamma M_z M_{x,y} - \frac{1}{T_2} M_{x,y} ~, 
	\label{eqn:SM:NMRvsSR:Blochxy} \\
	\frac{\mathrm{d}}{\mathrm{d} t} M_z &= + \gamma (M_x^2 + M_y^2) - \frac{1}{T_1} \left( M_z - \frac{N}{2} \right) ~,
	\label{eqn:SM:NMRvsSR:Blochz}
\end{align}
where $T_1$ and $T_2$ are the relaxation and dephasing time, respectively, and $\gamma$ is the gyromagnetic ratio of the spins.
These equations are similar to the \emph{semiclassical} equations of motion for superradiant decay \cite{Rehler1971}, which can be obtained from Eq.~\eqref{eqn:system:QmeCollectiveAndLocalDissipation} of the main text by deriving the equations of motion for $S_k \equiv \langle \hat{S}_k \rangle$, $k \in \{x,y,z\}$, and factorizing all higher-order expectation values $\langle \hat{S}_j \hat{S}_k \rangle \approx \langle \hat{S}_j \rangle \langle \hat{S}_k \rangle$, 
\begin{align}
	\frac{\mathrm{d}}{\mathrm{d} t} S_{x,y} &= + \Gamma S_z S_{x,y} - \left( \frac{\Gamma}{2} + \gamma_\phi + \frac{\gamma_\mathrm{rel}}{2} \right) S_{x,y} ~, 
	\label{eqn:SM:NMRvsSR:XY}\\
	\frac{\mathrm{d}}{\mathrm{d} t} S_z &=  - \Gamma \left( S_x^2 + S_y^2 \right) - \Gamma S_z - \gamma_\mathrm{rel} \left( S_z + \frac{N}{2} \right) ~.
	\label{eqn:SM:NMRvsSR:Z}
\end{align} 
Note that the two sets of differential equations have opposite stable steady-state solutions: $M_{x,y}(t \to \infty) = 0$ and $M_z(t \to \infty) = +N/2$ as opposed to $S_{x,y}(t \to \infty) = 0$ and $S_z(t \to \infty) = - N/2$. 

\begin{figure*}
	\centering
	\subfigure[]{
		\includegraphics[width=0.48\textwidth]{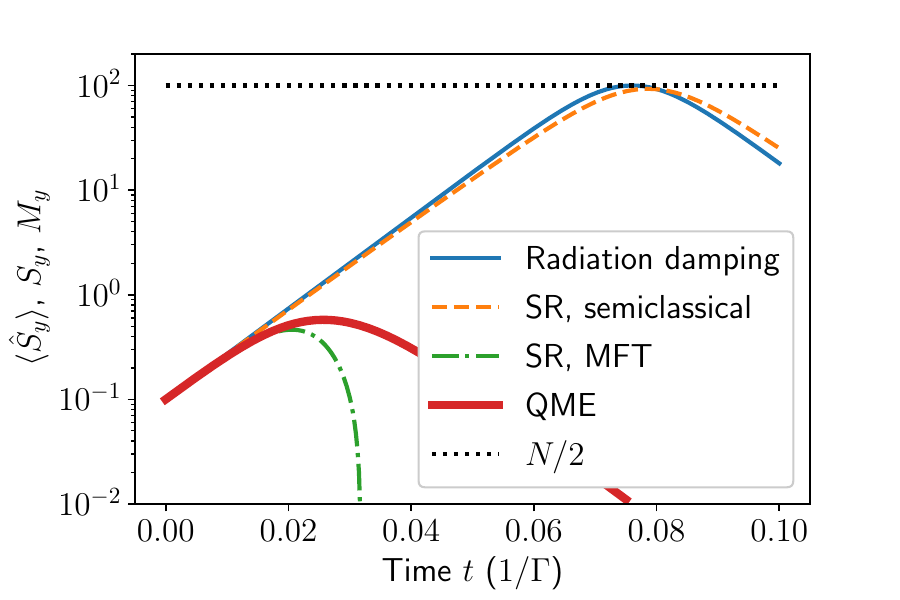}
	}
	\subfigure[]{
		\includegraphics[width=0.48\textwidth]{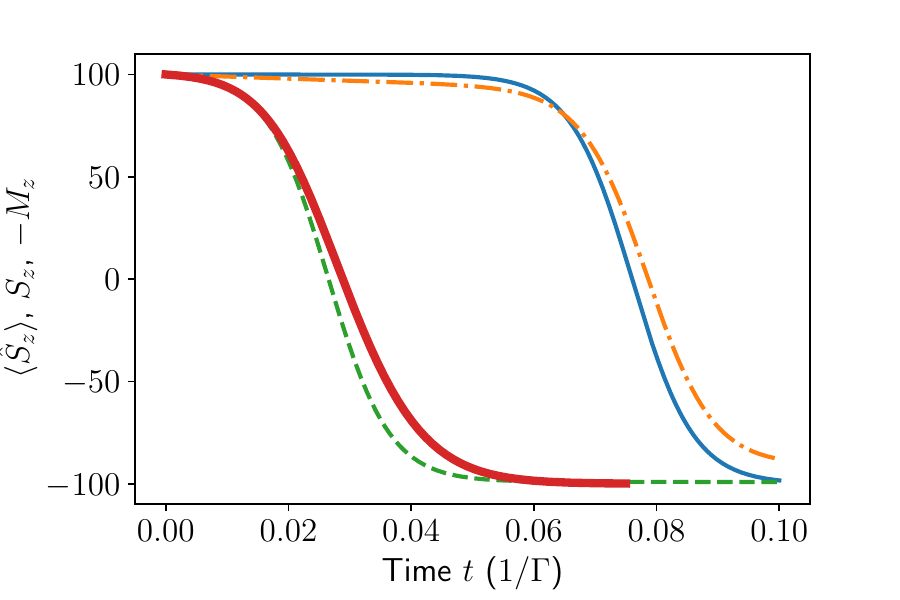}
	}
	\caption{
		(a) $y$ magnetization and (b) $z$ magnetization for radiation damping [Eqs.~\eqref{eqn:SM:NMRvsSR:Blochxy} and~\eqref{eqn:SM:NMRvsSR:Blochz}], a semiclassical treatment of superradiance (SR) [Eqs.~\eqref{eqn:SM:NMRvsSR:XY} and~\eqref{eqn:SM:NMRvsSR:Z}], a mean-field treatment of superradiance [Eqs.~\eqref{eqn:MFT:Sy} to~\eqref{eqn:MFT:Cyz}], and an numerically exact solution of the quantum master equation (QME) Eq.~\eqref{eqn:system:QmeCollectiveAndLocalDissipation}.
		The parameters are $N=200$, $\Gamma = \gamma = 1$, $\gamma_\phi = \gamma_\mathrm{rel} = 0$ and $T_1, T_2 \to \infty$.
		The initial state is a coherent spin state in the $z$-$y$ plane tilted away from a perfectly inverted state (i.e., a state pointing along the $+z$ direction for superradiance, along the $-z$ direction for radiation damping) by an angle $\phi = 0.001$. 
		While all methods describe the dynamics of the $z$ magnetization qualitatively correctly, radiation damping and a semiclassical treatment of superradiance fail to describe the dynamics of the $y$ magnetization correctly and predict a maximum transverse magnetization of $N/2$ (dotted black line). 
		The same is true in the case of a unitary Tavis-Cummings interaction, discussed in Sec.~\ref{sec:SM:TavisCummings} and shown in Fig.~\ref{fig:mftCoherentFullVsClass}.
	}
	\label{fig:SM:NMRvsSR}
\end{figure*}

For $S_z(0) \approx N/2 \gg 1/2$ one can neglect the term $-\Gamma S_{x,y}/2$ in Eq.~\eqref{eqn:SM:NMRvsSR:XY} at short times.
In the absence of local relaxation and dephasing, $\gamma_\mathrm{rel}, \gamma_\phi \to 0$ and $T_1, T_2 \to \infty$, one then finds that Eqs.~\eqref{eqn:SM:NMRvsSR:Blochxy} and~\eqref{eqn:SM:NMRvsSR:Blochz} are identical to Eqs.~\eqref{eqn:SM:NMRvsSR:XY} and~\eqref{eqn:SM:NMRvsSR:Z} upon the substitution $M_k \to S_k$, except for the second term $- \Gamma S_z$ in Eq.~\eqref{eqn:SM:NMRvsSR:Z}. 
This term captures spontaneous decay and is crucial to ``seed'' the superradiant decay dynamics out of a perfectly inverted state. 
Since is not present in the radiation damping equation~\eqref{eqn:SM:NMRvsSR:Blochz}, a magnetization aligned exactly along the $+z$ or $-z$ direction is a stable solution of Eqs.~\eqref{eqn:SM:NMRvsSR:Blochxy} and~\eqref{eqn:SM:NMRvsSR:Blochz} in the absence of $T_1$ relaxation. 
Seeding of the radiation damping dynamics must be introduced manually by considering thermal fluctuations of the current in the pickup coil \cite{Augustine2000}, experimental imperfections which cause a small deviations from a perfectly inverted state \cite{Augustine2000}, and dipole-dipole interactions between the spins \cite{Yukalov1995}. 
These effects will cause the magnetization to ultimately flip back to the stable orientation along the $+z$ direction and lead to a large transient magnetization in the $x$-$y$ plane, similar to a superradiant emission burst.

Walls \emph{et al.} proposed to use the time delay of this peak in the transverse magnetization for sensing \cite{Walls2007}.
They consider a system consisting of a solute in solution. 
The solvent spins are initialized in the metastable state, i.e., they are antialigned with the external magnetic field. 
The initial transverse magnetization of the solute spins triggers the solvent spins to flip back to the stable state. 
If the solute's magnetization is larger than the scale of the fluctuations around the metastable state, the delay time depends on the magnitude of the solute's initial magnetization \cite{Augustine2000,Walls2007}.
For a smaller solute magnetization, the delay time becomes independent of the state of the solvent spins and the radiation-damping-based scheme becomes insensitive.

In our scheme, such a situation would correspond to an initial tilt angle $\phi$ larger than the fluctuations of the sensing state. 
While one could in principle reduce the level of thermal fluctuations by cooling the setup, unavoidable quantum fluctuations of the sensing state will pose a strict lower bound on the minimum detectable angle $\phi$ in the quantum sensor.
However, quantum metrology protocols do operate in the regime where the angle $\phi$ is much \emph{smaller} than the scale of the quantum or thermal fluctuations.
In this regime, the delay time is \emph{independent} of $\phi$ [as shown in Eq.~\eqref{eqn:amplification:intuition:tmax} of the main text] and, thus, the radiation-damping-based amplification scheme is useless for quantum metrology.

Instead of focusing on the delay time, our scheme measures the amplitude of the transverse magnetization peak, which remains sensitive to the initial tilt angle $\phi$ [as shown in Eq.~\eqref{eqn:amplficiation:Sytmax} of the main text].
We stress that this amplitude dynamics \emph{cannot} be generated by the classical backaction due to radiation damping.
Figure~\ref{fig:SM:NMRvsSR} compares the decay dynamics due to radiation damping, the semiclassical equations of motion for superradiant decay, the mean-field theory for superradiant decay given by Eqs.~\eqref{eqn:MFT:Sy} to~\eqref{eqn:MFT:Cyz} of the main text, and results obtained by numerically exact integration of the quantum master equation ~\eqref{eqn:system:QmeCollectiveAndLocalDissipation} of superradiant decay.
As expected from the above discussion, the Bloch equations of radiation damping and the semiclassical equations of motion for superradiance predict very similar dynamics. 
While they manage to capture the dynamics of the $z$ component of the magnetization qualitatively, they fail completely to describe the dynamics of the transverse magnetiation in the $x$-$y$ plane, which is at the heart of our spin amplification scheme. 
This is due to the fact that Eqs.~\eqref{eqn:SM:NMRvsSR:Blochxy} and~\eqref{eqn:SM:NMRvsSR:Blochz} preserve the length of the magnetization vector (for $T_1, T_2 \to \infty$) and, thus, describe a \emph{rotation} of a \emph{pure state} on the surface of the collective Bloch sphere. 
Superradiant decay, however, creates a highly \emph{mixed} state at transient times, whose spin vector is in the interior of the Bloch sphere. 
In order to describe the transverse amplitude dynamics of superradiant decay, one must use at least a MFT approach, which takes quantum correlations into account.

In conclusion, while radiation damping in NMR has some similarity with a semiclassical analysis of superradiance, it leads to a completely different transient dynamics of the $x$-$y$ magnetization.
Therefore, it cannot be used to implement our spin-amplification scheme. 
While radiation damping dynamics could be used to infer a sufficiently strong initial transverse magnetization in NMR more efficiently from the peak delay time than from a direct measurement \cite{Walls2007} it cannot be used in the context of quantum metrology.
In contrast, our scheme is compatible with a standard quantum-metrology Ramsey sequence and allows one to approach the SQL even in the presence of extremely large readout noise, which is an aspect that has not been analyzed in the context of NMR.

\subsection{OAT amplification with single-spin dissipation using mean-field theory}

In the main text we showed that the performance of the OAT amplification protocol is particularly sensitive to noise, both collective decay (governed by the cavity relaxation rate $\kappa$) as well as single-spin dephasing and single-spin relaxation (governed by the rates $\gamma_{\phi}$ and $\gamma_\mathrm{rel}$ respectively). 
Here, we present additional results obtained using MFT simulations, which explore this effect in more detail.

\begin{figure*}
	\centering
	\subfigure[]{
        \includegraphics[width=0.45\textwidth]{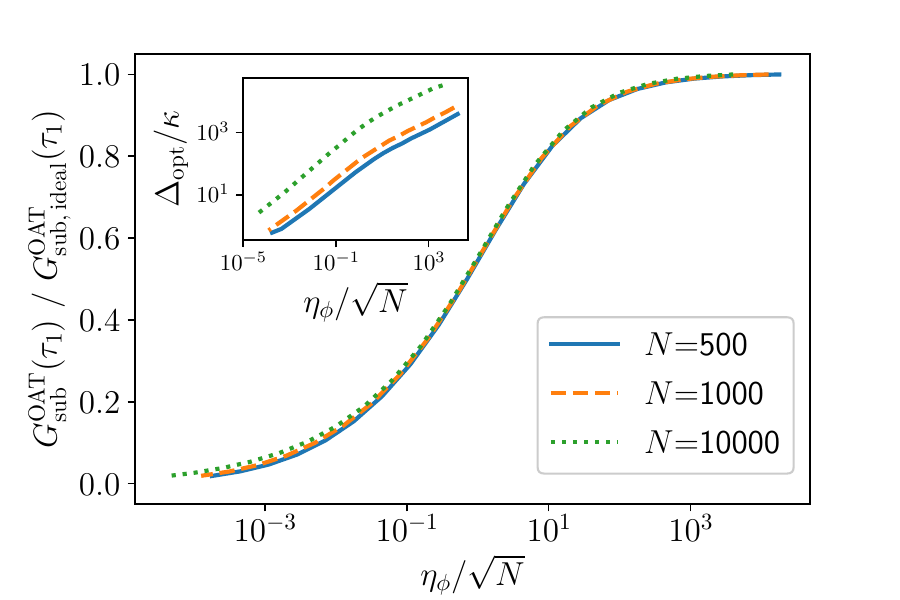}
	}
	\subfigure[]{
        \includegraphics[width=0.45\textwidth]{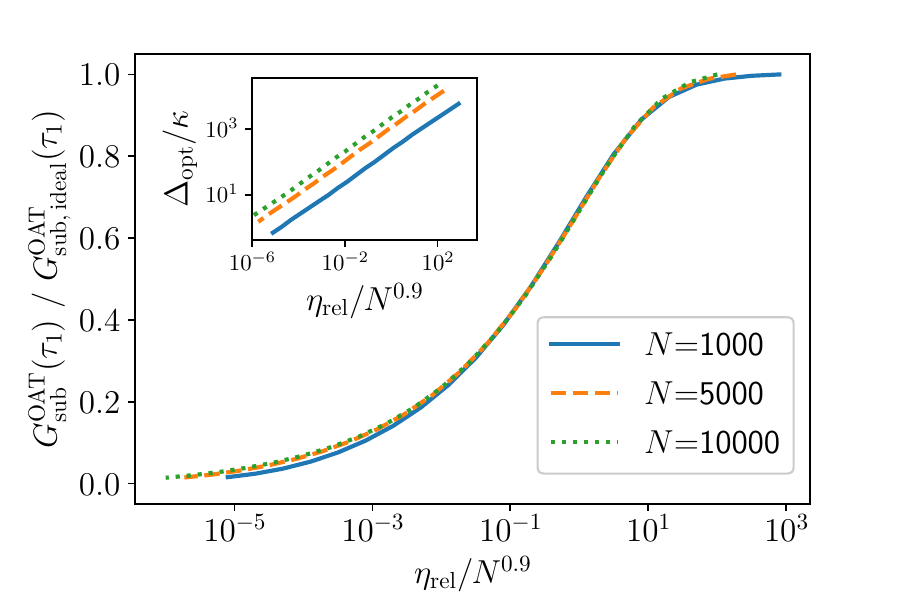}
	}
    \caption{
        Performance of the OAT spin amplification protocol proposed in Ref.~\cite{Davis2016} in the presence of \textbf{(a)} single-spin dephasing and \textbf{(b)} single-spin relaxation. 
        Each plot shows the optimal gain $G_\mathrm{sub}^\mathrm{OAT}(t)$ after background subtraction (evaluated at the time $\tau_1$ of the first peak) as a function of the single-spin cooperativity $\eta_k$ (with $k = \{ \phi, \mathrm{rel} \}$), normalized to the gain calculated in the limit $\eta_k \to \infty$. 
        Simulations were done using MFT for $N=1000$, $5000$, and $10000$ spins, and the spin-cavity detuning was optimized for each value of $\eta_{k}$ (see insets). 
        These plots suggest that, to achieve significant performance, the \emph{single-spin} cooperativity must be substantially large and satisfy $\eta_\phi \gg \sqrt{N}$ and $\eta_\mathrm{rel} \gg N^{0.9}$. 
    }
	\label{fig:nonmarkovianGvseta}
\end{figure*}

As discussed in the main text, collective decay generates a large background that needs to be subtracted to extract the amplified signal. 
Therefore, we consider the gain $G_\mathrm{sub}^\mathrm{OAT}(t)$ defined in Eq.~\eqref{eqn:Discussion:GOATsub} of the main text, which is the signal after background subtraction, $\delta \langle \hat{S}_y(t) \rangle = \langle \hat{S}_y(t,\phi) \rangle - \langle \hat{S}_y(t,0) \rangle$, normalized to the initial signal $\langle \hat{S}_z \rangle = N \phi/2$.
Single-spin decay decreases the gain over time and therefore limits the maximum possible amplification time. 
To determine the maximum gain, one thus has to optimize both the detuning (which determines the ratio between the OAT strength $\chi$ and the collective decay rate $\Gamma$, see main text), and the amplification time $t_\mathrm{max}$.
The result of this optimization is shown in Fig.~\ref{fig:nonmarkovianGvseta}, where we compare $G_\mathrm{sub}^\mathrm{OAT}(t)$ to its ideal value obtained in the limit $\eta_k \to \infty$ (i.e., no local dissipation). 
The gain in this limit is equivalent to the gain in the absence of any dissipation, since the condition $\Delta_\mathrm{opt} \gg \kappa$ holds, i.e., collective dissipation is strongly suppressed. 
Similar to Fig.~\ref{fig:OAT_plus_decay}(a) of the main text, we evaluate the maximum gain at the time $t_\mathrm{max} = \tau_1$ of the first peak of $G_\mathrm{sub}^\mathrm{OAT}(t)$. 
Figures~\ref{fig:nonmarkovianGvseta}(a) and (b) show data for single-spin dephasing and single-spin relaxation, respectively, and the insets show the corresponding values of the optimal spin-cavity detuning $\Delta_\mathrm{opt}$.
In both scenarios, very high \emph{single-spin} cooperativities are required to result significant gain: one needs $\eta_{\phi} \gg \sqrt{N}$ and $\eta_\mathrm{rel} \gg N^{0.9}$. 

\subsection{More general model of the readout process}
\label{sec:SM:ReadoutProcess}

In this section, we provide an alternative derivation of the readout model.
In contrast to the discussion in App.~\ref{sec:App:Sensitivity} of the  main text, we consider a more generic situation where the overall measurement result $n \equiv \sum_{j=1}^N n_j$ is the sum of $N$ independent random measurement results $n_j$ which are collected in parallel from each spin. 
The probability distribution $\mathcal{P}_{\sigma_j}(n_j)$ of each individual result $n_j$ depends on the quantum state of the corresponding spin $j$ in the measurement basis $\vert \sigma_j \rangle$.
In contrast to App.~\ref{sec:App:Sensitivity}, we leave the measurement basis and the properties of the probability distribution $\mathcal{P}_{\sigma_j}(n_j)$ completely general for now, and we will only specialize the final result to the case of fluorescence readout. 
Moreover, we do not use the language of POVMs, since the addition of readout noise is a classical process and a POVM is not necessarily needed to model it.

For a general pure single-spin state $\vert \psi_j \rangle = \sum_{\sigma_j} c_{\sigma_j} \vert \sigma_j \rangle$, we assume that the measurement result $n_j$ will be distributed according to the weighted sum of the probability distributions of the corresponding basis states, $\mathcal{P}_{\vert \psi_j \rangle}(n_j) = \sum_{\sigma_j} \vert c_{\sigma_j} \vert^2 \mathcal{P}_{\sigma_j}(n_j)$.
Given an ensemble of $N$ spins, the probability distribution of the overall measurement result $n = \sum_{j=1}^N n_j$ for a product state $\vert \sigma_1,\dots,\sigma_N \rangle$ in the measurement basis is the convolution of all single-spin probability distributions, $\mathcal{P}_{\sigma_1,\dots,\sigma_N}(n) = \left( \ast_{j=1}^N \mathcal{P}_{\sigma_j} \right)(n)$.
Similar to the single-spin case, we assume that the probability distribution of $n$ for a general pure $N$-spin state
\begin{align}
	\vert \psi \rangle = \sum_{\{\sigma_j\}} c_{\sigma_1,\dots,\sigma_N} \vert \sigma_1,\dots,\sigma_N \rangle
	\label{eqn:methods:sensitivity:state}
\end{align}
is given by the average
\begin{align}
    \mathcal{P}_{\vert \psi \rangle}(n) = \sum_{\{\sigma_j\}} \vert c_{\sigma_1,\dots,\sigma_N} \vert^2 \mathcal{P}_{\sigma_1,\dots,\sigma_N}(n)~.
    \label{eqn:methods:sensitivity:probabilitydistribution}
\end{align}

We are now interested in the fluctuations of $n$ with respect to the probability distribution~\eqref{eqn:methods:sensitivity:probabilitydistribution}, 
\begin{align}
    (\mathbf{\Delta}n)^2 
    &\equiv \sum_n n^2 \mathcal{P}_{\vert \psi \rangle}(n) - \Big( \sum_n n \mathcal{P}_{\vert \psi \rangle}(n) \Big)^2 \nonumber \\
    &= \sum_n \sum_{\{\sigma_j\}} n^2 \vert c_{\sigma_1,\dots,\sigma_N}\vert^2 \mathcal{P}_{\sigma_1,\dots,\sigma_N}(n) \nonumber \\
    &\phantom{=}\ - \Big( \sum_n \sum_{\{\sigma_j\}} n \vert c_{\sigma_1,\dots,\sigma_N}\vert^2 \mathcal{P}_{\sigma_1,\dots,\sigma_N}(n) \Big)^2  ~.
    \label{eqn:methods:sensitivity:variancenDefinition}
\end{align}
The second line shows that calculating a moment $n^m$ of the probability distribution~\eqref{eqn:methods:sensitivity:probabilitydistribution} involves two different averages: 
First, a $n$-average with respect to the classical conditional probability distribution $\mathcal{P}_{\sigma_1,\dots,\sigma_N}(n)$ describing the readout for a particular spin configuration $\{ \sigma_j \}$ in the measurement basis.
We will denote this average by 
\begin{align}
    \mathbf{E}_{\{\sigma_j\}}[n^m] \equiv \sum_n n^m \mathcal{P}_{\sigma_1,\dots,\sigma_N}(n)~.
\end{align}
Second, an average of the classical expectation values $\mathbf{E}_{\{\sigma_j\}}[n^m]$ with respect to the probabilities $\vert c_{\sigma_1,\dots,\sigma_N}\vert^2$ to obtain a certain spin configuration $\{\sigma_j\}$ in the quantum state~\eqref{eqn:methods:sensitivity:state}.
This is the step where the properties of the quantum state $\vert \psi \rangle$ enter and we will denote this average by
\begin{align}
    \langle f_{\{\sigma_j\}} \rangle_{\vert \psi \rangle} \equiv \sum_{\{\sigma_j\}} \vert c_{\sigma_1,\dots,\sigma_N} \vert^2 f_{\{\sigma_j\}}~,
\end{align}
where $f_{\{\sigma_j\}}$ is a function that depends on the spin configuration $\{\sigma_j\}$.
Note that this does not look like the typical quantum expectation value of an observable with respect to the quantum state $\vert \psi \rangle$. 
However, for a specific readout model, the moments of the probability distribution $\mathcal{P}_{\sigma_1,\dots,\sigma_N}(n)$ will be related to moments of an observable of the quantums state: for instance, in the case of fluorescence readout discussed below, this will be the spin component $\hat{S}_z$ [see also Eq.~\eqref{eqn:Methods:AveragePhotonNumber} of the main text]. 
Therefore, the expectation value $\langle f_{\{\sigma_j\}} \rangle_{\vert \psi \rangle}$ will turn into a familiar quantum expectation value.

With these definitions at hand, the variance of $n$ given by Eq.~\eqref{eqn:methods:sensitivity:variancenDefinition} can be rewritten as follows:
\begin{align}
	(\mathbf{\Delta}n)^2 
	&= \langle \mathbf{E}_{\{\sigma_j\}}[n^2] \rangle_{\vert\psi\rangle}  - \langle \mathbf{E}_{\{\sigma_j\}}[n] \rangle^2_{\vert\psi\rangle} \nonumber \\
	&= \Big( \langle \mathbf{E}_{\{\sigma_j\}}[n^2]\rangle_{\vert\psi\rangle} - \langle \mathbf{E}_{\{\sigma_j\}}[n]^2  \rangle_{\vert\psi\rangle} \Big) \nonumber \\
	&\phantom{=}\ + \Big( \langle \mathbf{E}_{\{\sigma_j\}}[n]^2 \rangle_{\vert\psi\rangle}  - \langle \mathbf{E}_{\{\sigma_j\}}[n] \rangle^2_{\vert\psi\rangle}  \Big)~,
	\label{eqn:Methods:Sensitivity:DeltanSquared}
\end{align}
where we added a zero in the last line.
Similar to Eq.~\eqref{eqn:Methods:DeltanSquared} of the main text, the first term in Eq.~\eqref{eqn:Methods:Sensitivity:DeltanSquared} describes the classical noise which is added by the detector due to the fact that $\mathcal{P}_{\sigma_1,\dots,\sigma_N}(n)$ has a finite variance for each basis state $\vert \sigma_1,\dots,\sigma_N\rangle$. 
The second term represents the variance of $\mathbf{E}_{\{\sigma_j\}}[n]$ due to the intrinsic fluctuations of the state $\vert \psi \rangle$, i.e., its intrinsic spin-projection noise expressed in terms of the measured quantity $n$.

The average measurement result can be expressed as follows:
\begin{align}
    \bar{n} 
    &\equiv \sum_n n \mathcal{P}_{\vert \psi \rangle}(n)
    = \langle \mathbf{E}_{\{\sigma_j\}}[n] \rangle_{\vert \psi \rangle}~.
\end{align}
The change of $\bar{n}$ with respect to the signal $\phi$, $\partial_\phi \bar{n}$, is the transduction factor that we need to refer the measurement error $(\mathbf{\Delta}n)^2$ back to the signal:
\begin{align}
	(\mathbf{\Delta}\phi)^2 
	&= \frac{(\mathbf{\Delta}n)^2}{\vert \partial_\phi \bar{n} \vert^2} \nonumber \\
	&= \frac{\langle \mathbf{E}_{\{\sigma_j\}}[n^2] \rangle_{\vert \psi \rangle} - \langle \mathbf{E}_{\{\sigma_j\}}[n]^2 \rangle_{\vert\psi\rangle}}{\vert \partial_\phi \langle \mathbf{E}_{\{\sigma_j\}}[n] \rangle_{\vert \psi \rangle} \vert^2} \nonumber \\
	&\phantom{=}\ + \frac{\langle \mathbf{E}_{\{\sigma_j\}}[n]^2 \rangle_{\vert\psi\rangle}  - \langle \mathbf{E}_{\{\sigma_j\}}[n] \rangle^2_{\vert\psi\rangle}}{\vert \partial_\phi \langle \mathbf{E}_{\{\sigma_j\}}[n] \rangle_{\vert \psi \rangle} \vert^2}~.
	\label{eqn:methods:sensitivity:generalFormulaDeltaPhiSquared}
\end{align}
Note that both the numerator and denominator in the second term are expressed using only $\mathbf{E}_{\{\sigma_j\}}[n]$, i.e., if $n$ is related to some spin observable $\hat{O}$ by a linear transformation [e.g., $\hat{S}_z$ as shown in Eq.~\eqref{eqn:Methods:AveragePhotonNumber} of the main text], the conversion factors will drop out and the second term will become the bare spin-projection noise with respect to $\hat{O}$.

Finally, we specialize this result to the case of fluorescence readout \cite{Barry2020}. 
In this case, the quantity $n$ denotes the number of detected photons and $\mathcal{P}_{\vert \sigma_j \rangle}(n_j)$ is a Poissonian distribution with mean $n_\mathrm{b}$ ($n_\mathrm{d}$) if spin $j$ is in the bright (dark) state. 
The overall measurement result $n$ also follows a Poissonian distribution with expectation value
\begin{align}
	\mathbf{E}_{\{\sigma_j\}}[n] = N_\mathrm{b} n_\mathrm{b} + N_\mathrm{d} n_\mathrm{d}~,
\end{align}
where $N_\mathrm{b}$ ($N_\mathrm{d}$) denotes the number of spins in the bright (dark) state. 
Using the basis $\vert j,m \rangle$ of simultaneous eigenstates of $\mathbf{\hat{S}}^2$ and $\hat{S}_z$, one can rewrite the state~\eqref{eqn:methods:sensitivity:state} as $\vert \psi \rangle = \sum_j \sum_{m=-j}^j c_m^j \vert j,m \rangle$. 
Assuming the ground state of each spin is the bright state, we then have $N_\mathrm{b} = N/2 - m$ and $N_\mathrm{d} = N/2 + m$ and obtain
\begin{align}
	\langle \mathbf{E}_{\{\sigma_j\}}[n] \rangle_{\vert \psi\rangle} = N n_\mathrm{avg} \left[ 1 - \frac{2}{N} \langle \hat{S}_z \rangle \tilde{C} \right]~,
\end{align}
where $n_\mathrm{avg} = (n_\mathrm{b} + n_\mathrm{d})/2$ is the average number of emitted photons and $\tilde{C} = (n_\mathrm{b} - n_\mathrm{d})/(n_\mathrm{b} + n_\mathrm{d})$ is the contrast between the bright and the dark state  \cite{Taylor2008,Shields2015,Barry2020}.
Evaluating Eq.~\eqref{eqn:methods:sensitivity:generalFormulaDeltaPhiSquared}, we find
\begin{align}
	(\mathbf{\Delta}\phi)^2 
	&= \frac{\frac{N}{4} \frac{1 - 2 \tilde{C} \langle \hat{S}_z \rangle /N}{\tilde{C}^2 n_\mathrm{avg}}}{\vert \partial_\phi \langle \hat{S}_z \rangle \vert^2} + \frac{\langle \hat{S}_z^2 \rangle - \langle \hat{S}_z \rangle^2}{\vert \partial_\phi \langle \hat{S}_z \rangle \vert^2}~.
\end{align}
This is the same result as Eq.~\eqref{eqn:Methods:Sensitivity:DeltaPhiSquared} of the main text. 


\end{document}